\documentclass[twocolumn]{aastex631}
\usepackage{hyperref}
\usepackage{rotating}
\usepackage{amsmath}
\usepackage{relsize}
\usepackage{enumitem}

\newcommand\logten{\ensuremath{\log_{10}}}
\usepackage{bm}
\usepackage{graphicx}
\usepackage[caption=false]{subfig}

\shorttitle{SED Fitting Systematics in the CLUES Sample}
\shortauthors{Mizener et al.}

\received{October 9 2024}
\revised{March 24 2025}
\accepted{June 2 2025}

\begin{document}

\title{CLUES III: Do User Choices Impact The Results of SED Fitting? Tests of ``Off-The-Shelf" Stellar Population and Dust Extinction Models in the CLUES Sample}

\author[0000-0002-0786-7307]{Andrew Mizener}
\affiliation{Department of Astronomy, University of Massachusetts Amherst MA 01003, USA: amizener@umass.edu}

\author[0000-0002-5189-8004]{Daniela Calzetti}
\affiliation{Department of Astronomy, University of Massachusetts Amherst MA 01003, USA}

\author[0000-0002-8192-8091]{Angela Adamo}
\affiliation{Department of Astronomy, Oskar Klein Centre, Stockholm University, AlbaNova University Centre, SE-106 91, Sweden}

\author[0000-0001-8289-3428]{Aida Wofford}
\affiliation{Instituto de Astronomía, Universidad Nacional Autónoma de México, A.P. 106, Ensenada 22800, BC, México}
\affiliation{Department of Astronomy and Astrophysics, University of California, San Diego, 9500 Gilman Drive, La Jolla, CA 92093, USA}

\author[0000-0001-8587-218X]{Matthew J. Hayes}
\affiliation{Department of Astronomy, Oskar Klein Centre, Stockholm University, AlbaNova University Centre, SE-106 91, Sweden}

\author[0000-0002-0302-2577]{John Chisholm}
\affiliation{Department of Astronomy, University of Texas at Austin, 2515 Speedway, Austin, Texas 78712, USA}

\author[0000-0001-6676-3842]{Michele Fumagalli}
\affiliation{Dipartimento di Fisica ``G. Occhialini'', Universit\`a degli Studi di Milano-Bicocca, Piazza della Scienza 3, I-20126 Milano, Italy}
\affiliation{NAF – Osservatorio Astronomico di Trieste, Via G. B. Tiepolo 11, I-34143 Trieste, Italy\label{oats}}

\author[0000-0003-4857-8699]{Svea Hernandez}
\affiliation{AURA for ESA, Space Telescope Science Institute, 3700 San Martin Drive, Baltimore, MD 21218, USA}

\author[0000-0003-1427-2456]{Matteo Maria Messa}
\affiliation{Università degli Studi di Milano, Dipartimento di Fisica, Via Celoria 16, 20133 Milano (MI), Italy}

\author[0000-0002-0806-168X]{Linda J. Smith}
\affiliation{Space Telescope Science Institute, 3700 San Martin Drive, Baltimore, MD 21218, USA}

\author[0000-0001-8068-0891]{Arjan Bik}
\affiliation{Department of Astronomy, Stockholm University, Oscar Klein Centre, AlbaNova University Centre, 106 91 Stockholm, Sweden}

\author[0000-0002-3247-5321]{Kathryn Grasha}
\altaffiliation{ARC DECRA Fellow}
\affiliation{Research School of Astronomy and Astrophysics, Australian National University, Canberra, ACT 2611, Australia}   
\affiliation{ARC Centre of Excellence for All Sky Astrophysics in 3 Dimensions (ASTRO 3D), Australia}   
\affiliation{Visiting Fellow, Harvard-Smithsonian Center for Astrophysics, 60 Garden Street, Cambridge, MA 02138, USA}   

\author{Mattia Sirressi}
\affiliation{Department of Astronomy, Stockholm University, Oscar Klein Centre, AlbaNova University Centre, 106 91 Stockholm, Sweden}

\begin{abstract}
The simple stellar population models produced by stellar population and spectral synthesis (SPS) codes are used as spectral templates in a variety of astrophysical contexts. In this paper, we test the predictions of four commonly used stellar population synthesis codes (YGGDRASIL, BPASS, FSPS, and a modified form of GALAXEV which we call GALAXEVneb) by using them as spectral templates for photometric SED fitting with a sample of 18 young stellar clusters. All clusters have existing HST COS FUV spectroscopy that provide constraints on their ages as well as broadband photometry from HST ACS and WFC3. We use model spectra that account for both nebular and stellar emission, and additionally test four extinction curves at different values of $R_V$. We find that for individual clusters, choice of extinction curve and SPS model can introduce significant scatter into the results of SED fitting. Model choice can introduce scatter of 34.8 Myr in age, a factor of 9.5 in mass, and 0.40mag in extinction. Extinction curve choice can introduce scatter of up to a factor of 32.3 Myr in age, a factor of 10.4 in mass, and 0.41mag in extinction. We caution that because of this scatter, one-to-one comparisons between the properties of individual objects derived using different SED fitting setups may not be meaningful. However, our results also suggest that SPS model and extinction curve choice do not introduce major systematic differences into SED fitting results when the entire cluster population is considered. The distribution of cluster properties for a large enough sample is relatively robust to user choice of SPS code and extinction curve. 

\end{abstract}

\section{Introduction} \label{sec:intro}
One of greatest successes of modern astrophysics is our fairly good understanding of the formation, evolution, and death of stars. Though naturally valuable in its own right, this knowledge is especially crucial to the study of galaxies and the complex stellar populations they contain. In this context, our understanding of stellar evolution is most often leveraged through the techniques of stellar population synthesis (SPS).

\begin{figure*}[htb!]
\includegraphics[width=0.95\textwidth]{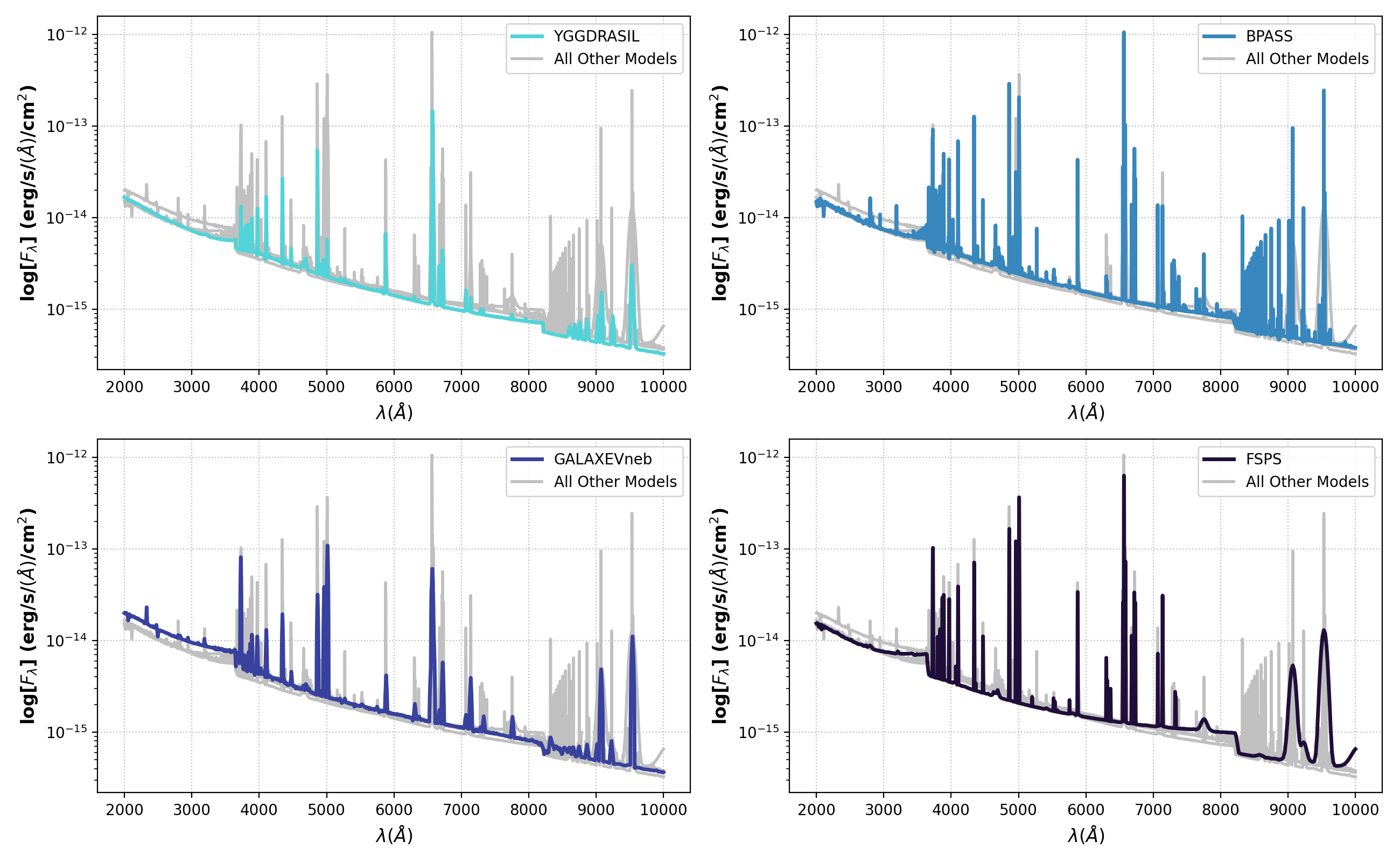}
\caption{Different SPS codes can produce different spectra when evaluated using the same input parameters. This plot shows spectra for each of the SPS models investigated in this work: YGGDRASIL, FSPS, BPASS (binary), and our modified version of GALAXEV (GALAXEVneb). In each panel, the spectrum produced by the code of interest is shown as in color, while the spectra produced by the other codes are shown in grey for comparison. The spectra shown here are evaluated at log(Age) = 6.5, log(Mass) = 6.0, E(V-B)=0.2, and a metallicity of 0.007. Despite the identical input parameters, differences in how each model handles stellar evolution, stellar atmospheres, and nebular emission produce differences in the output spectra. In this work, we test whether these differences are important to the results of broadband SED fitting.
}
\label{fig:example-spectra}
\end{figure*}

Stellar population synthesis in its modern form dates back to work done in 1970s by Beatrice Tinsley and her collaborators. Before Tinsley, attempts to interpret the stellar light emitted from galaxies typically involved constructing a linear combination of individual stellar spectra without physical constraints \citep[known today as the \textit{trial and error} approach; ][]{GALAXEV}. However, this technique suffered from strong degeneracies and was quickly abandoned. Instead, \citet{Tinsley1, Tinsley2, Tinsley3} pioneered the \textit{evolutionary population synthesis} approach, which computes the time evolution of a stellar population's spectrum by combining a stellar initial mass function (IMF) with stellar evolutionary tracks and modeled or observed stellar spectra \citep{Maraston2005}. In the intervening years, a rich library of SPS codes have been produced using some form of the evolutionary population synthesis approach \citep[e.g.][]{Sb99,GALAXEV,Conroy2009,YGGDRASIL,Eldridge2017,Orozco-Duarte2022b} and are in widespread use throughout the literature. 

\begin{figure*}[htb!]
\includegraphics[width=0.95\textwidth]{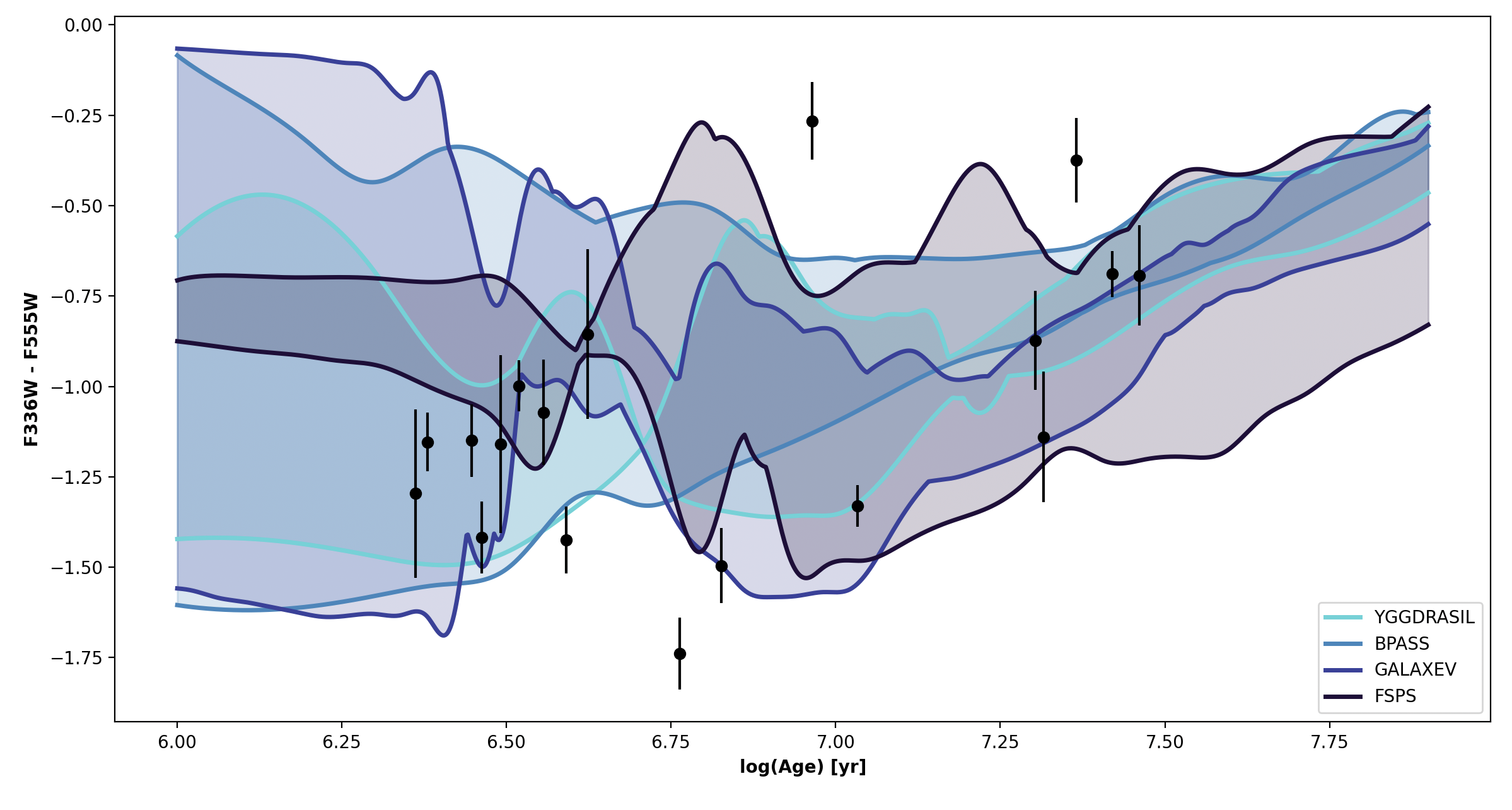}
\caption{ Different SPS models provide different dispersions in color as a function of metallicity and age. In this plot, we show the range in F336W - F555W color accessible by each of the models we tested over the range of metallicity covered by the model. For YGGDRASIL, the metallicity grid ranges from Z = 0.0004 to Z = 0.02. for BPASS, the metallicity grid ranges for Z = 0.001 to Z = 0.040. GALAXEV-neb covers a range from Z = 0.0001 to Z = 0.05. We tested FSPS on a metallicity grid covering the same range used by GALAXEV-neb. We show this metallicity-dependent color evolution as function of spectroscopic age from \citep{Sirressi2022}. We also plot the observed F336 - F555W colors of the YSCs used in this work. If the observed data falls outside of the region in color space accessible to a model, the fit will either adjust age and extinction of the model until appropriate colors are achieved or the fitted model will simply be a poor representation of the data.
}
\label{fig:colors}
\end{figure*}

The fundamental product of stellar population synthesis codes are single stellar populations (SSPs) and their spectra. An SSP represents an instantaneous burst stellar population with homogeneous chemical composition. By convolving SSPs with a star formation history and metallicity distribution, one may produce a model of the spectrum of a composite stellar population - making SSPs notable as a fundamental input to SED fitting and template-based photometric redshift estimation codes \citep{Conroy2013}. Contemporary SED fitting tools like MAGPHYS \citep{daCunha2008}, Prospector \citep{Johnson2021}, CIGALE \citep{Boquien2019}, SLUG \citep{daSilva2012}, and BAGPIPES \citep{Carnall2018} and photometric redshift estimators such as EAZY \citep{Brammer2008} vary in specific approach, but all use a combination of SSP spectra from various SPS models to represent the contribution from stellar light.

It is well known that existing SPS models are imperfect. Phenomena such as stellar rotation and binary effects make massive stars challenging to model. This in turn makes models somewhat more uncertain at young ages where these stars dominate the SED \citep{Walcher2011,Conroy2010,Conroy2013new}. There are also significant modeling challenges at low metallicity, as nearby metal-poor massive stars are rare \citep{Conroy2010,Conroy2013new}. Without sufficient empirical constraints, model grids that are complete in this region of parameter space are difficult to construct.

As SPS models are a fundamental input to the tools used to derive key galaxy parameters like stellar mass and redshift, it is important to verify that they behave in ways that we understand. Differences in how individual models handle e.g. stellar evolutionary physics, stellar atmospheric physics, or nebular emission can cause different codes to predict different spectra for an SSP at the same age, mass, extinction, and metallicity (Figure 1). These differences can produce significant differences in intrinsic color. Further, different models are differently sensitive to metallicity effects. Over a similar range in metallicity and age, different regions of color space are accessible as a function of model (Figure 2). In turn, this means that the SSP choice could in principle systematically bias the derived cluster parameters. 

Although work has been done to understand how these systematics propagate through to final derived galaxy parameters, the picture is not yet totally clear. It is evident that fits to the same data using different SPS models can produce discrepant results \citep[e.g.][]{Muzzin2009,Chen2010,Whitler2023,Tang2024,Wang2024}. However, it remains to be seen under exactly what circumstances these differences are significant, especially when convolved with observational uncertainties. Furthermore, SED fitting within large galaxy samples typically involves inference with sparse broadband photometry over a limited wavelength range. Since the results of such fitting are used infer broader astrophysical trends \citep[e.g.][]{Tachella2022,Gonzalez2023,Adams2023,Pacifici2023,Endsley2023}, it is sensible to investigate if the results obtained are self-consistent regardless of model choice. 

The most straightforward way to do this is through comparison to observations of young stellar clusters, or YSCs. YSCs, which form nearly instantaneously from a single parent cloud \citep{Zwart2010}, are the best physical analogues to SSPs that exist in nature, allowing for very direct comparisons between observations and models. Using YSCs to benchmark SSPs and stellar evolutionary models is a well established technique \citep[e.g. ][]{Renzini1988,Maraston2005,Chen2010,Wofford}.

Young clusters represent an excellent laboratory for benchmarking SPS codes not only because they are young, bright systems, but also because their FUV spectra are easily observed with the Cosmic Origins Spectrograph (COS) on the Hubble Space Telescope (HST). Fits to photospheric and P Cygni-like lines in FUV spectra provide robust constraints on stellar ages \citep[e.g.][]{Wofford2013,Chisholm2019,Sirressi2022}, which allows for the otherwise pernicious age/extinction degeneracy \citep[e.g.][]{Anders2004,Bridzius2008} to be broken. Additionally, using YSCs allows one to bypass the age/IMF degeneracy \citep{Wofford2011}.

It is also important to account for nebular gas, particularly when interpreting observations of young systems where the contribution to the SED by nebular emission can be very significant \citep{Osterbrock,Orozco-Duarte2022b,Wang2024}. The typical approach is to pass model stellar continuum spectra through a photoionization code like CLOUDY \citep{CLOUDY} or MAPPINGS-III \citep{Groves2004}, which predict emitted spectra by calculating the full radiative transfer through a specified cloud composition and geometry. Because such an approach can be computationally expensive, it is also somewhat common to interpolate from tabulated nebular emission predictions as an alternative \citep{Byler2017}. We note that is no one-size-fits-all approach to nebular emission modeling; the important parameters here (e.g. covering fraction, ionization parameter, hydrogen density, gas abundances) may be reasonable over a range of possible values, depending on the nature of the system at hand.

There are two key effects investigated in this paper. First, we wish to quantify the systematic effects (if any) introduced into the results of photometric SED fitting via the SPS choice. In particular, we investigate 4 commonly used SPS codes: YGGDRASIL (which is built around Starburst99) \citep{YGGDRASIL}, BPASS \citep{Eldridge2017}, FSPS \citep{Conroy2009,Conroy2010}, and a custom-modified version of GALAXEV \citep[][2016 version]{GALAXEV}. These codes are associated with publicly available model stellar continuum spectra that are usable ``off the shelf" in many applications; as such, we wish to check that fits performed on broadband phtotometry using different SSPs produce consistent results. We want to test both whether the results produced are self-consistent in a model-to-model sense, as well as whether they are consistent with more reliable cluster properties (predominantly age) obtained via FUV photospheric line fits. We also want to determine what sort of constraints we can place on age, mass, and extinction when only sparse photometry is available.   

Second, and similarly, we wish to investigate whether the adopted dust extinction curve (as parameterized by the specific-to-total extinction, $R_{V}$) has any systematic effect on the results of SED fitting. We use the SSPs produced by these 4 SPS codes, as well as a set of 4 extinction curves spanning a reasonable parameter space in $R_{V}$, as inputs for the SED fitting of 20 YSCs from the CLUES \citep{Sirressi2022} sample.

We describe the CLUES sample and our photometric procedure in Section 2. We describe in detail the models investigated in Section 3. We describe our model post-processing and fitting procedure in Section 4. The results of these fits are shown in Section 5, and we conclude with a discussion of our results in Section 6.

\section{DATA AND PHOTOMETRY} \label{sec:data}

\subsection{The CLUES Project}
This work is part of the larger CLUES (CLusters In the Uv as EngineS; ID 15627, PI: Adamo) program, which has used FUV spectroscopy from HST's COS instrument to investigate feedback from YSCs \citep{Sirressi2022,Sirressi2024}\footnote{ https://archive.stsci.edu/hlsp/clues}. A comprehensive description of the target selection criteria for CLUES may be found in \citet{Sirressi2022}. The CLUES sample consists of 20 young, UV bright star clusters drawn from sources in the LEGUS (Legacy Extragalactic UV Survey; IDs 10402 and 13364, PI: Calzetti) footprint \citep{Calzetti2015}. Clusters in CLUES were selected such that a wide range of host galaxy types (dwarfs, spirals, and interacting systems) were included. Individual clusters were drawn from within this larger sample with apparent UV magnitude brighter than $< 18.0$ mag in F275W, color excess E(B-V) $< 0.3$ mag, and age $< 30$ Myr in preliminary SED fitting. This FUV flux cutoff acts as a rough selection on mass, nominally limiting the sample to clusters above $\sim 10^4 M_{\odot}$. These clusters are sufficiently massive to avoid stochastic sampling of the IMF \citep{Krumholz2015,Orozco-Duarte2022b}, making deterministic models such as those tested here a reasonable assumption. 

We note that because these clusters are selected to be very young, they are reasonable SSP analogues. Though including older globular clusters would allow us to test model performance at larger log(Age), older clusters are also more susceptible to contamination by stellar populations of multiple ages. Older clusters have been used to benchmark the performance of stellar evolutionary and populations synthesis models in the literature \citep[e.g.][]{Marigo2008}, but their inclusion is out of the scope of this work. In this paper, we examine all YSCs in the CLUES sample except for those in NGC 5253. These clusters do not dominate the optical/NIR flux within their respective apertures and so are not appropriate targets for our analysis. 

\subsection{FUV Spectroscopic Fits with COS}
Fits to the FUV spectra for clusters in the CLUES sample were performed by \citet{Sirressi2022}. We defer to their paper for a complete description of their methods but describe them in brief here. They begin by pre-processing the observed UV spectra to match the resolution of the SB99 \citep[][]{Sb99,Leitherer2014} models used in their fits and use the observed C III 1176 $\text{\AA}$ and C III 1247 $\text{\AA}$ line widths to determine the (kinematic doppler) redshift and kinematic line broadening appropriate for each source. They used Geneva high-mass-loss stellar evolutionary tracks \citep{MeynetMassLoss}, a Salpeter IMF between stellar masses 0.1 and 120 $M_\odot$, and no stellar rotation. An instantaneous burst star formation history was assumed.

Parameters estimated through fitting to stellar P-Cygni lines are relatively robust. To summarize from \citet{Chisholm2019}, C IV P Cygni absorption depends on metallicity and the emission depends on age. The strength of the N V P Cygni profile is also very sensitive to stellar age. Old populations have strong C III and Si III photospheric absorption features, while these features are undetected in younger populations. Joint measurement of multiple features in absorption and emission is thus a powerful probe of stellar population properties. Model-related systematics are also worth noting. \citet{Berg2024} uses this method with both Sb99 and BPASS models. They find generally good agreement between the two, with typical age offsets $\sim1.6$ Myr and metallicity offsets around the $20\%$ level. Naturally, calibration of the measured relationship between age or metallicity and P-cygni line properties is only be as good as the calibration of the underlying models.

The fitting procedure follows \citep{Chisholm2019}. They fit out broad Ly$\alpha$ absorption with a Voigt profile and mask out bad regions of each spectrum (e.g. detector gaps) by eye. ISM absorption lines are also masked. Two fits were performed for each cluster: one using a single SSP, and another using two independent SSPs. Two SSPs were necessary to obtain good fits in situations where stellar populations separate from the target cluster contributed significantly to the measured spectrum. For each cluster, the results of each fit were evaluated with the Akaike information criterion estimator in order to determine which run (single or double-population) to adopt. For each cluster, the best-fit values of each component stellar population, as well as a light-weighted average, are reported. A table summarizing the location and host galaxy of each source is provided in \hyperref[tab:CLUESloc]{Table 1}. We also provide the metallicity of each cluster as reported by \citep{Sirressi2022}.

We use the results of these fits as loose baselines for model validation, and use the FUV-derived metallicities in our optical photometric fits. As such, it is reasonable to ask whether a direct comparison can truly be made between cluster parameters inferred via spectroscopic FUV-based fits and those from our optical photometric SED fits. Indeed, \citep{Sirressi2022} found poor agreement between their spectroscopic fits and preliminary photometric fits. In particular, they found that the spectroscopic and photometric ages were in agreement for only about $50\%$ of the clusters, with the performance generally worsening with increasing age. They find metallicities that tend towards solar or supersolar values. In the cases where FUV fits infer a low metallicity, the measured values tend to be in agreement with nebular abundance measurements from the literature.

It is suggested that the general age offset between the two methods, as well as the greater age sensitivity of the FUV fits, are a result of differences in how relevant spectral features evolve with time. Optical photometric colors can be relatively insensitive to age at some epochs (e.g. $>$ 3 Myr) and this technique generally suffers from the well-known age-extinction degeneracy \citep[e.g][]{Fouesneau2012}. By comparison, features in the FUV spectrum such as massive star P-Cygni lines can be strong and are strongly age sensitive. As such, age offsets between a photometric SED fits and FUV spectroscopic fits does not necessarily signal any problem with the models used. Rather, it is a signpost for challenges fundamental to broadband SED fitting. Although changing the underlying models used for the spectral fits would change the best-fit YSC properties, it is unlikely that any choice of model would produce significantly better agreement with the broadband fits.

\subsection{Data}
We obtained archival Hubble Space Telescope ACS and WFC3 data for each of the clusters in \hyperref[tab:CLUESloc]{Table 1} via the Mikulski Archive for Space Telescopes (MAST). All data used in this paper are part of the LEGUS \citep{Calzetti2015} program. We retrieved images of our target clusters in the WFC3 F275W, WFC3 F336W, ACS F435W or WCF3 F438W, ACS F606W or WCF3 F555W, and ACS or WFC3 F814W. We show these filter profiles in Figure 3.

\begin{figure}[htb!]
\includegraphics[width=0.5\textwidth]{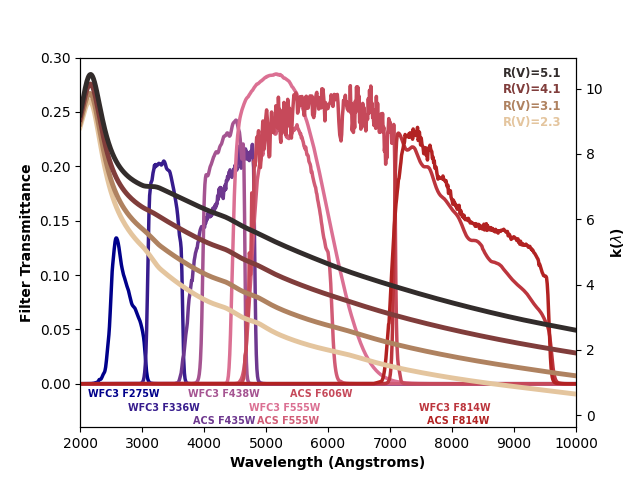}
\caption{The filters and extinction curves used in this work. For each individual cluster, we have photometry in exactly five of the filters shown. }
\label{fig:filters_curves}
\end{figure}

\begin{rotatetable*}
\begin{deluxetable*}{l||ccc|cc|ccc}
\tablecaption{Parent Cluster Sample}
\tablehead{\colhead{Cluster} & \colhead{RA} & \colhead{DEC} & \colhead{Dist.} & \colhead{E(V-B)$_{MW}$} & \colhead{Metallicity} & \colhead{Age (LW)} & \colhead{Mass} & \colhead{Extinction (LW)}\\\colhead{} & \colhead{} & \colhead{} & \colhead{Mpc} & \colhead{Mag} & \colhead{Z} & \colhead{Log[yr]} & \colhead{$10^6$ M$_{\odot}$} & \colhead{Mag}} 
\startdata
NGC1313-1 & 49.597 & -66.478 & 4.602 & 0.096 & ${0.007}_{-0.002}^{+0.000}$ & ${6.826}_{-0.019}^{+0.019}$ & ${0.15}_{-0.02}^{+0.03}$ & ${0.125}_{-0.002}^{+0.002}$\\
NGC1313-2 & 49.659 & -66.492 & 4.602 & 0.096 & ${0.004}_{-0.000}^{+0.008}$	 & ${7.461}_{-0.03}^{+0.03}$ & ${0.06}_{-0.0}^{+0.2}$ & ${0.197}_{-0.004}^{+0.004}$\\
M74-1 & 24.157 & 15.752 & 9.691 & 0.062 & ${0.033}_{-0.001}^{+0.006}$ & ${6.38}_{-0.0}^{+0.0}$ & ${0.07}_{-0.0}^{+0.03}$ & ${0.251}_{-0.002}^{+0.002}$\\
M74-2 & 24.198 & 15.774 & 9.691 & 0.062 & ${0.039}_{-0.001}^{+0.001}$ & ${6.362}_{-0.0}^{+0.0}$ & ${0.04}_{-0.0}^{+0.0}$ & ${0.205}_{-0.005}^{+0.005}$\\
NGC1512-1 & 60.948 & -43.364 & 11.506 & 0.009 & ${0.038}_{-0.002}^{+0.001}$ & ${6.491}_{-0.014}^{+0.014}$ & ${0.03}_{-0.0}^{+0.0}$ & ${0.126}_{-0.007}^{+0.007}$\\
NGC1512-2 & 60.976 & -43.35 & 11.506 & 0.009 & ${0.040}_{-0.003}^{+0.000}$	& ${6.462}_{-0.0}^{+0.0}$ & ${0.03}_{-0.0}^{+0.01}$ & ${0.101}_{-0.002}^{+0.002}$\\
NGC1566-2 & 65.012 & -54.941 & 17.89 & 0.008 & ${0.037}_{-0.001}^{+0.001}$ & ${7.316}_{-0.023}^{+0.023}$ & ${0.25}_{-0.0}^{+0.63}$ & ${0.163}_{-0.004}^{+0.004}$\\
NGC1566-1 & 64.98 & -54.931 & 17.89 & 0.008 & ${0.005}_{-0.000}^{+0.035}$ & ${6.623}_{-0.01}^{+0.01}$ & ${0.8}_{-0.0}^{+0.93}$ & ${0.096}_{-0.002}^{+0.002}$\\
M95-1 & 160.99 & 11.704 & 9.658 & 0.025 & ${0.040}_{-0.000}^{+0.000}$ & ${6.447}_{-0.0}^{+0.0}$ & ${0.72}_{-0.0}^{+0.31}$ & ${0.376}_{-0.002}^{+0.002}$\\
NGC4485-1 & 187.623 & 41.698 & 8.751 & 0.019 & ${0.008}_{-0.002}^{+0.000}$ & ${6.763}_{-0.03}^{+0.03}$ & ${0.05}_{-0.02}^{+0.0}$ & ${0.111}_{-0.002}^{+0.002}$\\
NGC4485-2 & 187.618 & 41.69 & 8.751 & 0.019 & ${0.005}_{-0.000}^{+0.001}$ & ${6.519}_{-0.0}^{+0.0}$ & ${0.01}_{-0.0}^{+0.01}$ & ${0.19}_{-0.002}^{+0.002}$\\
M51-1 & 202.482 & 47.197 & 8.589 & 0.031 & ${0.040}_{-0.004}^{+0.000}$	 & ${6.556}_{-0.012}^{+0.012}$ & ${0.07}_{-0.0}^{+0.05}$ & ${0.242}_{-0.002}^{+0.002}$\\
M51-2 & 202.507 & 47.23 & 8.589 & 0.031 & ${0.039}_{-0.010}^{+0.001}$ & ${7.303}_{-0.037}^{+0.037}$ & ${0.2}_{-0.14}^{+0.0}$ & ${0.22}_{-0.004}^{+0.004}$\\
NGC7793-1 & 359.453 & -32.582 & 3.403 & 0.017 & ${0.028}_{-0.000}^{+0.001}$	 & ${6.591}_{-0.0}^{+0.0}$ & ${0.02}_{-0.0}^{+0.0}$ & ${0.135}_{-0.002}^{+0.002}$\\
NGC7793-2 & 359.4 & -32.595 & 3.403 & 0.017 & ${0.004}_{-0.000}^{+0.000}$ & ${7.42}_{-0.025}^{+0.025}$ & ${0.01}_{-0.0}^{+0.02}$ & ${0.137}_{-0.003}^{+0.003}$\\
NGC4656-2 & 190.986 & 32.171 & 7.876 & 0.012 & ${0.001}_{-0.000}^{+0.000}$ & ${7.033}_{-0.052}^{+0.052}$ & ${0.45}_{-0.35}^{+0.0}$ & ${0.239}_{-0.003}^{+0.003}$\\
NGC4656-1 & 190.99 & 32.17 & 7.876 & 0.012 & ${0.001}_{-0.000}^{+0.000}$ & ${7.365}_{-0.043}^{+0.043}$ & ${0.7}_{-0.26}^{+0.18}$ & ${0.254}_{-0.005}^{+0.005}$\\
NGC4449-1 & 187.046 & 44.094 & 3.889 & 0.017 & ${0.001}_{-0.000}^{+0.000}$ & ${6.964}_{-0.052}^{+0.052}$	 & ${0.31}_{-0.05}^{+0.07}$ & ${0.266}_{-0.003}^{+0.003}$\\
\enddata
\tablecomments{This table shows the fundamental properties of our cluster sample. Ages, masses, extinctions, and metallicities are based on FUV spectroscopy and are taken from \citep{Sirressi2022}. We report the light-weighted age and extinction alongside the mass of the dominant stellar population; for a complete catalog of the spectroscopic fit results, see \citep{Sirressi2022}.}
\end{deluxetable*}
\end{rotatetable*}

\subsubsection{Photometry}
\label{sec:err}
We measured the flux of each source in each available filter via aperture photometry with a 2.5$^{\prime\prime}$ diameter aperture, selected to match the COS aperture. COS suffers from nontrivial vignetting \citep{Bethan2022}; in order to allow us to fairly compare the result of our photometric SED fits with the FUV spectral fits from \citet{Sirressi2022}, this must be taken into account. We thus weight the flux of each pixel within the aperture by the COS vignetting function before taking any measurement \citep{Goudfrooij2010}. Background subtraction was performed via standard sigma-clipping methods with $\sigma=3$ on an annulus with an outer radius of 1.875$^{\prime\prime}$, starting at the outer edge of the aperture. Two example photometric apertures are shown in Figure 4.

\begin{figure*}[htb!]
\subfloat[An aperture containing a relatively isolated cluster.]{%
  \includegraphics[width=\columnwidth]{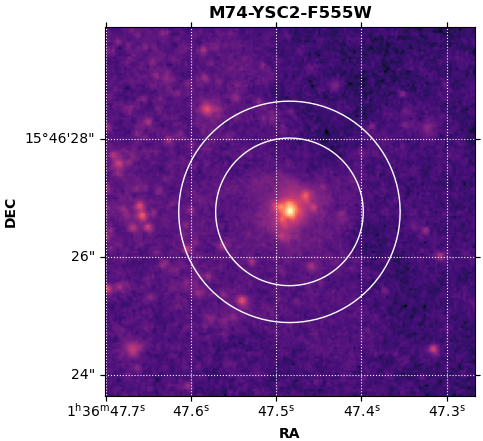}%
}
\subfloat[An aperture containing a significant amount of diffuse emission from the background galaxy.]{%
  \includegraphics[width=\columnwidth]{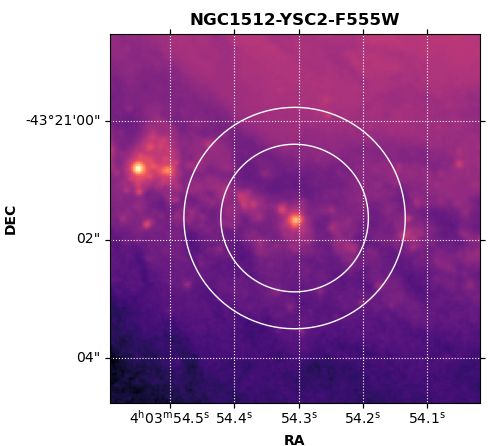}%
}
\caption{We use a large photometric aperture (necessary to match our results with COS FUV spectroscopy), which means that we must be careful to perform an appropriate background subtraction. Two example photometric apertures and annuli are shown, both in WFC3 F555W: one example where the contribution to the total aperture flux from the galaxy background is small (M74-YSC2) and one where the galaxy background is bright and spatially complex (NGC1512-YSC2).}
\end{figure*}

Also important is an estimate of how the (potentially bright and complex) local galaxy background impacts our measured photometry. In order to estimate this contribution, we calculate the sigma-clipped standard deviation in an annulus centered on each cluster with a starting width of 0.5 arcseconds. The width of this annulus is increased by 0.1 arcseconds until the standard deviation converges within $10\%$ for three consecutive iterations. For two sources (M95-1 and NGC1512-2), the size of the converged annulus includes significant surface brightness variation that artificially inflates the measured photometric uncertainty. To bring these estimates down into a more realistic range, we instead assign to these clusters a photometric uncertainty in each band equal to median relative uncertainty in that band averaged over the rest of the sample. Instrumental contributions to the photometric uncertainty are calculated using standard recipes.

Because the vignetting correction we apply to our photometry means that not all pixels in the aperture have equal weight, we correct the uncertainties in each band by the ratio of the vignetted aperture volume to the volume of an aperture with identical size but idealized uniform response \citep[analagous to the JvM correction commonly performed with interferometric data; e.g. ][]{Czekala2021}. We apply the same correction to our background subtraction in order to avoid over-subtraction. A typical correction factor is around $\sim 0.7$. This vignette and corresponding correction is not wavelength-dependent.

In \citet{Sirressi2022}, no background subtraction was performed in the FUV. Instead, their SED fits were performed using a combination of two independent stellar populations. To first approximation, the dominant stellar population of the two can be considered to represent the cluster, with the other taken to represent background young stars. We are unable to perform an analogous two-population fit in this paper, as LEGUS only includes five filters and such an analysis would be associated with six degrees of freedom. The background still exists, however, and must be accounted for. We note that because older stellar populations are more dominant in our optical photometry than they are in the FUV, our backgrounds are likely stronger than those encountered by \citep{Sirressi2022}. In the case of (for example) M95 YSC1, the ratio of background-subtracted to un-subtracted flux in our aperture is 0.96 in F150LP (analogous to the COS spectroscopic range) but just 0.23 in F814W. Our method is not a one-to-one equivalent to the procedure of \citet{Sirressi2022}, but it should at least minimize the impact of background stellar populations on our derived cluster quantities.  

All photometry was corrected for Milky Way foreground extinction following standard procedure \citep[e.g.][]{Calzetti2015b} using extinction maps from \citet{Schlafly2011}. The galaxies in which our clusters are located were bright enough to be masked during the creation of these extinction maps and so the Milky Way extinction estimates should not be biased by there presence. Measurements were performed using the AstroPy \citep{astropy} and PhotUtils \citep{photutils} packages. Typical photometric uncertainties are between $3\%$ and $7\%$ ; however, a number of measurements are significantly more uncertain ($>10\%$ uncertainty). These are generally driven by our unusually large apertures; in these cases, there is generally a gradient or complex structure in the background making the background subtraction highly uncertain (Figure 2, right panel). We remind the reader that we use a large, fixed aperture in order to match the COS aperture. Our measurements are given in Table 2.

\section{MODELS} \label{sec:models}
In order to determine to what extent the choice of spectral synthesis code has an effect on the results of SED fitting, we performed SED fitting using 4 common spectral synthesis models: namely, YGGDRASIL, BPASS, FSPS, and a modified form GALAXEV which we call GALAXEVneb. In the following sections we briefly summarize the approach taken by each spectral synthesis code used.

\subsection{YGGDRASIL}
The YGGDRASIL \citep{YGGDRASIL} spectral synthesis code was originally developed to model the SEDs of Population III galaxies. It is not a complete end-to-end isochrone synthesis code by itself; rather, it is designed to handle arbitrary star formation histories by modifying the SSPs derived from various other population synthesis models. In practice, however, the precomputed Population I SSPs produced based on Starburst99 with Padova evolutionary tracks \citep{Sb99,Vazquez2005} and a Kroupa IMF \citep{Kroupa2001} are commonly used. The mass-loss rates used are described in detail in \citep{Vazquez2005} but are generally a modified form of those produced by \citep{deJagerMassLoss}. The stellar atmospheres of \citep{Smith2002} are used for hot stars and the ATLAS9+Phoenix models are used for stars of spectral type B and below \citep{Lejeune1997,Lejeune1998}. YGGDRASIL has become a relatively common choice of spectral synthesis model for those studying star formation and young stellar populations \citep[e.g.][]{Linden2023}, as it provides easily accessible SSPs including nebular emission predictions at a range of metallicites and gas covering fractions. 

The contribution of nebular emission to the output SED is computed following the approach of \citep{Zackrisson2001}. In brief, the input stellar continuum spectrum is passed through the photoionization code CLOUDY \citep{CLOUDY} at each age step. A spherical, ionization-bounded, constant density nebula is assumed. Rather than specifying an ionization parameter $U$ directly, it is instead derived from the state of the input stellar population. The ionizing photon production rate, Q(H), is set by the age of the stellar population and the inner radius of the cloud is set to \[ R_{in} = 100 R_{\odot} (\frac{L}{L_{\odot}})^{1/2}\] A hydrogen density $n(H) = 100$ cm$^{-3}$ is used. Combined, these three parameters are sufficient to specify an ionization parameter. We use the version of YGGDRASIL that assumes a gas covering fraction $f_{COV} = 0.5$. The nebular gas is assumed to be dust-free. For the purposes of this work, we used the standard YGGDRASIL data products described above, based on Starburst99 and a Kroupa IMF with an upper mass limit of 100 $M_\odot$ and a lower mass limit of 0.1 $M_\odot$.
	
\subsection{BPASS}
Most existing population synthesis models use evolutionary tracks / isochrones that follow the evolution of single, isolated stars. Up to 70\% of massive stars will exchange mass with a companion at some point during their evolution \citep{SanaReview}, and this may have important consequences for the shape of stellar population SEDs and their time evolution. 

The Binary Population and Spectral Synthesis (BPASS) \citep{Eldridge2017} suite of binary stellar evolution models and synthetic stellar populations was designed from the ground-up to improve upon existing population synthesis codes by including interacting binary effects. Because most existing isochrones do not include binary effects, the authors of BPASS calculate stellar evolutionary tracks themselves. The core of BPASS is its single-star stellar evolution code, the history of which can be traced back to the 1971 Cambridge STARS code \citep{EggletonCode}. It uses the standard methods of \citet{HenyeyMethod} to solve for the detailed stellar structure using an adaptive numerical mesh and time-step. In all models, the authors apply the stellar wind mass-loss rates of \citep{deJagerMassLoss} unless the star is of spectral type O or B, in which case the mass-loss rates of \citet{VinkMassLoss} are used. All models are evolved from the zero-age main sequence (ZAMS) up until the end of core carbon burning, or neon ignition for the most massive models. 
	
The BPASS binary-star models are identical to single-star models in most respects, except for the fact that the authors allow for additional mass-loss or gain via binary interactions. During the evolution of a binary, only one star is followed in detail at a time. After Roche lobe overflow, the code calculates a mass-loss rate following \citep{HurleyBinaries}. If the mass transfer from a companion exceeds 5\% of a star’s initial mass, it is assumed to either rejuvenate the star or lead to quasi-homogeneous evolution (QHE). Rejuvenation is modeled by resetting the star to the ZAMS and evolving it normally thereafter. QHE is modeled by assuming the relevant stars are fully mixed throughout their main-sequence lifetimes, and is assumed to occur in low-metallicity stars with $M > 20M_\odot$ after mass transfer. Spectral synthesis is performed by combining the aforementioned population synthesis results with the BaSeL v3.1 \citep{WesteraLib} stellar atmosphere library, with a few minor corrections. The fiducial BPASS models use a Kroupa-like IMF \citep{Kroupa1993} with an upper mass limit of 300 $M_\odot$ and a lower mass limit of 0.5 $M_\odot$.

We use the 2.2.1 version of BPASS (2018 release) with nebular emission predictions by \citep{Xiao2018}. When calculating nebular emission, nebulae were assumed to be spherical, ionization-bounded, and at a constant hydrogen density. For our analysis, we use the model grid calculated using hydrogen density log($cm^{-1}$) = 2.3, a 1.0 covering fraction, and an ionization parameter at the Str{\"o}mgen radius of -2. The nebular gas is assumed to be dust-free.

\subsection{GALAXEV}
GALAXEV is a library of evolutionary stellar population synthesis models computed using the isochrone synthesis code of \citep{GALAXEV}. The earliest iteration of these models were notable at the time of their initial release (1993) \citep{BC_old} for providing spectra at $3\AA$ resolution between $3200\AA$ and $9500\AA$. Spectral evolution was also computed at a lower resolution between $91\AA$ and $160\mu m$ across the same grids. 

As these models have been continually updated with improved stellar physics, including detailed modeling of TP-AGB and post-AGB stars, they have remained quite popular and are used widely throughout the literature. For this work, we use the 2016 revision of the GALAXEV SSPs produced with a Kroupa IMF \citep{Kroupa2001} with an upper mass limit of 100 $M_\odot$ and a lower mass limit of 0.1 $M_\odot$, the STELIB stellar spectral library \citep{LeBorgne} and Padova 1994 stellar evolutionary tracks \citep{Bressan1993,Alongi1993,Fagotto1994a,Fagotto1994b}. As the public version of these models do not include nebular emission, we have performed our own nebular emission predictions as described in the next section.

\subsection{FSPS}
The Flexible Stellar Population Synthesis (FSPS) code \citep{Conroy2009,Conroy2010} was originally developed to address perceived deficiencies in how existing SPS models handled uncertainties in stellar evolutionary physics and the IMF. Like most other models used in this paper, it uses the standard isochrone synthesis approach to generate SSPs. The original models used the Padova \citep{Girardi2000} stellar evolutionary tracks and BaSeL \citep{LejeuneBaSeL, WesteraLib} spectral libraries. From the end-user perspective, FSPS is particularly attractive because its outputs are accessible through a Python interface \citep{FSPS_python} and are easily customized. Users can, for example, select from a range of different stellar evolutionary tracks, spectral  libraries, and IMFs. 

They also use a model for nebular emission that computes continuum and line emission for arbitrary stellar populations without requiring the computational expense of a full radiative-transfer treatment like CLOUDY \citep{Byler2017} at run-time. By default, a spherical nebula at the metallicity of the underlying stellar population with ionization parameter -2 and constant hydrogen density 100 $cm^{-1}$ is assumed. They assume an ionizing photon escape fraction of zero (i.e. a covering fraction of 1.0). The only user-tunable parameters are the ionization parameter and gas-phase metallicity. We keep these at their defaults (logU = -2 and a metallicity matched to the underlying stellar population). The nebular gas is assumed to be dust-free. We use the default Kroupa IMF with with an upper mass limit of 120 $M_\odot$ and a lower mass limit of 0.08 $M_\odot$ We also use the MIST \citep{Choi2016} isochrones and MILES \citep{Vazdekis2010} stellar spectral library, which are the FSPS defaults.

Following our general approach of using SPS models as they would be encountered by a typical user, we adopt the default evolutionary tracks \citep[MIST,][]{Choi2016} and default spectral library \citep[MILES,][]{Vazdekis2010} adopted by \citet{FSPS_python}, as well as a Kroupa IMF \citep{Kroupa2001}.
 
\section{MODEL POST-PROCESSING AND PHOTOMETRIC SED FITTING} 

\begin{rotatetable*}
\tablenum{2}
\movetableright=1mm
\begin{deluxetable*}{ccccccccccc}
\tablecaption{Photometry (All Clusters)}
\tablewidth{0pt}
\tablehead{
\colhead{Cluster} &
\colhead{F275W} &
\colhead{$\sigma_{F275W}$} &
\colhead{F336W} &
\colhead{$\sigma_{F336W}$} &
\colhead{F43(5/8)W} &
\colhead{$\sigma_{F43(5/8)W}$} &
\colhead{F(555/606)W} &
\colhead{$\sigma_{F(555/606)W}$} &
\colhead{F814W} &
\colhead{$\sigma_{F814W}$}
}
\startdata
M95-1 & 14.699 (W) & 0.072 (W) & 15.124 (W) & 0.07 (W) & 15.672 (W) & 0.065 (W) & 16.273 (W) & 0.072 (W) & 17.692 (W) & 0.084 (W) \\
NGC4449-1 & 14.857 (W) & 0.083 (W) & 15.115 (W) & 0.075 (W) & 15.121 (A) & 0.069 (A) & 15.38 (A) & 0.076 (A) & 15.792 (A) & 0.049 (A) \\
NGC4485-2 & 16.383 (W) & 0.064 (W) & 16.865 (W) & 0.062 (W) & 17.477 (A) & 0.033 (A) & 17.864 (A606) & 0.032 (A606) & 19.126 (W) & 0.064 (W) \\
NGC4485-1 & 16.134 (W) & 0.057 (W) & 16.81 (W) & 0.06 (W) & 17.522 (A) & 0.044 (A) & 18.55 (A606) & 0.08 (A606) & 19.647 (W) & 0.114 (W) \\
NGC1512-2 & 17.063 (W) & 0.072 (W) & 17.656 (W) & 0.07 (W) & 18.47 (W) & 0.065 (W) & 19.074 (A) & 0.072 (A) & 20.16 (A) & 0.084 (A) \\
NGC1512-1 & 18.069 (W) & 0.185 (W) & 18.666 (W) & 0.206 (W) & 19.274 (W) & 0.245 (W) & 19.825 (A) & 0.135 (A) & 20.936 (A) & 0.269 (A) \\
NGC7793-1 & 15.7 (W) & 0.047 (W) & 16.34 (W) & 0.051 (W) & 17.076 (W) & 0.061 (W) & 17.765 (W) & 0.076 (W) & 19.172 (A) & 0.149 (A) \\
M51-1 & 15.377 (W) & 0.047 (W) & 15.888 (W) & 0.053 (W) & 16.42 (A) & 0.106 (A) & 16.959 (A) & 0.136 (A) & 18.041 (A) & 0.207 (A) \\
NGC7793-2 & 16.38 (W) & 0.062 (W) & 16.842 (W) & 0.057 (W) & 16.947 (W) & 0.043 (W) & 17.531 (W) & 0.029 (W) & 18.342 (A) & 0.03 (A) \\
M51-2 & 16.863 (W) & 0.121 (W) & 17.296 (W) & 0.102 (W) & 17.636 (A) & 0.086 (A) & 18.169 (A) & 0.091 (A) & 19.065 (A) & 0.105 (A) \\
NGC1313-1 & 15.055 (W) & 0.065 (W) & 15.594 (W) & 0.06 (W) & 16.413 (A) & 0.074 (A) & 17.089 (A) & 0.085 (A) & 18.308 (A) & 0.114 (A) \\
NGC1313-2 & 16.566 (W) & 0.113 (W) & 17.043 (W) & 0.098 (W) & 17.291 (A) & 0.08 (A) & 17.736 (A) & 0.099 (A) & 18.284 (A) & 0.067 (A) \\
M74-2 & 17.812 (W) & 0.161 (W) & 18.277 (W) & 0.162 (W) & 19.032 (A) & 0.126 (A) & 19.573 (A) & 0.168 (A) & 20.751 (A) & 0.2 (A) \\
NGC4656-2 & 14.834 (W) & 0.03 (W) & 15.347 (W) & 0.027 (W) & 15.961 (W) & 0.032 (W) & 16.678 (W) & 0.051 (W) & 17.886 (W) & 0.053 (W) \\
NGC1566-1 & 17.152 (W) & 0.139 (W) & 17.652 (W) & 0.152 (W) & 17.953 (W) & 0.169 (W) & 18.507 (W) & 0.179 (W) & 19.346 (W) & 0.253 (W) \\
M74-1 & 16.938 (W) & 0.089 (W) & 17.333 (W) & 0.064 (W) & 17.937 (A) & 0.042 (A) & 18.487 (A) & 0.05 (A) & 19.556 (A) & 0.059 (A) \\
NGC4656-1 & 16.527 (W) & 0.095 (W) & 16.909 (W) & 0.09 (W) & 16.927 (W) & 0.082 (W) & 17.283 (W) & 0.076 (W) & 17.798 (W) & 0.062 (W) \\
NGC1566-2 & 16.649 (W) & 0.091 (W) & 17.226 (W) & 0.099 (W) & 17.763 (W) & 0.132 (W) & 18.366 (W) & 0.151 (W) & 19.537 (W) & 0.253 (W) \\
\enddata
\tablecomments{This table reports the logarithm of the photometric flux in each band \textbf{in STmag}. The (W) or (A) after each entry denotes whether the image used was taken using WFC3 or ACS. For ACS F435W and WFC3 F438W, this letter also denotes which filter is used. We also use this notation to specify 2 clusters (in NGC4485) for which ACS F606W measurements were used instead of WFC3 or ACS F555W.}
\end{deluxetable*}
\end{rotatetable*}  

\subsection{Model Post-Processing}
We cannot use raw stellar continuum spectra in our SED fits directly. We must first add nebular emission to GALAXEV and include dust extinction in all of our model spectra.

\subsubsection{Nebular Emission Predictions for GALAXEV}
\label{sec:neb}

Stars younger than about 10 Myr and more massive than about 15 M$_\odot$ produce significant ionizing flux, which in turn produces nebular continuum and line emission upon contact with the surrounding gas \citep{Osterbrock}. Different ways of handling the nebular emission can produce significantly different predicted fluxes, even in broadband filters \citep{Zackrisson2001,Wofford,Byler2017,Orozco-Duarte2022b,Wang2024}. Metallicity can have a strong impact on nebular emission, as the ionizing output of a stellar population is metallicity-dependent. When using clusters to benchmark SPS model performance, it is common to handle nebular emission uniformly such that only differences in the underlying stellar continuum spectra drive differences in inferred cluster properties \citep[e.g.][]{Wofford}. This is a necessary approach, as it can (in principle) tease out how different assumptions about stellar evolutionary physics affect one's results. However, we do not adopt it in this paper. Instead, we are interested in comparing these models as they exist ``off the shelf", as they would be encountered by a typical end user, in order to understand the systematics introduced by their use. 

This includes systematics introduced by nebular emission predictions, which may not be consistent between models even if the assumptions made by any individual model are physically reasonable. We note that default model parameters are not always used and may not be appropriate for every problem; regardless, testing them remains a valuable exercise. Three of the four SPS codes we examine in this paper come with nebular emission predictions included; however, the GALAXEV models do not. Because we are studying young clusters in this analysis, we must produce nebular emission predictions for these models before proceeding.  

We used CLOUDY \citep{CLOUDY} to produce nebular emission predictions for the GALAXEV SSPs, attempting to roughly compromise between the choices made by the other models we test. The key nebular parameters are a covering fraction of 1.0, hydrogen density of 100 cm$^{-1}$, CLOUDY default HII region abundances, and a default spherical geometry. The stopping criteria were a temperature of 100K or if the log of the ratio of electron to total hydrogen densities fell below -2. An ionization parameter of -2 was used to match the existing predictions for BPASS and FSPS. Though some work suggests a lower ionization parameter around $-3$ is preferred \citep[e.g.][]{Gutkin2016}, particularly when trying to reproduce the properties of high-redshift systems, it has been found that ionization parameter does not have a strong effect on fit quality within the range -2 to -3 \citep{Chisholm2019}. 

We rescaled our CLOUDY output such that the total power of the nebular continuum and line emission was equal to the total power below 912A in the input stellar continuum spectra. This is a simplifying assumption that adopts a 100\% absorption fraction of the ionizing photons. While some ionizing photons may escape the immediate surroundings of a YSC, this fraction is generally $<$50\% \citep{Oey}. The nebular gas is assumed to be dust-free.

\subsubsection{Dust Extinction}
Dust extinction must be accounted for in order to make any meaningful comparison between model and observed SEDs. The shape of the extinction curve along any given line of sight is not known \textit{a priori}; rather, a curve must be adopted as an assumption during SED fitting. Extinction curves shapes are strongly wavelength dependent and vary from sight-line to sight-line; in particular, extinction curves diverge strongly at and blueward of the 2175\AA{} feature, as well as at and redward of the 10 $\mu$m silicate feature. Between these bumps, extinction curves are nearly linear \citep[e.g.][]{Fitzpatrick2007} but vary in slope and overall normalization, which can generally be described in tandem via the V-band selective-to-total extinction: \[R_{V} = A(V) / E(B-V)\] The Milky Way average value of $R_{V}$ = 3.1 is commonly adopted \citep[e.g.][]{Fitz99}, but individual lines of sight vary greatly with $R_{V}$ = 2.5-5.5 typically reported \citep{Massa2020,Gordon2023}. 

In order to explore the effects, if any, different dust extinction curves have on the results of broadband optical SED fitting, we test a set of extinction curves from \citet{Gordon2023} selected to span this parameter space in $R_{V}$. Namely, we test four different values of $R_{V}$: 2.3 (probing the low end of the distribution), 3.1 (matching the Milky Way value), 4.1 (approximately matching the canonical \citep{CalzettiDust} value), and 5.1 (probing the high end of the distribution). We follow the approach of \citep{Calzetti2015} in our handling of dust by adopting a foreground screen of the form \[F(\lambda)_{out} = F(\lambda)_{\text{model}}10^{-0.4E(B-V)k(\lambda)}\] where $k(\lambda)$ is the extinction curve and $E(B-V)$ is the color excess in magnitudes. The four \citet{Gordon2023} extinction curves we test are shown in Figure 2. As provided, these curves were  normalized such that $A(\lambda)/A(V) = 1.0$ regardless of their actual $R_{V}$; we thus multiply these input curves by their $R_V$ such that they have the proper selective-to-total extinction before use. We show these extinction curves in Figure 3.

\subsection{SED Fitting with Monte Carlo Markov Chains}
The probability surfaces associated with SED fitting in YSCs can be complicated and are often multimodal \citep[e.g.][]{Krumholz2015}, so we want to explore the posterior distributions of our fits and any degeneracies between parameters as completely as possible. We thus constrain model parameters (age, mass, E(B-V)) from the photometry with via a Monte Carlo Markov Chain approach with emcee \citep{emcee}. We perform a fit for each unique combination of cluster, SPS model, and extinction curve. Our fitting procedure is as follows:
\begin{enumerate}
  \item Maximum-likelihood estimators require an initial guess to function properly. Because the probability surfaces we are working with can be complex, it is important to have a reasonable initial guess. Thus, we begin by performing a brute-force initial fit. We evaluate over 200-step uniform grids from 0 to 1.0 in E(B-V) and 6.0 to 7.9 in log(Age). We do not have enough data to constrain metallicity, so it is kept constant at the closest model grid point to the cluster`s (FUV measured) metallicity. This is an important limitation of our analysis and means that our uncertainties are probably somewhat underestimated. However, we note that it is a common approach when fitting star clusters with sparse photometry \citep[e.g.][]{Calzetti2015b}. We construct an effective mass grid by scaling model flux in the reddest available filter (ACS or WFC3 F814W) to the observed flux in that filter, and assembling a 300-step uniform grid from 0.5 times the scaled value to 5.0 times the scaled value. We then calculate the $\chi^2$ at each grid point, selecting the solution corresponding to the minimum value. 
  \item Once an initial approximate solution has been identified, we calculate a refined maximum likelihood solution using the SciPy Nelder-Mead solver. Instead of using a pre-calculated grid, we now interpolate over age (using a cubic spline), extinction, and mass such that the solver can move smoothly through the parameter space.  We assume an uninformative uniform prior covering the same parameter space as the previous step. We apply the same overall bounds to this solver that we used to construct the grids in the previous step. For both this maximum likelihood calculation and the MCMC analysis described below, we use the likelihood function:
  \[ \mathcal{L} = -\frac{1}{2}  \mathlarger{\mathlarger{\sum}}_{i=1}^{5}\left(\frac{\logten(\bm{d}_i) - \logten(\bm{m}_i)}{\logten(e) \times (\bm{\sigma}_i / \bm{d}_i)}\right)^2 \] \\ 
  where $\bm{d}_i$ is the measured flux in filter $i$, $\bm{m}_i$ is the model photometry in filter $i$, and $\bm{\sigma}_i$ is the photometric uncertainty in filter $i$, summed over all 5 filters. 
  \item We then use MCMC to explore the posterior using emcee \citep[][]{emcee}. Because multimodal posteriors are a possibility, we use a combination of differential evolution \citep{DEMove} and Snooker differential evolution \citep{snookerMove} moves at 80\% and 20\% probability respectively. We initialize 50 walkers in a tight Gaussian ball around the maximum likelihood solution and use uninformative uniform priors corresponding to the same bounds described in earlier steps. In order to ensure sampling of the equilibrium distribution we run 30,000 steps of burn-in before computing the chain for 7500 steps. This is sufficient to ensure that all chains have run for at least 50 autocorrelation times, as recommended by emcee. We verify the autocorrelation time of each chain hits this benchmark after fitting is complete. We save the entire computed chain (excluding burn-in) for further analysis. 
\end{enumerate}

\begin{figure*}[htb!]
\subfloat[Reduced $\chi^2$ as a function of cluster. The horizontal line denotes $\chi^2/\nu = 10.0$, our cutoff for inclusion.]{%
  \includegraphics[width=\columnwidth]{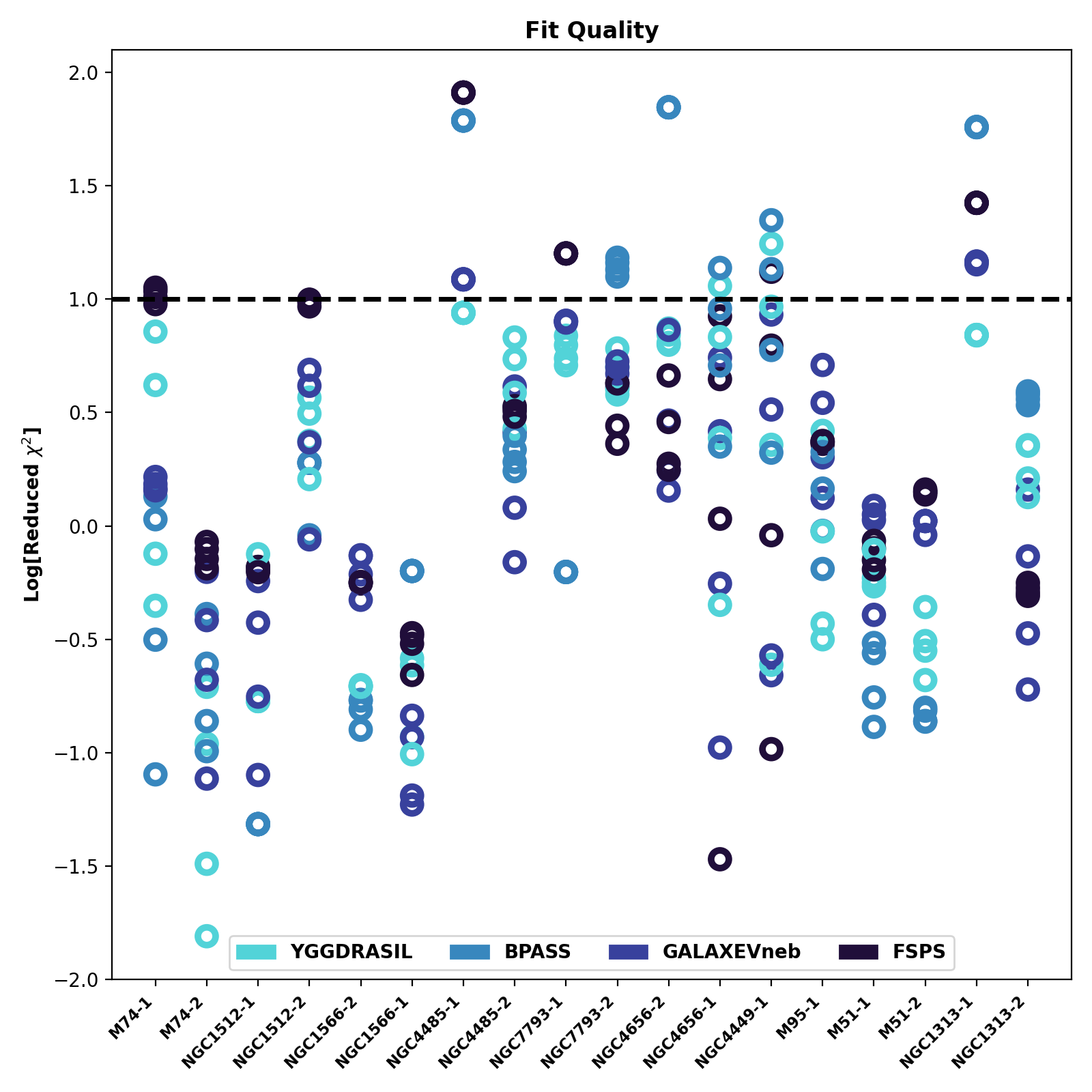}%
}
\subfloat[Reduced $\chi^2$ as a function of metallicity, as measured in the FUV]{%
  \includegraphics[width=\columnwidth]{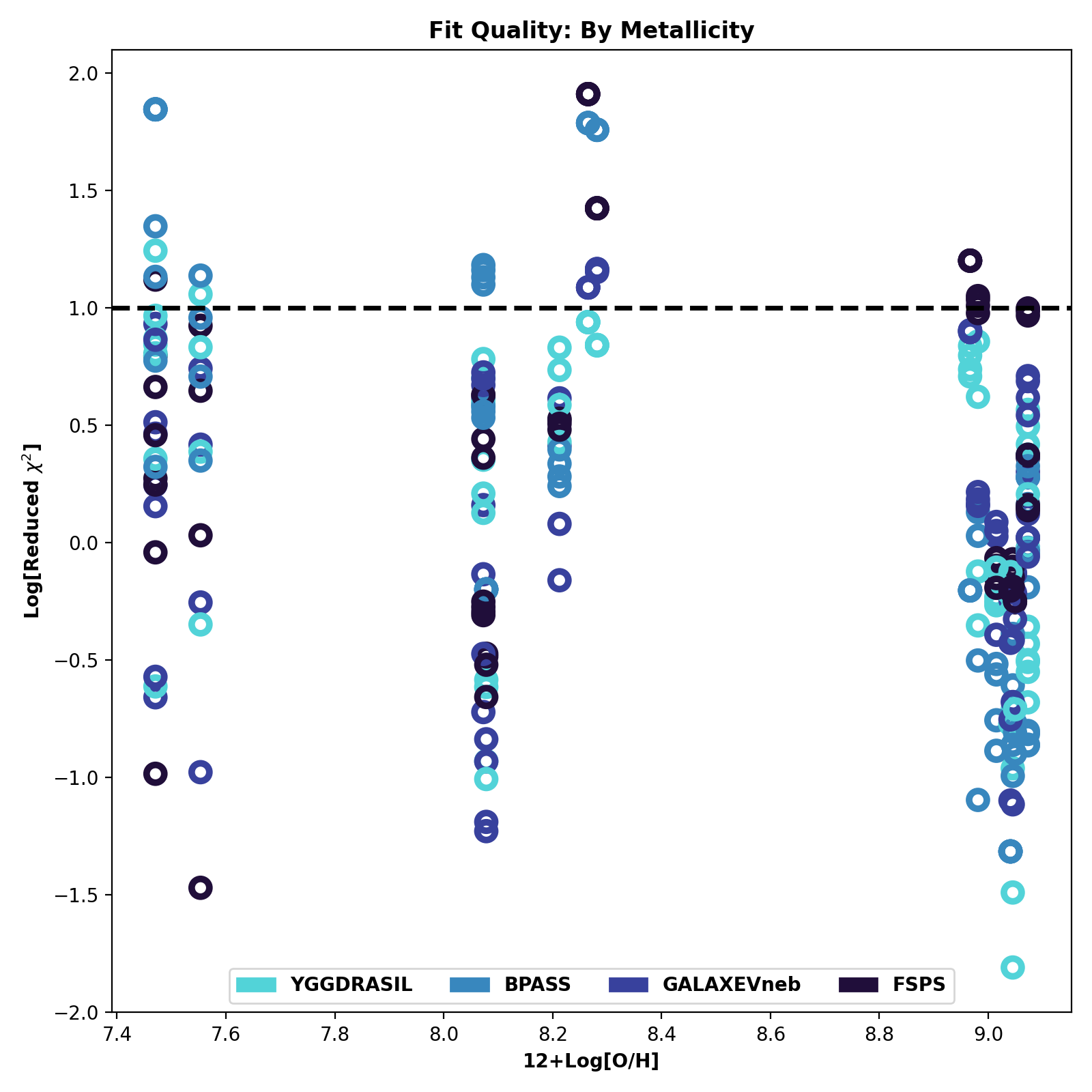}%
}
\caption{This plot shows the $\chi^2$ for each fit of each cluster. 16 fits are performed per cluster, one for each unique combination of SPS model and extinction curve. The left panel shows all fits as a function of cluster; the right panel shows all fits as a function of metallicity. We reject 45 of 288 fits due to unacceptable reduced $\chi^2$. The horizontal dashed line represents the maximum $\chi^2$ we consider acceptable. Poor fits partially a consequence of our large apertures; if an aperture contains multiple bright star clusters, the assumption of a single stellar population breaks down. We also note that all models struggle to reproduce the measured photometry at low metallicity. The metallicities shown here are those measured by \citet{Sirressi2022}, associated with the dominant stellar population. }
\label{fig:chiSq_dist}
\end{figure*}

 Finally, we generate the following products for each unique combination of cluster, SPS  model, and extinction curve:
\begin{itemize}
\item Maximum likelihood solutions for age, E(B-V), and stellar mass (i.e. the ``most probable'' values, following  \citep{Wofford}).
\item The full 3-D posterior distribution of these parameters, given the SPS model and extinction curve.
\item The median (50th percentile) age, E(B-V), and stellar mass in the posterior (i.e. the ``most typical'' values from \citep{Wofford}).
\item The $1\sigma$ upper and lower errors (16th and 84th percentiles) on the median.
\end{itemize}

Note that in this paper, we define stellar mass as the total formed stellar mass rather than the surviving stellar mass at a given age.

\section{RESULTS}
The results of our fits are provided in Appendix A, where we list the most probable and most typical ages, masses, and extinctions for each fit. We also provide the reduced $\chi^2$ as described below.

\subsection{Fit Quality}
Fits with a large reduced $\chi^2$ are not physically meaningful. This occurs when the model is an inappropriate description of the data - in our sample, two phenomena are likely responsible for these poor fits. First, this can occur when multiple bright star clusters contribute meaningfully to the flux within the aperture, breaking the assumption of a single stellar population. Second, and perhaps more importantly, it appears that all models perform poorly at low metallicity (as measured in the FUV). We note that we were unable to attempt fits using multiple stellar populations or free metallicity, as we only have photometry in five bands. We reject all fits with $\chi^2 / \nu > 10$ before proceeding with any further analysis. This threshold allows us to remove the worst fits from further consideration and roughly symmetrizes our $\chi^2$ distribution. We plot the reduced $\chi^2$ of all fits performed in Figure 5.

Out of 288 fits performed, we reject 45 ($\sim 16\%$) on this basis. Dividing these rejected fits by SPS model we find that 19 of the 45 used BPASS, 16 used FSPS, 2 used YGGDRASIL, and 8 used GALAXEVneb. Dividing them by $R_{V}$ we find a more uniform picture, with 15 rejected fits having an $R_{V}$ of 2.3, 11 having an $R_{V}$ of 3.1, 10 having an $R_{V}$ of 4.1, and 9 having an $R_{V}$ of 5.1. We plot all fits performed as a part of this work in Figure 5 as a function of cluster and SPS model. We note that fitting performance is generally poor at lower metallicities. We also note that the rejected fits are not necessarily associated with the clusters with the lowest photometric uncertainty.

We also want to understand how well each model performs as a function of filter. We want to ensure that our fits are assigning each filter approximately equal weight - if, for example, the fits in bluer filters were in agreement with the data significantly more often than the redder filter, there would be cause for concern. To test this, we plot the total number of observations in each band and the number of (acceptable) fits that successfully reproduced the flux in that band as a function of SPS model in Figure 6. We find no strong evidence of color bias - each model performs similarly well across the whole spectral range. We confirm this by checking the reduced chi-square as a function of filter, as described in Appendix B.

\begin{figure*}
    \centering
    \subfloat[Performance of YGGDRASIL as a function of band]{\includegraphics[width=0.49\textwidth]{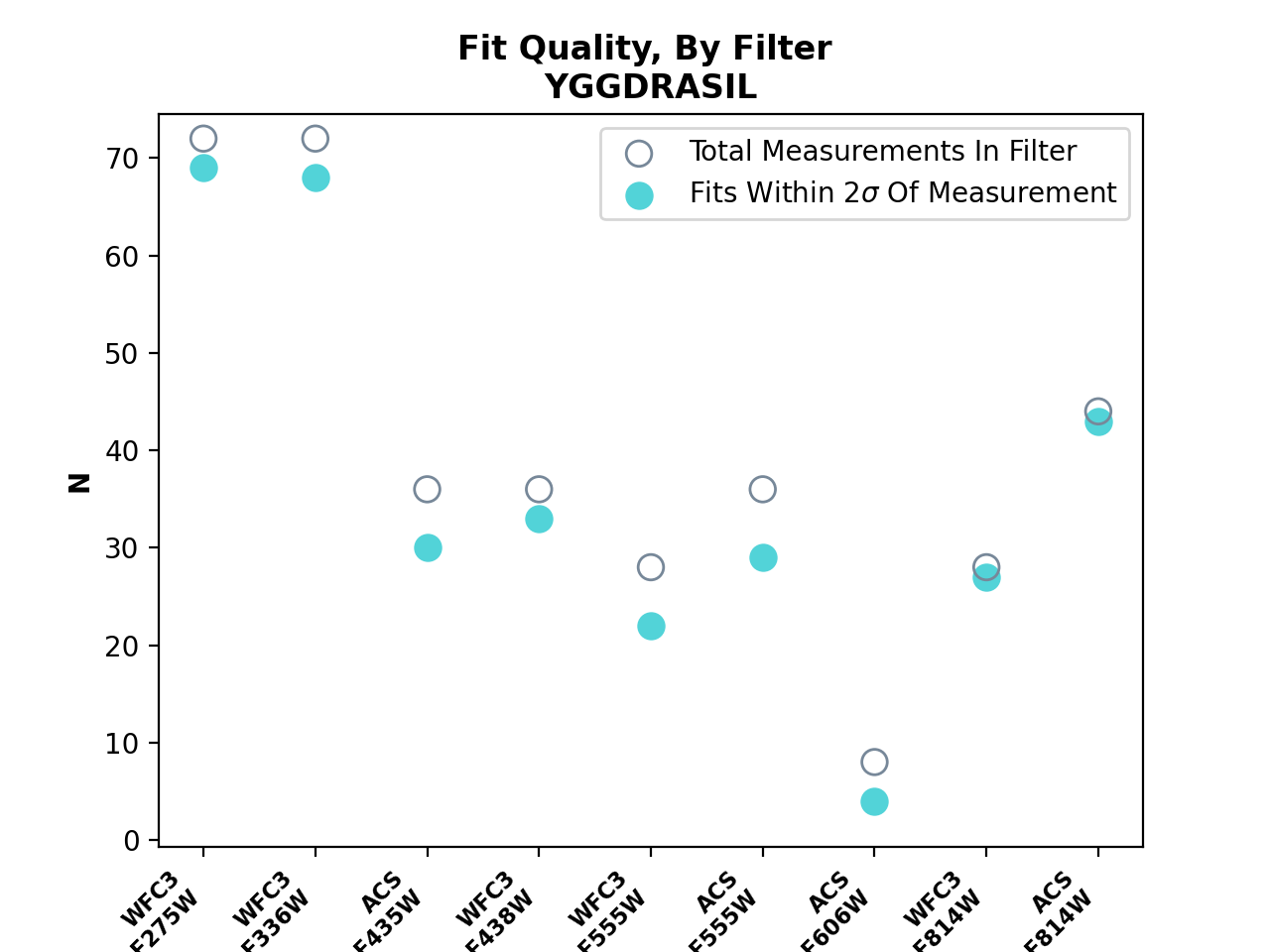}}
    \hfill
    \subfloat[Performance of BPASS as a function of band]{\includegraphics[width=0.49\textwidth]{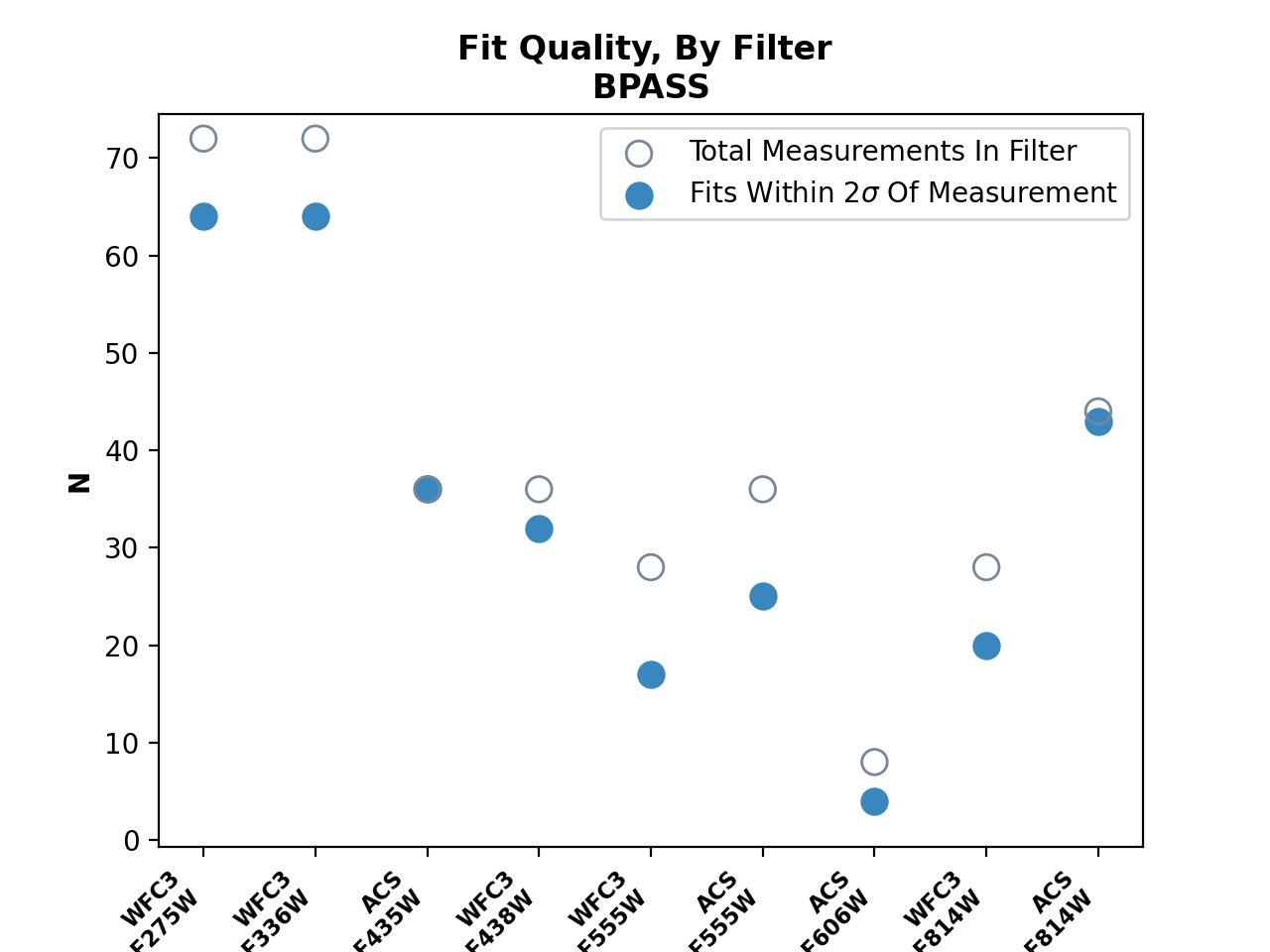}}
    
    \vspace{0.5cm} 
    
    \subfloat[Performance of GALAXEVneb as a function of band]{\includegraphics[width=0.49\textwidth]{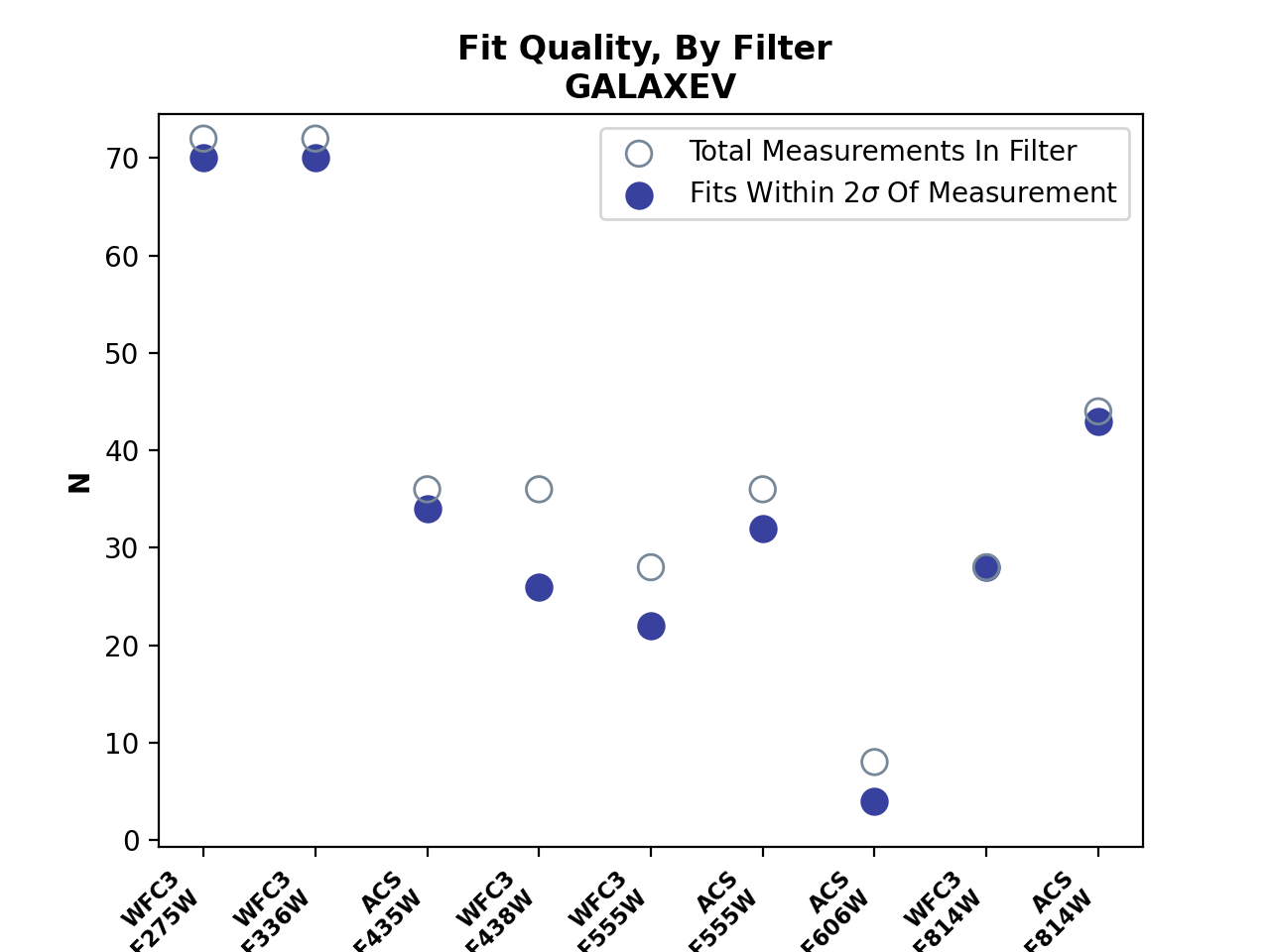}}
    \hfill
    \subfloat[Performance of FSPS as a function of band]{\includegraphics[width=0.49\textwidth]{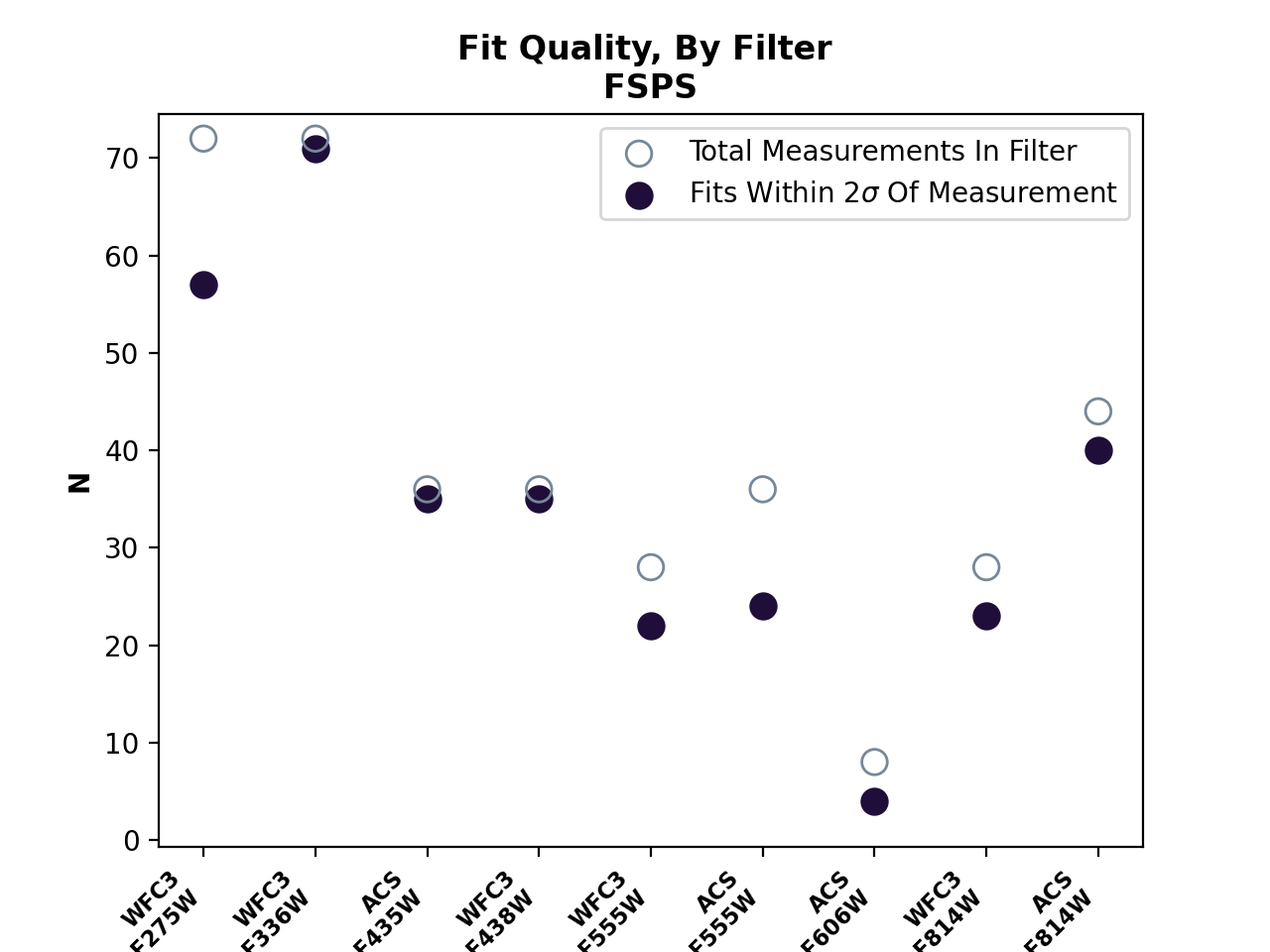}}
    \caption{These plots show, for every accepted fit, the number of attempts to fit each band (open circles) and the number of fits that successfully reproduced the flux in that band (solid circles). All of the models we tested perform similarly as a function of band. That is, we do not find any cases in which (for example) a model performs significantly better in the red part of the spectrum. }
\end{figure*}


\begin{figure*}
    \centering
    \subfloat[CDFs and histograms for all accepted fits in $\chi^2$, by SPS model.]{\includegraphics[width=0.49\textwidth]{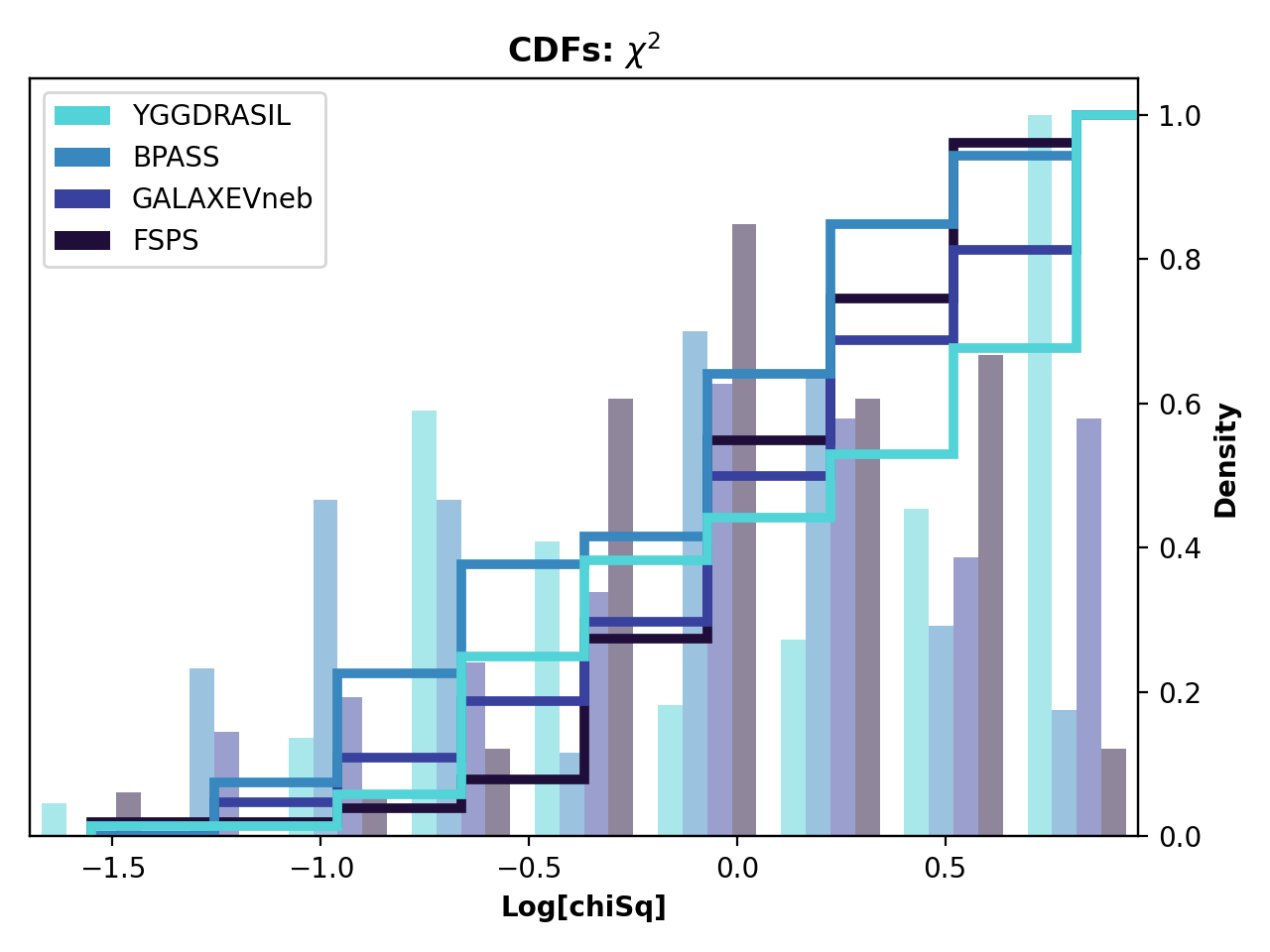}}
    \hfill
    \subfloat[CDFs and histograms for all accepted fits in Log(Age), by SPS model]{\includegraphics[width=0.49\textwidth]{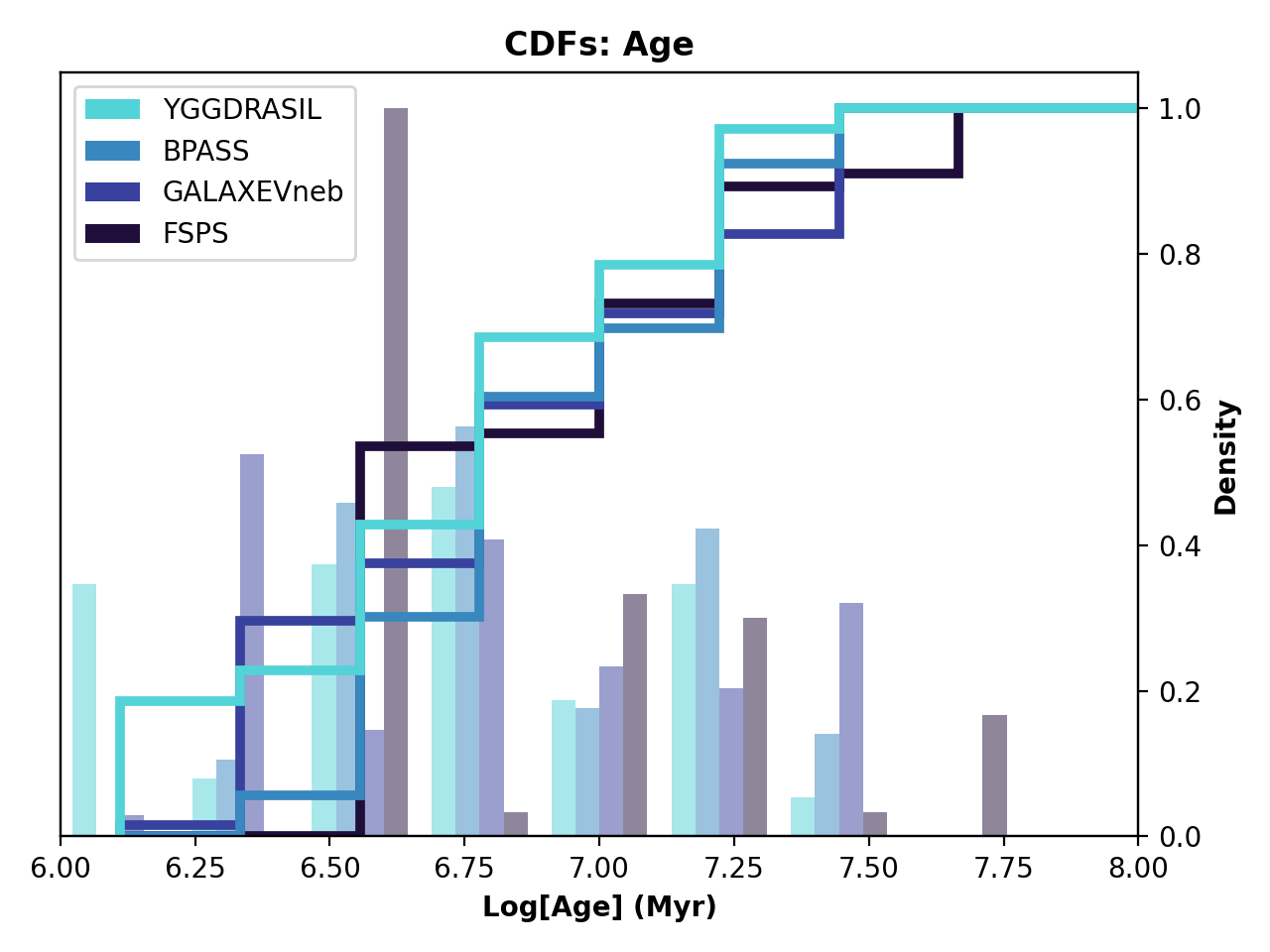}}
    
    \vspace{0.5cm} 
    
    \subfloat[CDFs and histograms for all accepted fits in Log(Mass), by SPS model]{\includegraphics[width=0.49\textwidth]{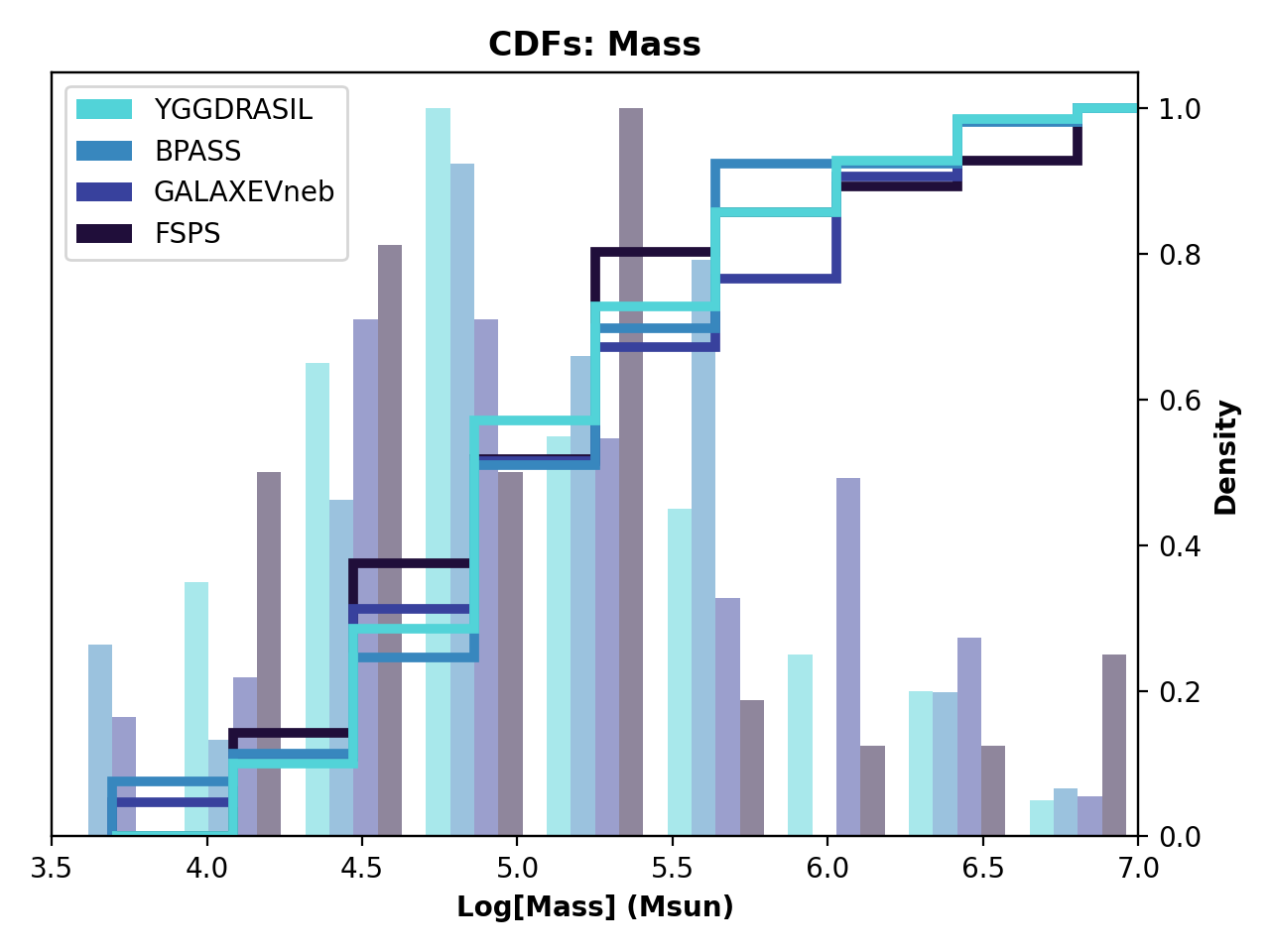}}
    \hfill
    \subfloat[CDFs and histograms for all accepted fits in E(B-V, by SPS model]{\includegraphics[width=0.49\textwidth]{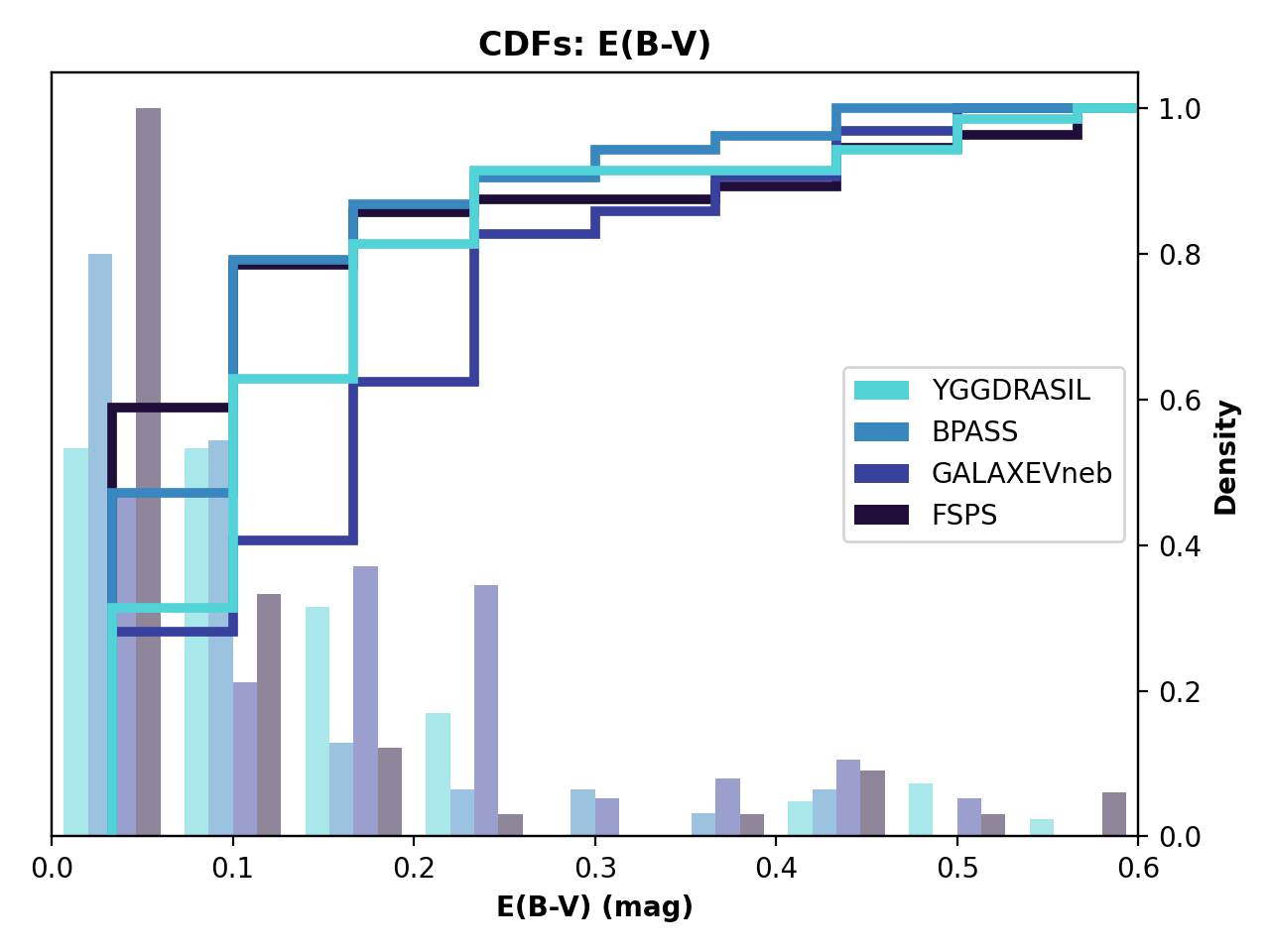}}
    \caption{This plot shows CDFs and histograms of maximum-likelihood fitted parameters, as well as best-fit $\chi^2$ values, divided by SPS model. In many cases, comparing the CDFs of fitted parameters on a model-model basis using the KS test demonstrates that they were drawn from different parent distributions. This suggests that systematic differences between the results produced using different SPS models may exist. From this perspective, we find that age and color excess are most sensitive to model choice, while mass is relatively insensitive.}
\end{figure*}

\subsection{Model-Model Comparison Across All Clusters}
Before analyzing our results on a cluster-by-cluster basis, we first examine our fitted results as a population. We want to test whether our samples of fitted parameters produced using each SPS model (across \textit{all} clusters) are drawn from different parent distributions. If true, this would suggest that there is a systematic offset between the results inferred using any given pair of models - the models would essentially be ``seeing" different cluster populations.  

To test this, we calculate the two-sample Kolmogorov-Smirnov statistic between the samples of maximum-likelihood $\chi^2$, age, mass, and E(V-B) for each unique pair of SPS models. Only fits with a reduced $\chi^2 < 10$ are included in this analysis. We note that the KS test \textit{can only demonstrate inconsistency}; alternatively stated, just because we cannot reject the null hypothesis does not necessarily mean there is no difference between the results of a given model-model or $R_V$-$R_V$ pair. The corresponding p-values are shown in Table 3. We further show the CDFs and histograms for reduced $\chi^2$, age, mass, and E(B-V) as a function of SPS model in Figure 7. In these plots (as well as those in Figure 7), we use bins of uniform size in $\chi^2$, mass, age, and E(B-V). Note that this binning is performed for visualization purposes only; it has no impact on the KS statistic or median properties across the sample of clusters. We reject the null hypothesis wherever $p < 0.05$ (i.e. at the $2\sigma$ level). 

\begin{deluxetable}{c|cccccc}[htb!]
\tablenum{4}
\tablecaption{Kolmogorov–Smirnov P-Values: Model-Model}
\tablehead{Param. & Y-B & Y-G & Y-F & B-G & B-F & G-F } 
\startdata
$\chi^2$ & \textbf{4.61e-4} & 0.149 & 0.053 & 0.39 & \textbf{4.52e-3} & \textbf{0.035} \\
Age & 0.1309 & 0.079 & \textbf{0.006} & \textbf{6.06e-4} & \textbf{0.015} & \textbf{0.027} \\
Mass & 0.213 & 0.507 & 0.360 & 0.123 & 0.113 & 0.294 \\
E(B-V) & \textbf{0.022} & \textbf{0.005} & \textbf{6.99e-4} & \textbf{4.65e-6} & \textbf{0.019} & \textbf{9.93e-5} \\
\enddata
\tablecomments{KS-test p-values comparing on a model-model basis the overall distributions of $\chi^2$, age, mass, and color excess while marginalizing over $R_{V}$. Situations where the null hypothesis is rejected are bolded. In the column headers, `Y' refers to YGGDRASIL, `B' refers to BPASS, `G' refers to GALAXEV-neb, and `F' refers to FSPS.}
\vspace{-20pt}
\end{deluxetable}
It is clear from Figure 7a that some models perform better in $\chi^2$ terms than others, with YGGDRASIL and GALAXEVneb in particular under-performing. BPASS is associated with more fits at low $\chi^2$ than the other models, despite being associated with the greatest number of rejected fits overall. The fact that the KS tests suggest the $\chi^2$ samples are drawn from different parent distributions in 3/6 cases reinforces this conclusion - on average, these models are not equally successful at reproducing the measured photometry.

In many cases, the samples of fitted parameters inferred using different SPS models were drawn from different parent distributions. As demonstrated in Figure 7c, mass is the most robust against choice of SPS model. No model-model pairs have mass samples drawn from different parent distributions. This is true despite the fact that, at young ages, FSPS tends to find lower stellar masses than the other models by a factor of $\sim 0.2$ dex (see section 6.1 and Figure 13). It's also evident that the extinction distributions are quite different (Figure 7d).

\begin{deluxetable}{c|cccc}[]
\tablenum{5}
\tablecaption{Population Median Fitted Parameters: By Model}
\tablehead{Param. & YGGDRASIL & GALAXEVneb & BPASS & FSPS} 
\startdata
Age & 6.72 & 6.72 & 6.74 & 6.59 \\
Mass & 4.91 & 5.03 & 5.04 & 5.04 \\
E(V-B) & 0.12 & 0.17 & 0.08 & 0.04 \\
\enddata
\tablecomments{We find that median fitted parameters within our cluster sample differ somewhat as a function of SPS model. The median age across all clusters varies by up to a factor of $\sim1.4$ on a model-to-model basis, the mass varies by up to a factor of $\sim 3.2$, and E(V-B) can vary by up to 0.15mag. Age and mass are reported in log units. }
\vspace{-20pt}
\end{deluxetable}

By comparison, age and extinction are less robust against model choice. The fitted parameter samples differ significantly in 4/6 possible model-model comparisons for age and for 6/6 in extinction. Although the age samples inferred by BPASS and GALAXEV were drawn from different parent distributions, the median of these age samples are very similar (6.74 and 6.72 respectively). We note that different models have different accumulation points in age; for example, GALAXEVneb finds only a few systems younger than logAge $\simeq$ 6.4 while FSPS finds no systems younger than logAge $\simeq$ 6.55. This is our first hint that SPS model choice can introduce meaningful scatter into the results of SED fitting.  

We also compare the median fitted parameters across all clusters (i.e. the medians of the distributions shown in Figure 7), which are shown in Table 4. We find that model choice can introduce offsets in median age up to 1.6 Myr, in mass up to a factor of $\sim 1.3$, and extinction of up to 0.13 mag. Though these offsets are modest, their existence when combined with evidence of model-model inconsistency from the KS test demonstrates that the choice of SPS model is not necessarily trivial. Using different off-the-shelf SPS models to fit the same objects may produce statistically distinct results. However, as we will discuss in Section 6.1, the model-model differences are more important for individual objects than they are for the sample as a whole. 

\subsection{$R_V$ - $R_V$ Comparison Across All Clusters}

We compare the results inferred using extinction curves with different values of $R_V$ via the same method we used to compare SPS models. We find no evidence that any $R_V$-$R_V$ pair produced results drawn from different parent distributions in terms of fit quality, age, or cluster stellar mass. From the perspective of the KS test, the only parameter that appears sensitive to extinction curve choice is color excess. We find that in 2 out of 6 cases (2.3 vs. 4.1, 2.3 vs. 5.1) the samples of maximum-likelihood color excesses were likely drawn from different parent distributions. As discussed below, these differences can be explained by differences in the median fitted values, which themselves differ by construction due to the selection of different values of $R_V$. The $R_V$-$R_V$ KS p-values are shown in Table 5. We show the CDFs and histograms for $\chi^2$, age, mass, and E(B-V) as a function of extinction curve in Figure 8.

\begin{deluxetable}{c|cccccc}[htb]
\tablenum{6}
\tablecaption{Kolmogorov–Smirnov P-Values: $R_V$-$R_V$}
\tablehead{Param. & 2.3-3.1 & 2.3-4.1 & 2.3-5.1 & 5.1-4.1 & 5.1-3.1 & 3.1-4.1 } 
\startdata
$\chi^2$ & 0.979 & 0.555 & 0.426 & 0.672 & 0.373 & 0.814 \\
Age & 0.999 & 0.978 & 0.688 & 0.938 & 0.650 & 0.988 \\
Mass & 0.926 & 0.398 & 0.203 & 0.848 & 0.376 & 0.875 \\
E(V-B) & 0.388 & \textbf{0.002} & \textbf{0.001} & 0.941 & 0.102 & 0.311 \\
\enddata
\tablecomments{KS-test p-values comparing on a $R_{V}$-$R_{V}$ basis the overall distributions of $\chi^2$, age, mass, and color excess while marginalizing over choice of SPS model. We find that the distributions of fitted parameters are consistent within 3$\sigma$ for each possible pair of $R_V$ values tested.}
\vspace{-20pt}
\end{deluxetable}

As with the model-model comparisons, we compare the medians of the total age, mass, and color excess distributions on a $R_V$-$R_V$ basis in Table 6. We find that in terms of these medians, $R_V$-$R_V$ differences are small as a function of cluster age, around 0.9 Myr. Curve choice can introduce offsets up to a factor of $\sim 1.7$ in mass and 0.06mag in extinction. Inferred mass and extinction consistently rise as one pushes towards higher values of $R_V$, while age tends to decrease. This is not surprising - after all, $R_V$ is just the amount of attenuation in V per magnitude of color change in B-V. If $R_V$ increases, one infers a higher $A_V$ at fixed color change which translates to higher masses and extinctions. 

\begin{deluxetable}{c|cccc}[htb]
\tablenum{7}
\tablecaption{Population Median Fitted Parameters: By Curve}
\tablehead{Param. & $R_V$=2.3 & $R_V$=3.1 & $R_V$=4.1 & $R_V$=5.1} 
\startdata
Age & 6.74 & 6.73 & 6.73 & 6.66 \\
Mass & 4.87 & 4.95 & 5.06 & 5.11 \\
E(V-B) & 0.07 & 0.08 & 0.13 & 0.13 \\
\enddata
\tablecomments{Median age, mass, and E(V-B) over all accepted fits as a function of extinction curve $R_V$. From the perspective of the overall distribution of fitted parameters, curve choice can introduce up to a factor of $\sim 66\%$ in mass and $\sim 3$ in extinction.}
\vspace{-20pt}
\end{deluxetable}

\begin{figure*}
    \centering
    \subfloat[CDFs and histograms for all accepted fits in $\chi^2$, by extinction curve]{\includegraphics[width=0.49\textwidth]{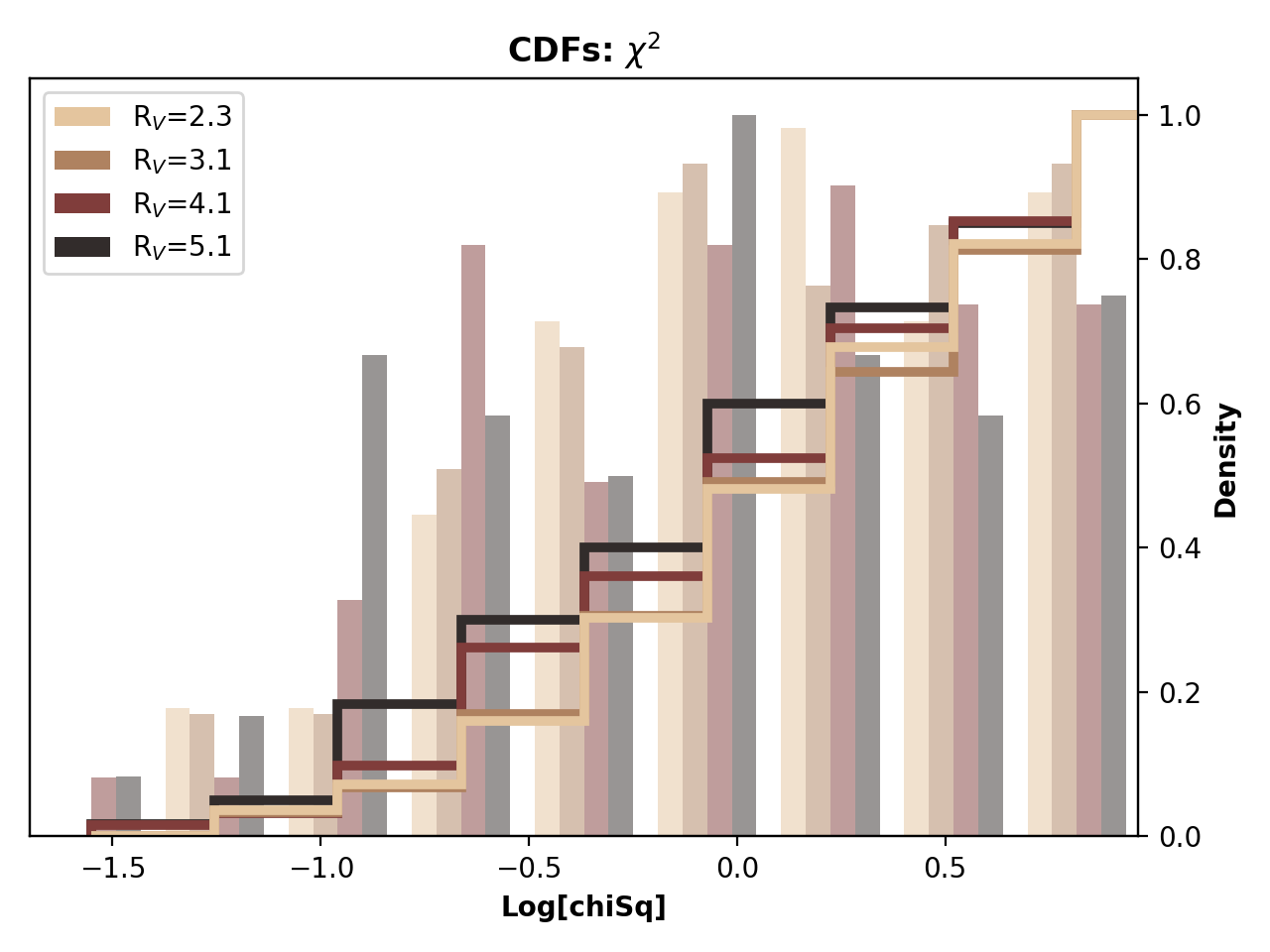}}
    \hfill
    \subfloat[CDFs and histograms for all accepted fits in Log(Age), by extinction curve]{\includegraphics[width=0.49\textwidth]{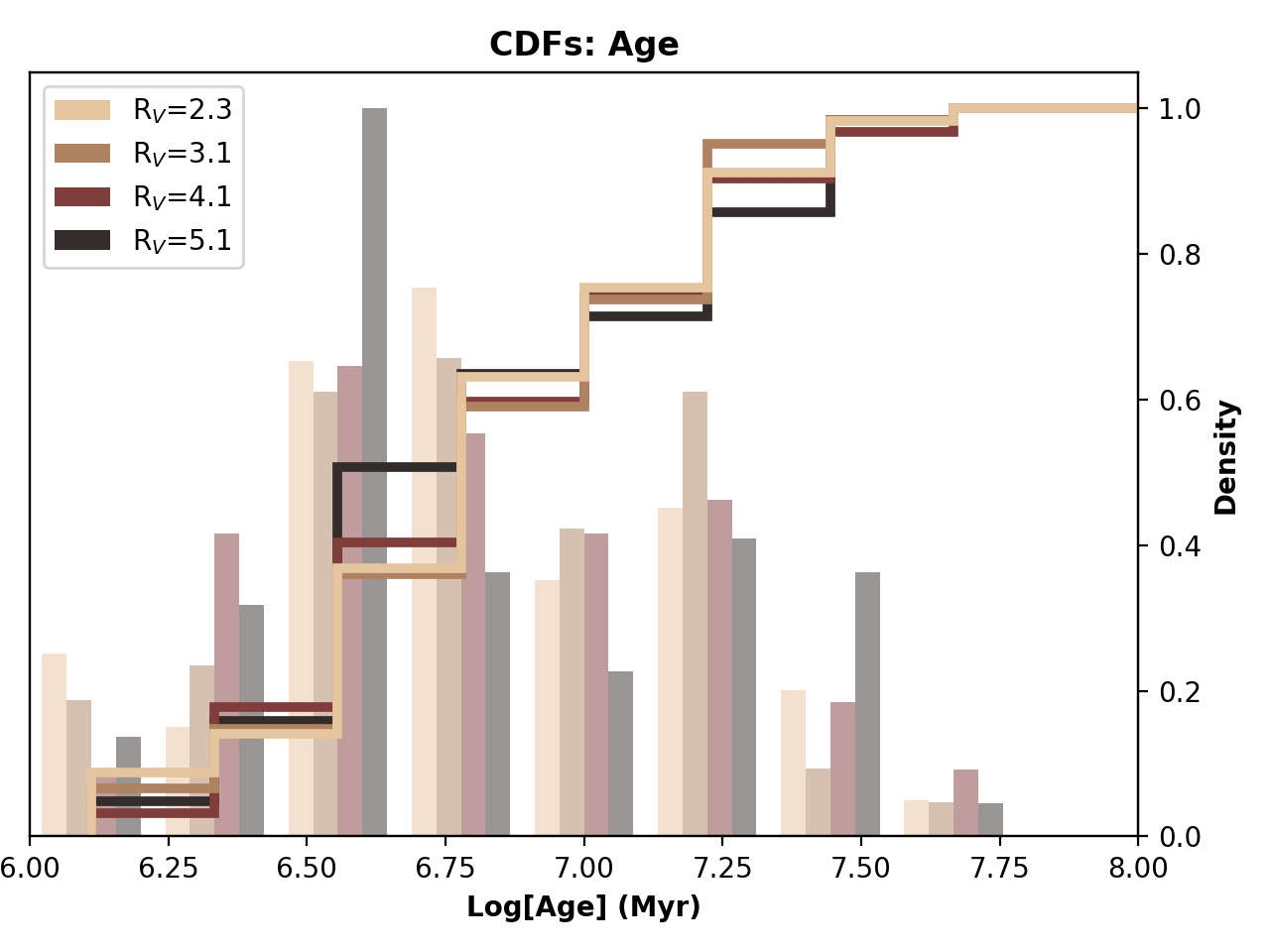}}
    
    \vspace{0.5cm} 
    
    \subfloat[CDFs and histograms for all accepted fits in Log(Mass), by extinction curve]{\includegraphics[width=0.49\textwidth]{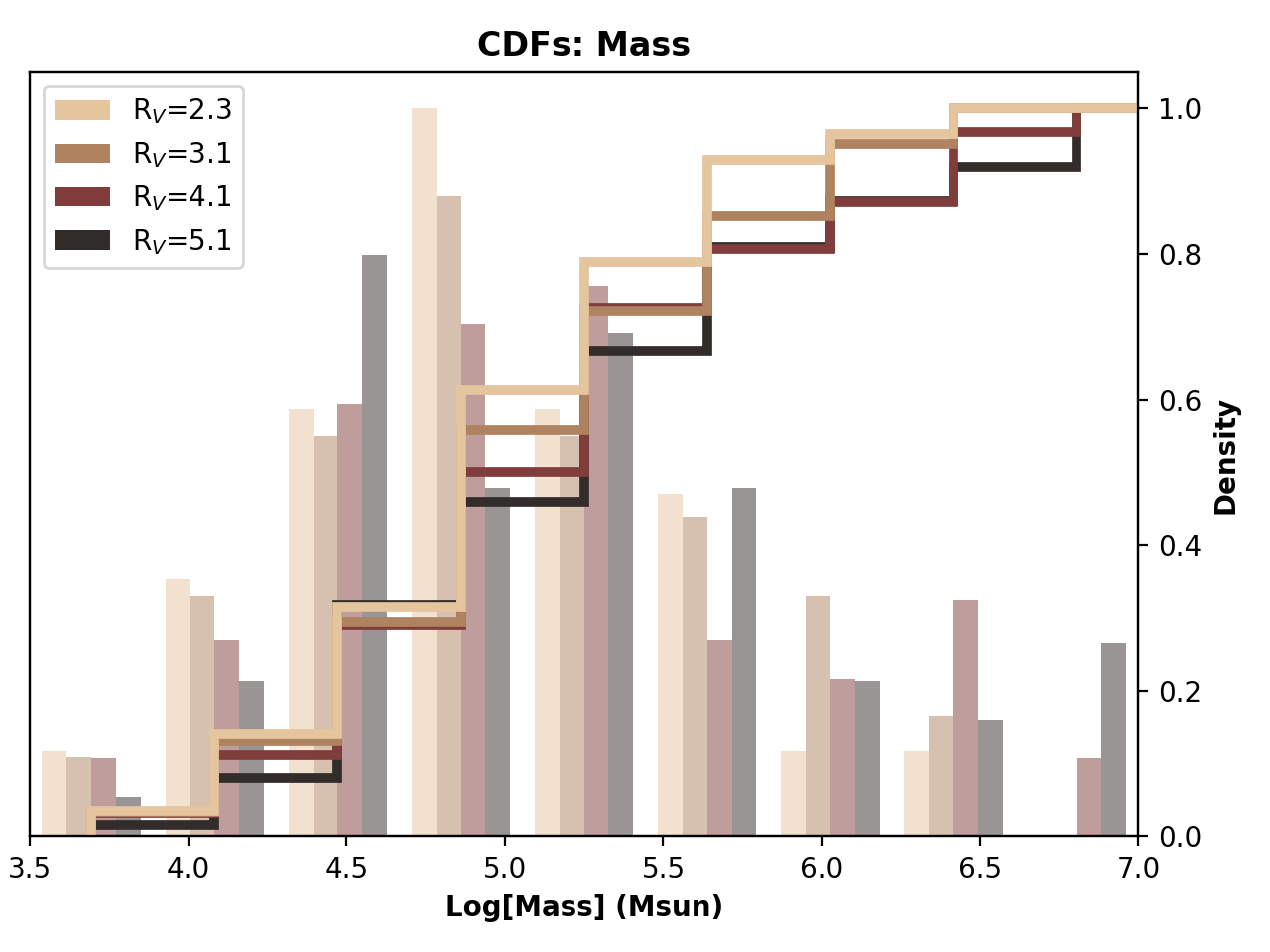}}
    \hfill
    \subfloat[CDFs and histograms for all accepted fits in E(V-B), by extinction curve]{\includegraphics[width=0.49\textwidth]{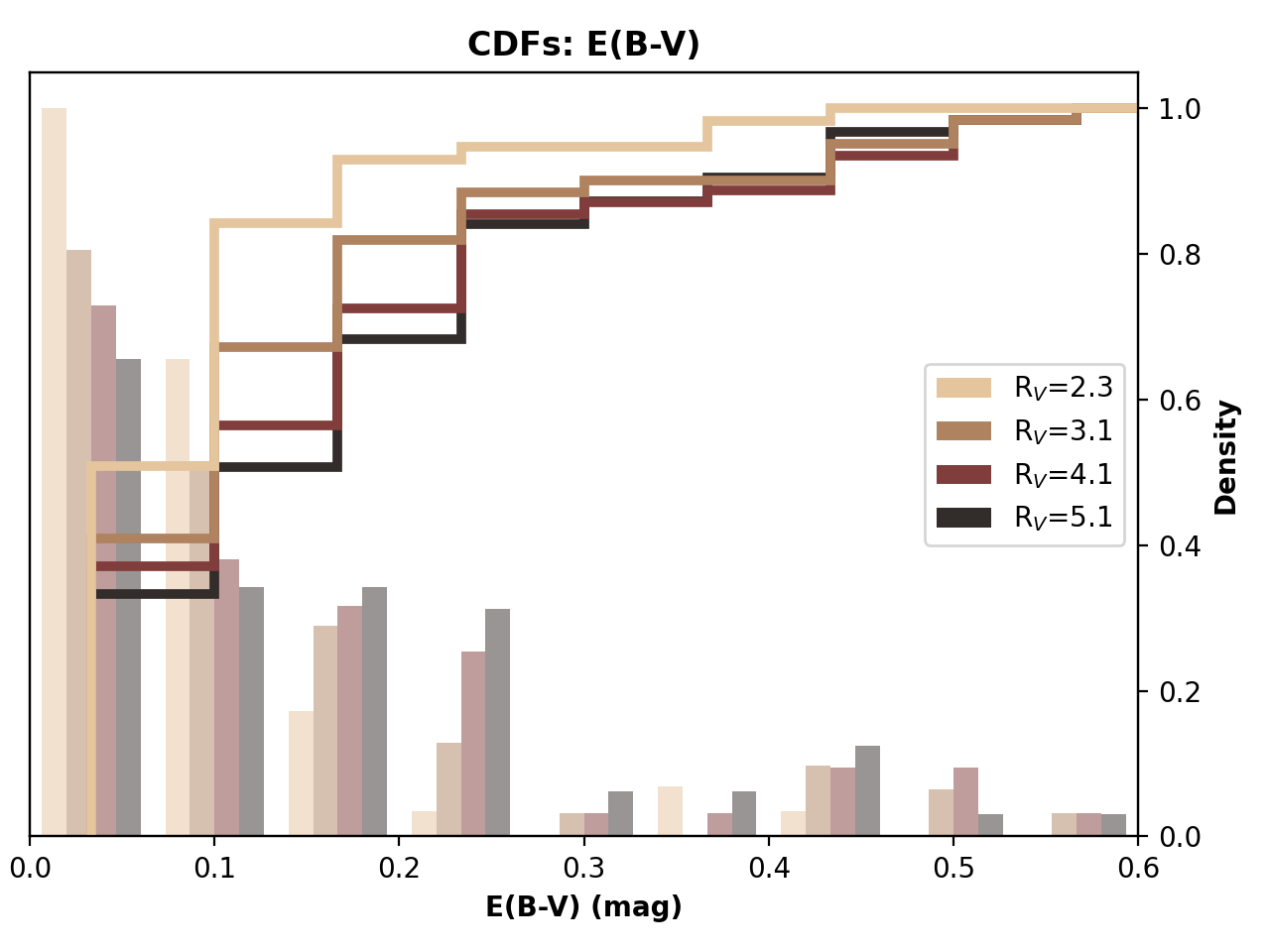}}
    \caption{Unlike SPS model choice, the choice of extinction curve $R_V$ does not appear to strongly affect derived cluster parameters outside of total extinction. Here, we show CDFs and histograms of maximum-likelihood fitted parameters, as well as best-fit $\chi^2$ values, divided by extinction curve. As a function of $R_v$, the overall sample of fitted parameters are not consistent with being drawn from different parent distributions on a $R_V$-$R_V$ basis. From the perspective of the KS-test, the only potentially significant $R_V$-$R_V$ differences are between the distributions of E(B-V).}
\end{figure*}

\subsection{1-D Posterior Distributions}
Although examining the overall distribution of fitted parameters can provide insight into whether model-model and $R_V$-$R_V$ differences exist, it is difficult to understand the nature of these differences without examining these posteriors on a cluster-by-cluster basis. We plot the 1-D posteriors in age, mass, and E(B-V) for each cluster as a function of SPS model and $R_V$. A ``good'' example (NGC4485-YSC2), where there appears to be decent agreement in terms of cluster properties regardless of extinction curve or SPS model used, is shown in Figure 9. A ``bad'' example (M51-YSC1), where there appears to be a significant systematic offset in these properties as a function of $R_V$ and model choice, is shown in Figure 10. In these plots, rows denote model choice, and curve colors denote $R_V$. The black dashed line is the \citet{Sirressi2022} ``light-weighted'' solution, the dotted blue line is the major population in the ``two-population'' solution, and the dotted red line is the minor population in that solution. \citet{Sirressi2022} does not provide a ``light-weighted'' mass solution. In these plots, fits rejected due to an unacceptable reduced $\chi^2$ are shown using a dotted line rather than a solid one.

A few interesting trends are visible in these plots. First, it is clear that the posterior distributions are not always well-described by a Gaussian and are often multimodal. This is reflective of the complex probability surfaces associated with SED fitting. We note that in some cases, as demonstrated in M51-1, \textit{the choice of extinction curve can change which peak of a multimodal distribution is associated with the maximum-likelihood solution.} 

Second, different models can provide constraints on cluster properties of different precision. Take, for example, the age solutions for M74-YSC1 (Figure 11). BPASS and YGGDRASIL constrain the cluster age weakly, with typical one-sided uncertainties across $R_V$ in Log[Age] $\sim 0.1$ to  $\sim 0.4$. GALAXEVneb and FSPS, however, provide much stronger constraints on the order $\sim 0.01$ to  $\sim 0.02$. This should serve as a stark reminder that the uncertainties associated with SED fitting include both propagated observational uncertainty as well as a contribution from the underlying models themselves, representing regions of the parameter space between which a given model cannot confidently distinguish. 

\begin{figure*}[]
\centering
\subfloat[1-D PDFs in Log(Age)]{%
  \includegraphics[width=1.8\columnwidth]{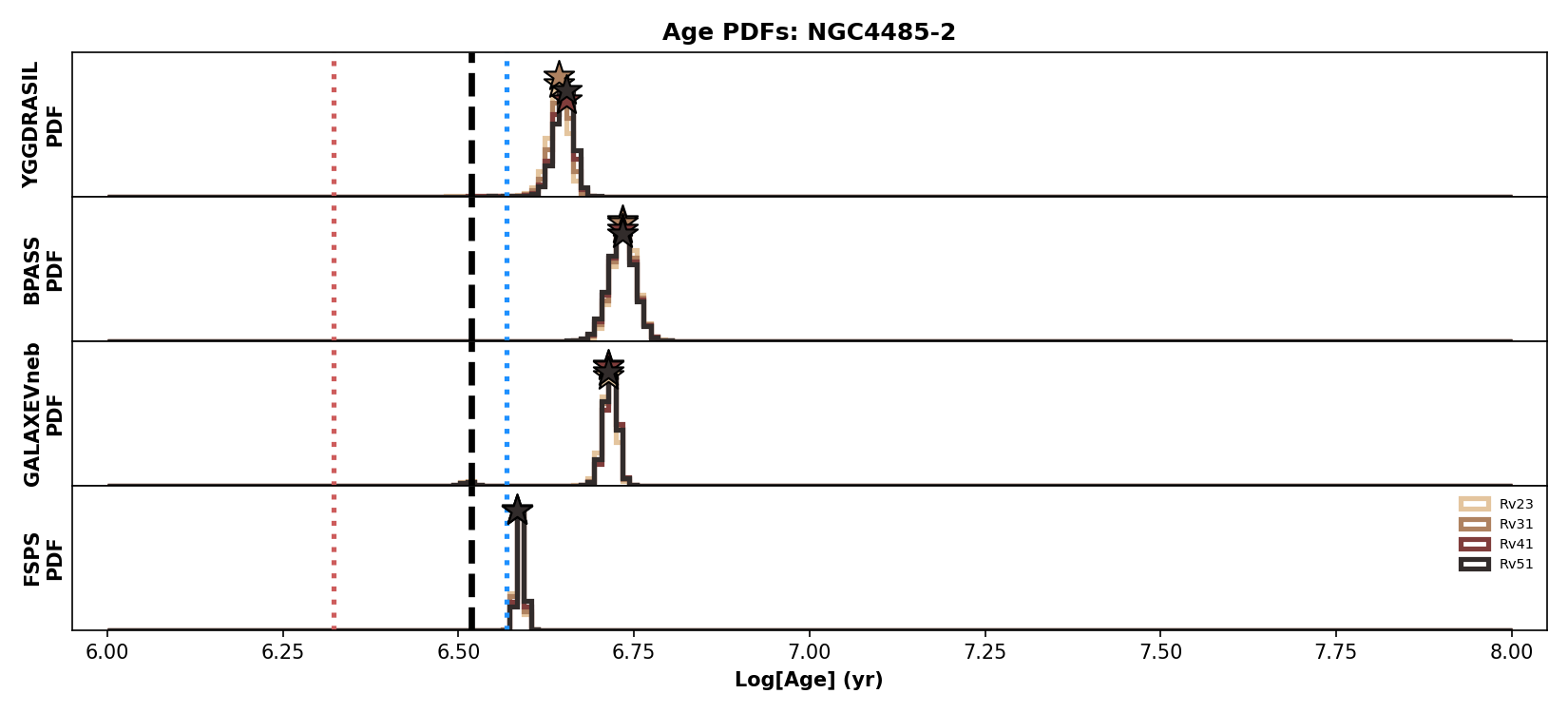}%
}

\subfloat[1-D PDFs in E(V-B)]{%
  \includegraphics[width=1.8\columnwidth]{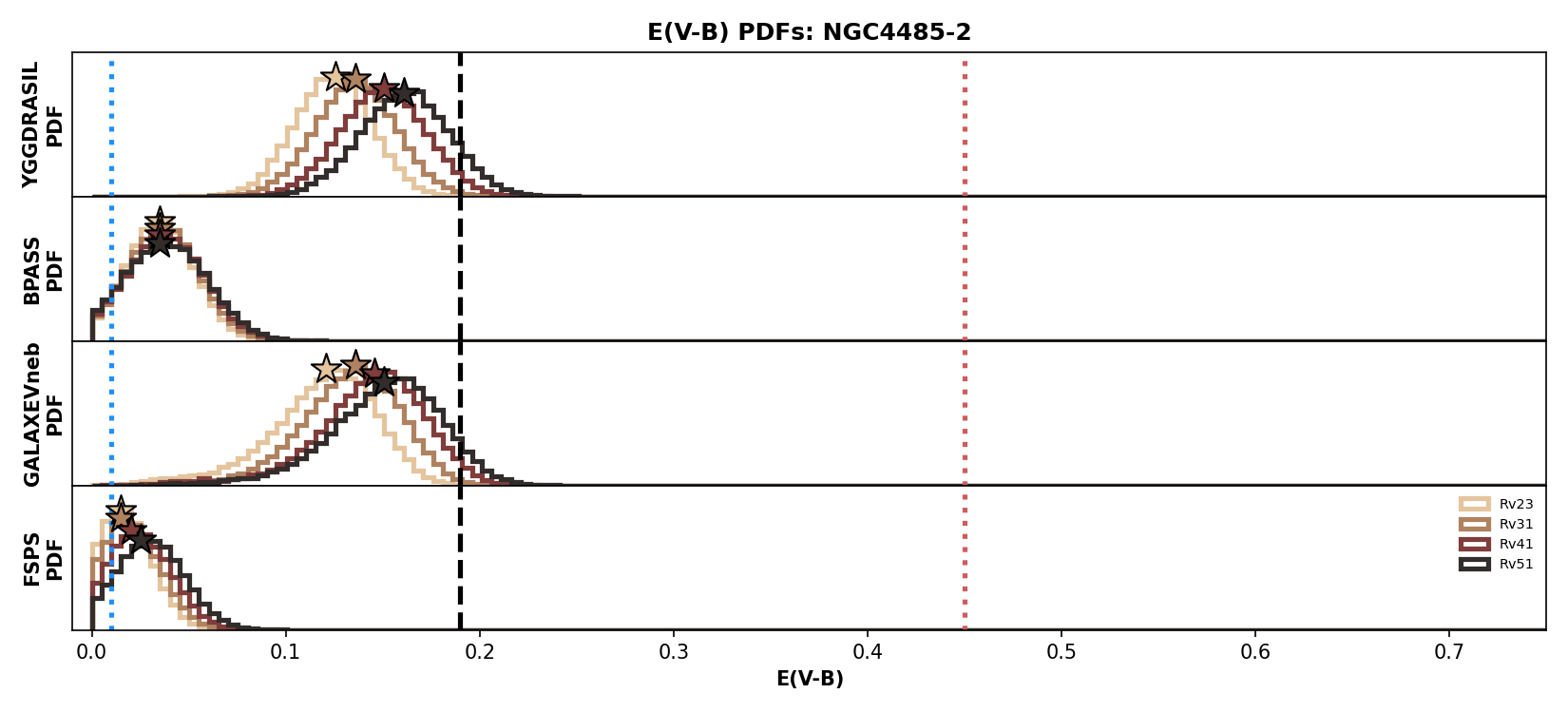}%
}

\subfloat[1-D PDFs in Log(Mass)]{%
  \includegraphics[width=1.8\columnwidth]{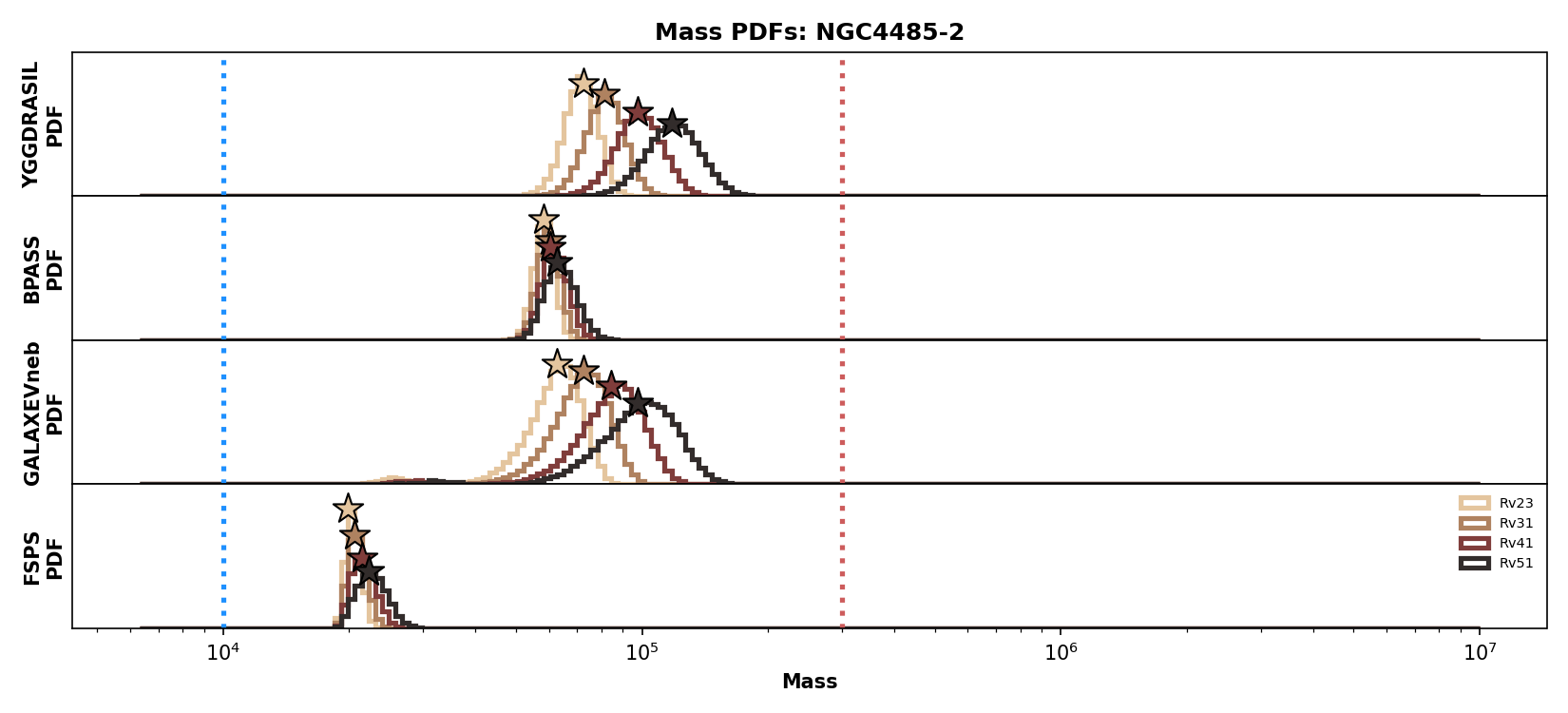}%
}

\caption{NGC4485-2 - a relatively well-behaved example. In this plot, the black dashed line is the \citet{Sirressi2022} ``light-weighted'' solution, the dotted blue line is the major population in the ``two-population'' solution, and the dotted red line is the minor population in that solution. Here, the choice of extinction curve affects derived cluster properties only minimally. Cluster properties as a function of SPS model are also broadly similar. }

\end{figure*}

\begin{figure*}[]
\centering
\subfloat[1-D PDFs in Log(Age)]{%
  \includegraphics[width=1.8\columnwidth]{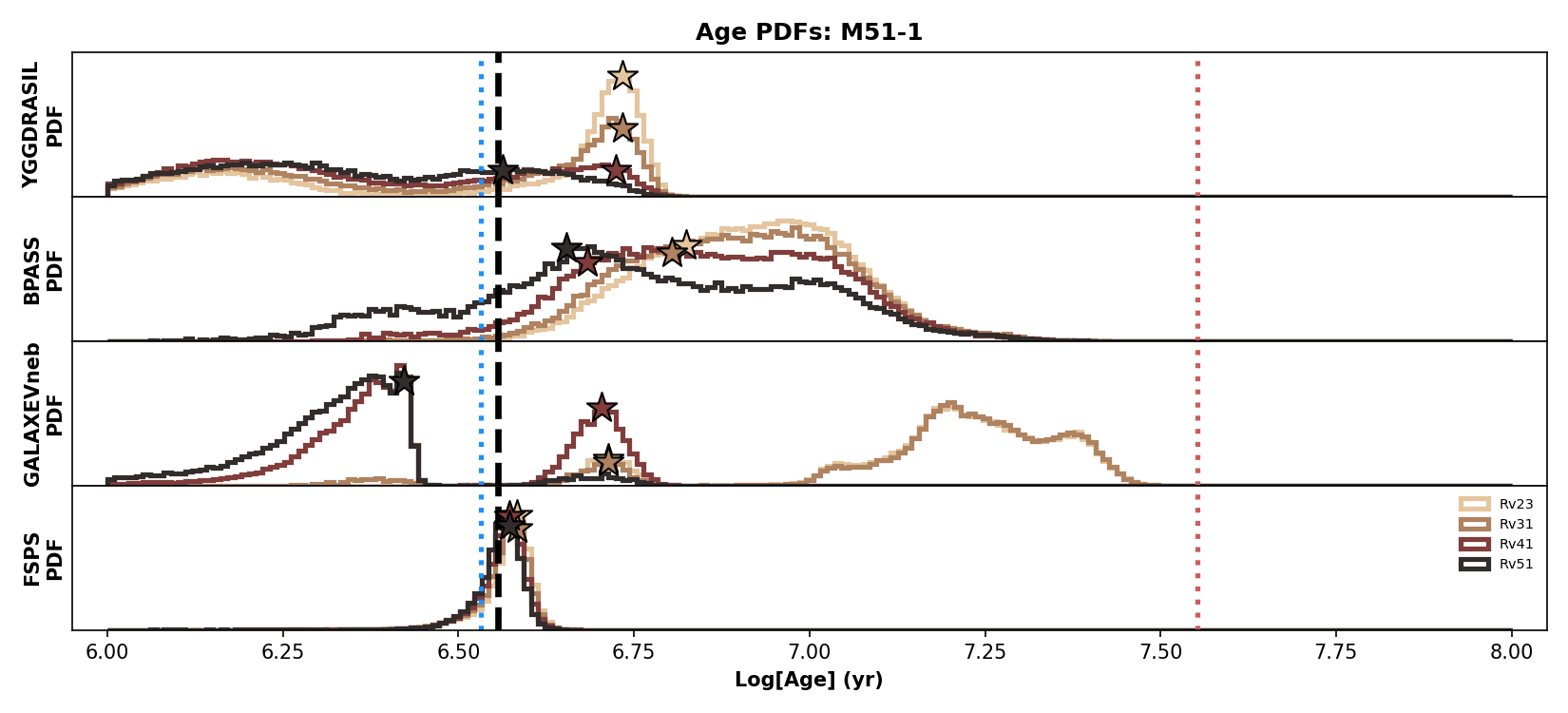}%
}

\subfloat[1-D PDFs in E(V-B)]{%
  \includegraphics[width=1.8\columnwidth]{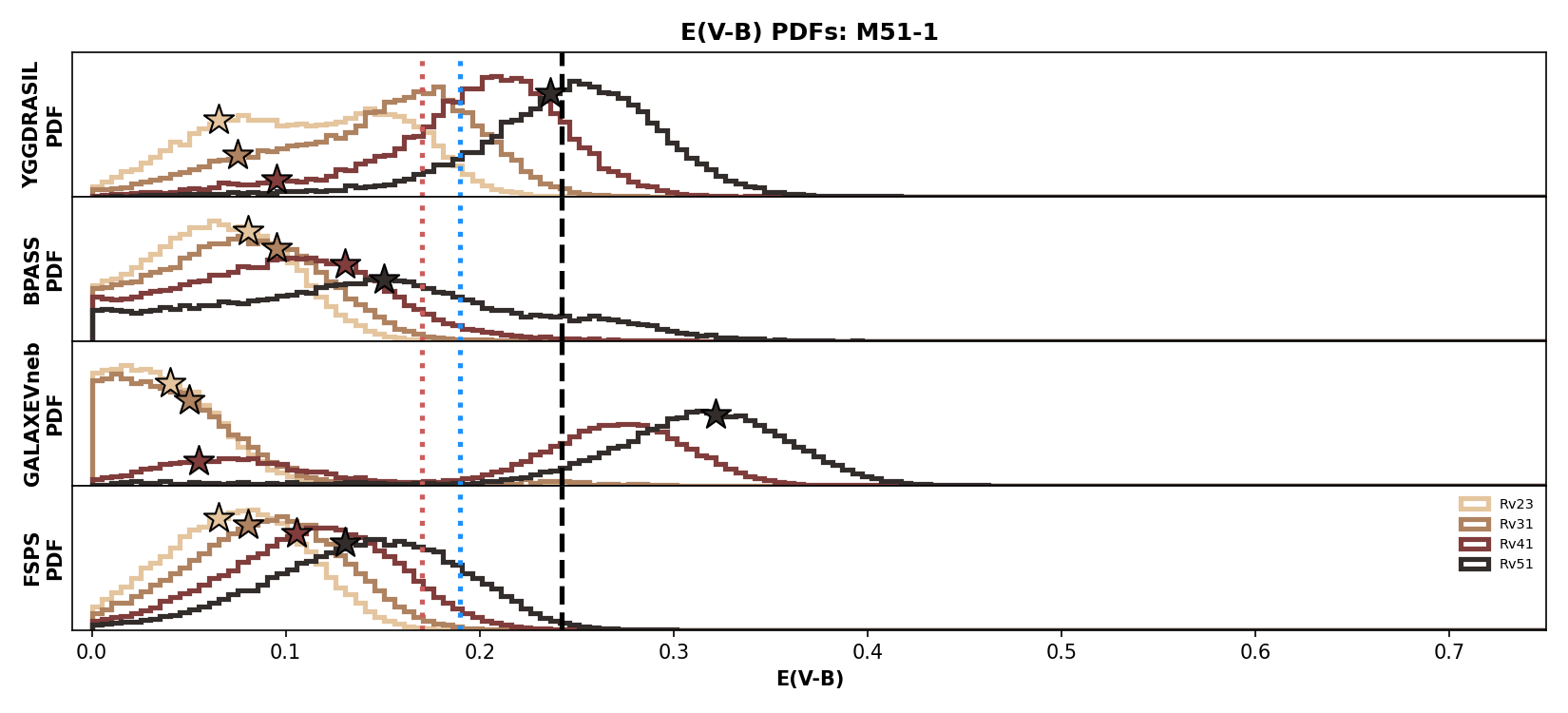}%
}

\subfloat[1-D PDFs in Log(Mass)]{%
  \includegraphics[width=1.8\columnwidth]{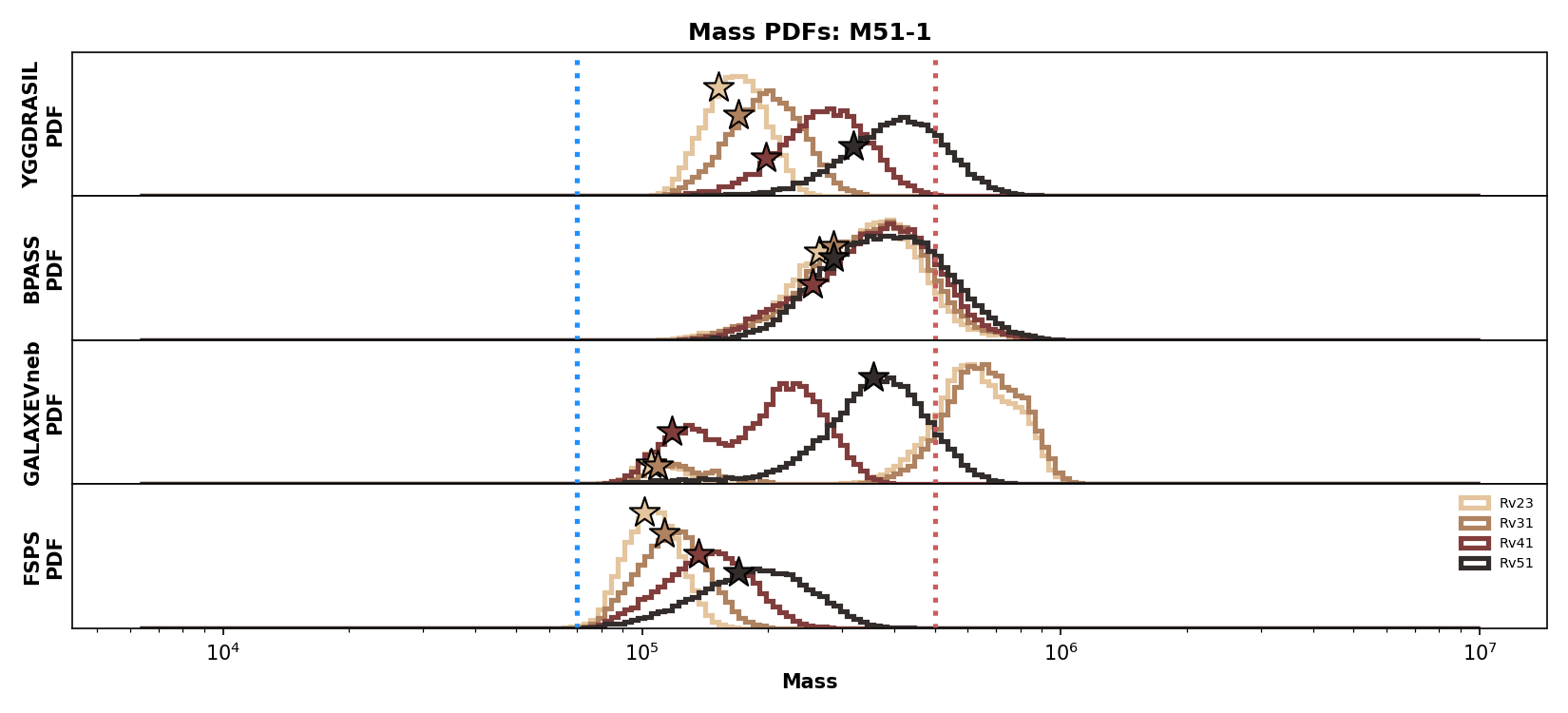}%
}

\caption{M51-1 - a relatively poorly-behaved example.  Here, the choice of extinction curve strongly affects derived cluster properties, and most models produce a multimodal posterior. Model-model differences are also strong. In this plot, the black dashed line is the \citet{Sirressi2022} ``light-weighted'' solution, the dotted blue line is the major population in the ``two-population'' solution, and the dotted red line is the minor population in that solution. Some fits in GALAXEVneb and FSPS were rejected due to unacceptable $\chi^2$ for this cluster despite being close to the spectroscopic results in age.}

\end{figure*}

\begin{figure}[htb!]
\includegraphics[width=0.45\textwidth]{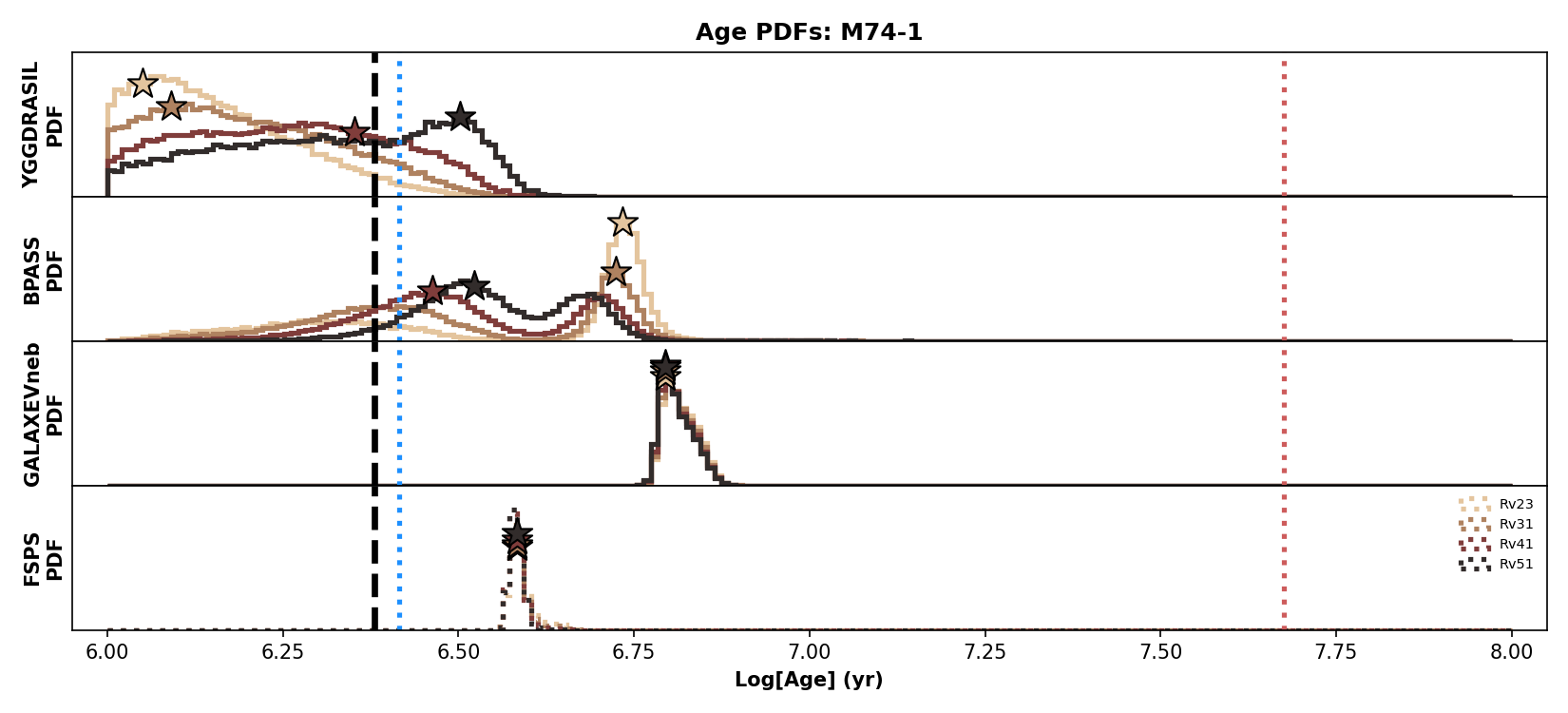}
\caption{Sometimes, fitting the same data with different models produces results that differ not only in the sense of absolute offsets but also in precision. Here, we show the 1d posteriors in age for M74 YSC-1. Age is poorly constrained for YGGDRASIL and BPASS but well constrained for GALAXEVneb and FSPS, even though the input data are identical.}
\label{fig:m74age1d}
\end{figure}

It is also clear from these plots that on a model and $R_V$ basis, there is sometimes significant disagreement between fits. On a model-to-model basis, posterior shapes can vary wildly. $R_V$ choice can also strongly impact the posterior shape. On average, however, using an extinction curve with a higher $R_V$ tends to produce solutions that are associated with an older age, more stellar mass, and more total extinction. We discuss this result further in Section 6.2.

\subsection{How Well Do Broadband SED Fits Agree With FUV Spectral Fits?}
The 1-D posteriors are also a good benchmark for the overall performance of broadband SED fitting compared to FUV spectral fits. We consider 3 different ways to define agreement between our broadband fits and the \citet{Sirressi2022} FUV fits, which we denote Case A, Case B, and Case C:

\begin{enumerate}[label=(\Alph*)]
\item The median of the broadband posterior is within $1\sigma$ of the FUV light-weighted solution. This comparison is ill-defined for mass, as the FUV fits provide no light-weighted stellar mass.
\item The median of the broadband posterior is within $1\sigma$ of the major population of the FUV 2-population solution.
\item The broadband maximum-likelihood solution lies somewhere between the major and minor population solutions of the FUV fit. This comparison is ill-defined in situations where only a single stellar population was required to properly fit the FUV spectrum.
\end{enumerate}

We calculate the number of fits meeting each criteria, both overall and on a model-by-model basis. As expected, we find that there is relatively poor agreement between cluster properties as estimated by broadband SED fitting and by FUV photospheric line fits. Assuming Case A, we find that 79/243 (32.5$\%$) of all fits are in agreement with the FUV in terms of age, and 73/212 (30.0$\%$) are in agreement in terms of extinction. Looking for Case B agreement produces a similar picture; we find agreement in 78/243 (32.1$\%$) of fits for age, 50/243 (20.59$\%$) for mass, and 45/243 (18.5$\%$) of fits for extinction. Case C is the least conservative metric we tested; from this perspective, 107/167 (64.1$\%$) of fits (ignoring clusters without a valid 2-population FUV solution) are in age agreement alongside 107/132 (64.1$\%$) in mass and 83/167 (49.7$\%$) in extinction. On average, our photometric analysis agreed with the spectroscopic analysis less often than in \citep{Sirressi2022}, who found agreement for $\sim 50\%$ of clusters. The difference is likely due to a combination of factors - our large photometric apertures, more diverse set of models, and our more conservative definition of FUV/broadband consistency may all contribute here.

We also note that there is no strong age or mass bias in terms of FUV/broadband agreement. Our broadband fits returned ages older than the FUV fits in 126/243 (51.9$\%$) of cases and masses larger than the dominant FUV population mass in 134/243 (55.1$\%$) of cases. Broadband fits tended to find less total extinction than the FUV fits, with the broadband value exceeding the FUV value in 54/243 (22.2$\%$) cases. The results on a model-by-model basis are summarized below.

\subsubsection{Agreement with FUV Fits: YGGDRASIL}
We find that YGGDRASIL has average performance among models we tested in terms of agreement with the FUV fits. Although both YGGDRASIL and the \citet{Sirressi2022} fits are based on the Starburst99 \citep{Sb99,Sb99Params} models, the fact that one set of fits is based on broadband NUV-to-I photometry and the other is based on FUV spectroscopy means that agreement is not guaranteed. We also note that YGGDRASIL used Starburst99 with Padova stellar libraries, while the FUV fits used Starburst99 with Geneva stellar libraries. Particularly at young ages, the differences between these evolutionary tracks are significant. Out of 70 accepted fits performed using YGGDRASIL, we find Case A agreement for 20/70 (28.6$\%$) in age and 29/70 (41.4$\%$) in extinction. The Case B picture is similar, with age agreement in 23/70 (32.9$\%$) of cases, mass agreement in just 11/70 (15.7$\%$), and extinction agreement in 13/70 (18.6$\%$). As expected, Case C presents the most optimistic picture. 50 YGGDRASIL fits remain after filtering out clusters with only single-population FUV fits. Out of these, 30/50 (60.0$\%$) fulfill Case C in age, 33/50 (66.0$\%$) fulfill it in mass, and 27/50 (54.0$\%$) fulfill it in extinction. Comparing these results to the average performance across all models, YGGDRASIL tends to under-perform in terms of its ability to predict cluster ages but is decent at predicting masses and extinctions. 

\subsubsection{Agreement with FUV Fits: BPASS}
We find that results derived with BPASS are in reasonable agreement with the FUV fits, and is perhaps the most accurate model we tested by this (limited) metric. Out of 53 accepted fits performed using BPASS, we find Case A agreement for 23/53 (43.4$\%$) in age and 11/53 (20.8$\%$) in extinction. As with YGGDRASIL, Case B paints a similar picture to Case A, with age agreement in 18/53 (34.0$\%$), mass agreement in 23/53 (43.4$\%$), and extinction agreement in 11/53 (20.8$\%$) of fits. After filtering out FUV single-population clusters to enable Case C comparison, we are left with 33 accepted BPASS fits. Out of these, 24/33 (72.7$\%$) fulfill Case C in age, 21/33 (63.6$\%$) fulfill it in mass, and 15/33 (45.5$\%$) do in extinction. Comparing these results to those obtained across all models, BPASS performs better than average in age and mass terms but sometimes struggles to recover the extinction. 

\subsubsection{Agreement with FUV Fits: GALAXEVneb}
We find that our modified version of GALAXEVneb has an overall accuracy generally comparable to that of YGGDRASIL. We find Case A agreement here in 19/64 (29.7$\%$) of fits for age and 22/64 (34.4$\%$) for extinction. GALAXEVneb is in Case B agreement with the FUV in 18/64 (28.1$\%$) fits in age terms, 10/64 (15.6$\%$) in mass terms, and 14/64 (21.9$\%$) in extinction terms. We once again filter out FUV single-population clusters, leaving us with 44 usable fits for Case C comparison. Here, 25/44 (56.8$\%$) agree with the FUV in age, alongside 30/44 (68.2$\%$) in mass and 21/44 (47.7$\%$) in extinction. Comparing the overall acceptance fractions, we find that GALAXEVneb is better than average at recovering extinctions but suffers in terms of accurately recovering ages.

\subsubsection{Agreement with FUV Fits: FSPS}
We find that FSPS performed reasonably well in terms of agreement with the FUV fits. The accuracy picture drawn by Case A is average, with agreement in 17/56 (30.4$\%$) of accepted FSPS fits in age terms, alongside 11/56 (19.6$\%$) in extinction. The Case B results are unimpressive; 19/56 (33.9$\%$) of fits agree in age, alongside a mere 6/56 (10.7$\%$) in mass and 7/56 (12.5 $\%$) in extinction. FSPS is associated with 40 fits appropriate for Case C comparison, of which 28/40(70.0$\%$) agree in age, 23/40 (57.49$\%$) agree in mass, and 20/40 (50.0$\%$) agree in extinction. Comparing these results to the fits across all models, FSPS performed particularly well in terms of recovering ages, inferior in this respect only to BPASS.


\section{Discussion}

\subsection{Model-Model Comparison: Cluster-by-Cluster}
In this work, we demonstrate that there is sometimes a statistically significant difference between the parent distributions of cluster properties derived using different SPS models, which implies that different SPS models may ``see'' different cluster populations given the same data. We also show that different SPS models perform differently in terms of how well their results agree with fits of the same objects in the FUV. Taken together, these suggest that SPS model choice is not trivial. We caution that making comparisons between the results of SED fitting performed using different input SPS models may not be meaningful, at least for individual objects. With this in mind, a few questions still remain:

\begin{enumerate}[]
\item Are the model-model differences in derived cluster properties random, or are some SPS models consistently biased (e.g. towards older ages or higher extinctions)?
\item Do any models show biases as a function of cluster properties (e.g. do some models have issues that only occur in a certain age range)?
\item If no strong biases exist, how extreme can the scatter introduced by model choice be?
\end{enumerate}

To answer these questions, we plot fitted parameters on a cluster-by-cluster basis for every combination of SPS model and extinction curve (Figure 12), normalized to the median value of that parameter over all models and curves in a given cluster. That is, each point on this plot represents the offset between the results of an individual fit and the median result for that cluster, averaged over every possible extinction curve and SPS model. We show the median of these offsets (as a function of SPS model) over the entire cluster sample as the horizontal dotted lines. To search for systematics related to cluster age, we also plot age, mass, and extinction offsets as a function of median age (Figure 13).

We find a picture similar to that found in Section 5.2 -- there are, on average, modest offsets as a function of SPS model. Importantly, we note that the lines of median offset shown here are not always consistent with the population medians shown in Table 5 in terms of their relative magnitude. For example, we can see in Figure 12b that FSPS is associated with slightly lower masses relative to the other models, while in Table 5 YGGDRASIL had the lowest median mass. This is evidence that no strong model-dependent bias exists for these parameters -- if such a bias existed, it would be visible both in the population medians as well as the individual offsets. Instead, this suggests any systematic bias introduced by model choice is relatively weak and the model-model inconsistencies we find are primarily the result of scatter between individual fits of the same object. We discuss this apparent mass offset in more detail below.

We find no strong evidence of age-dependent bias for most parameters, though we note that many of the FSPS solutions with low mass relative to other models were found in systems with log[Age] $<$ 7.0 (Figure 13b).  It is clear that model-model differences can be quite strong for individual clusters. When performing SED fits of the same cluster and using the same extinction curve, SPS model choice can introduce offsets of up to 34.8 Myr in age, a factor of 9.5 in mass, and 0.40 mag in extinction (Figure 12). Averaged over all models, the median offset between any individual fit and the median result for a given cluster was 2.98 Myr in age, a factor of 2.1 in mass, and 0.13 mag in extinction. 


\begin{figure*}[htb!]\centering
\subfloat[Cluster ages relative to the median, for all clusters]{\label{a}\includegraphics[width=.49\linewidth]{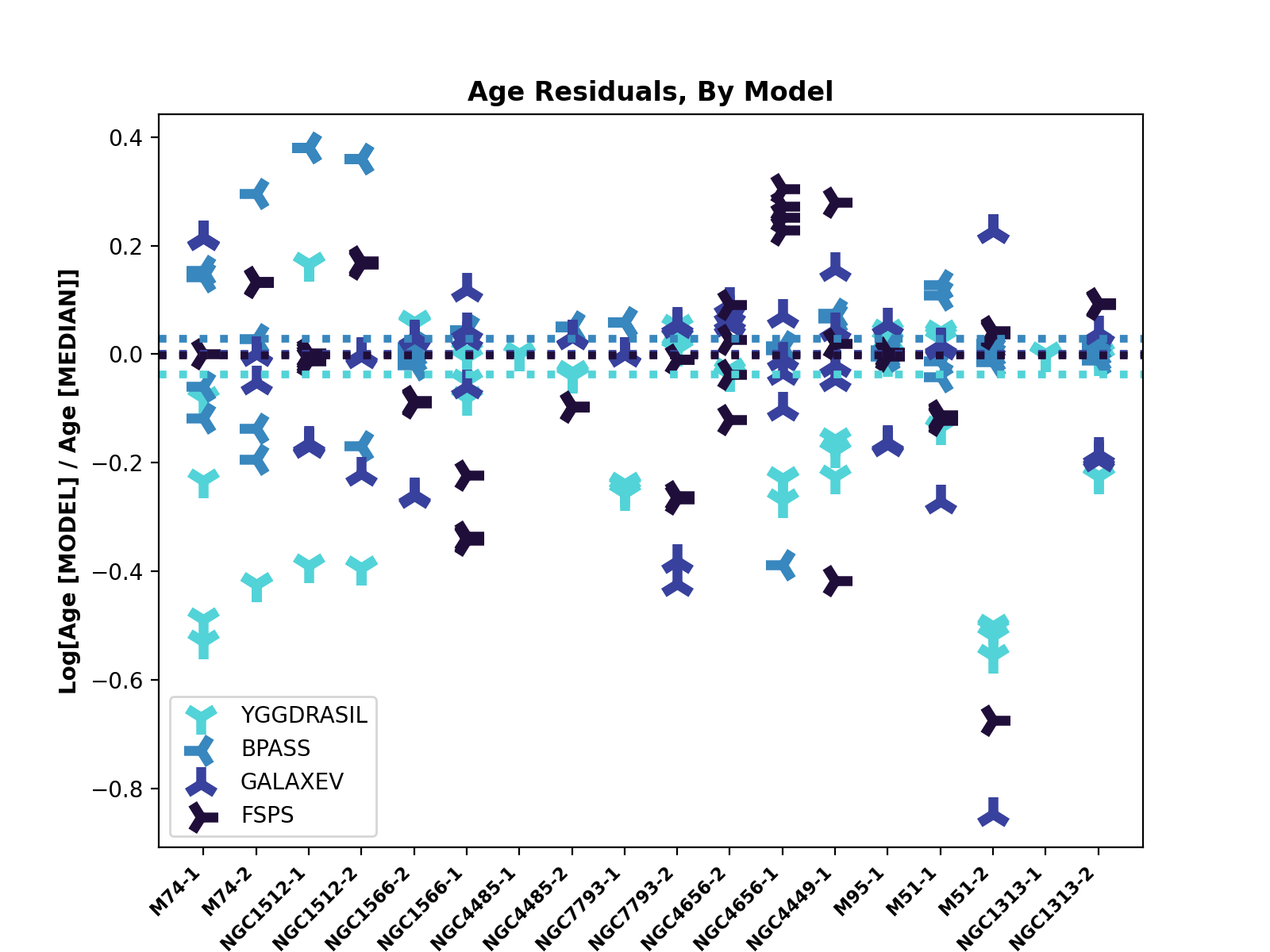}}\hfill
\subfloat[Cluster masses relative to the median, for all clusters]{\label{b}\includegraphics[width=.49\linewidth]{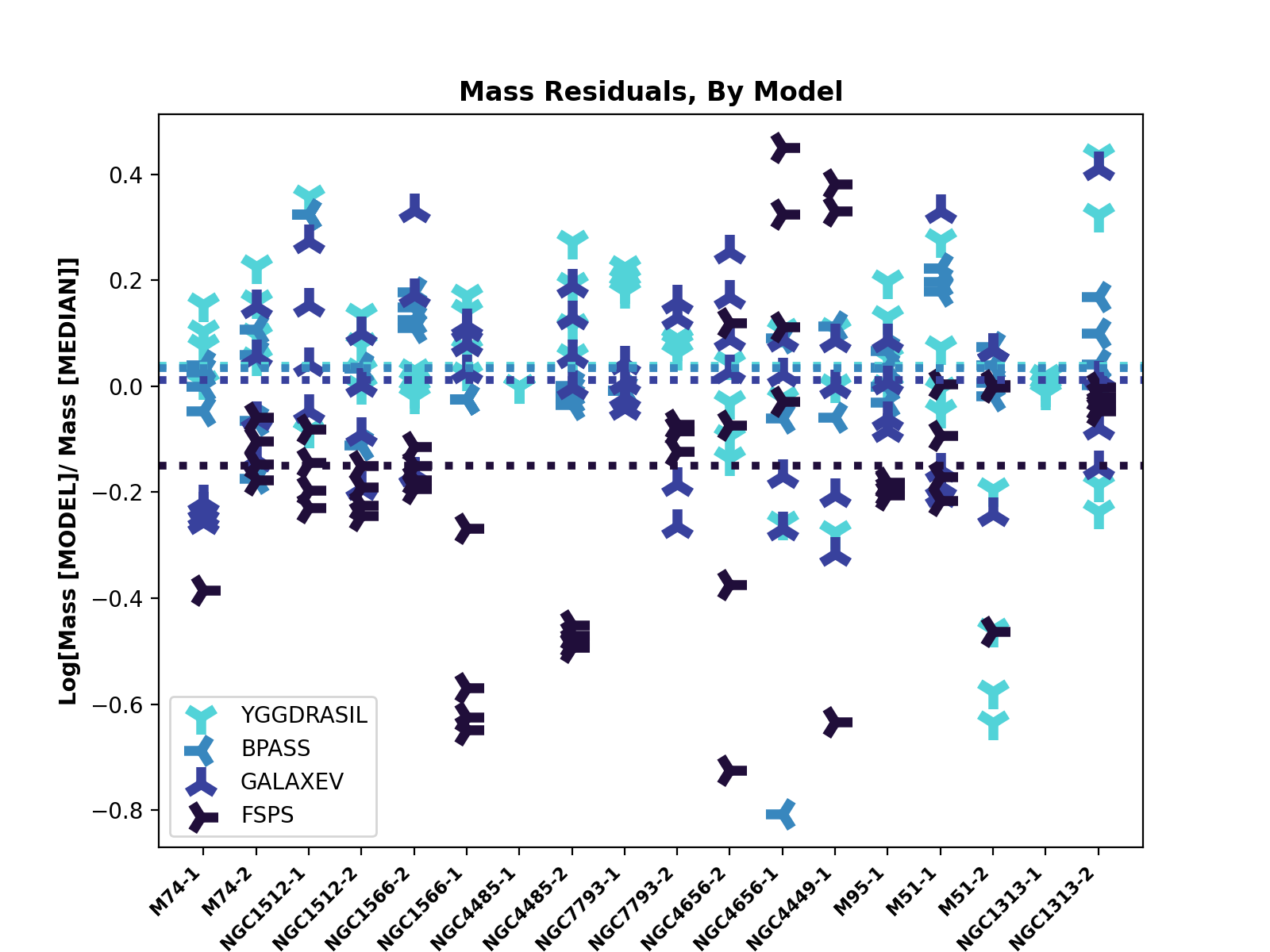}}\par 
\subfloat[Cluster E(B-V)s relative to the median, for all clusters]{\label{c}\includegraphics[width=.49\linewidth]{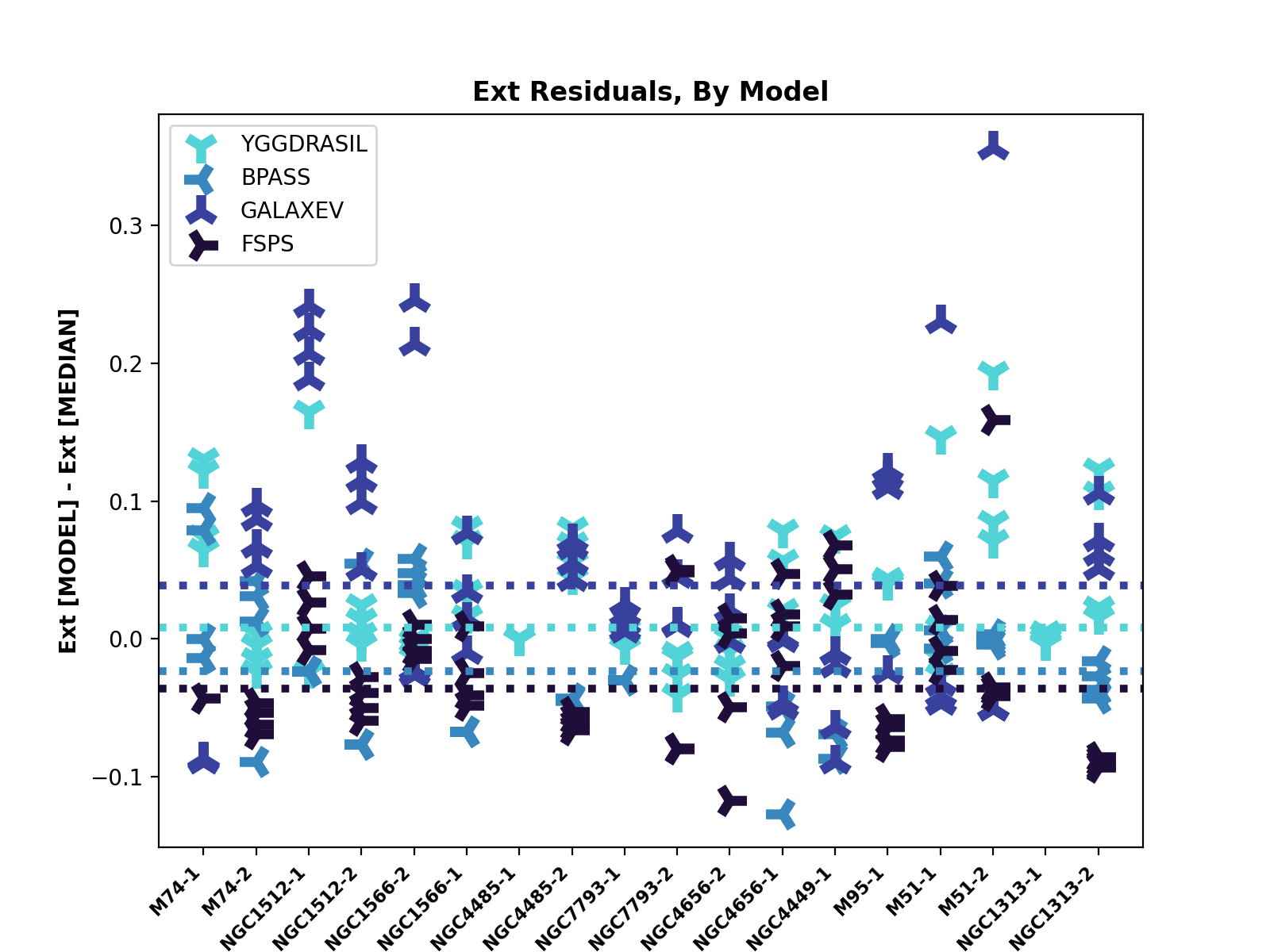}}
\caption{The difference between cluster parameters inferred using different SPS codes can be large for individual clusters. However, the systematic offset between models is small. Panel (a) shows cluster ages for individual fits relative to the median age determined for each cluster, marginalized over $R_v$. Similarly, panel (b) shows cluster masses relative to the median mass determined for each cluster. Panel (c) is the same, this time in terms of extinction. Only accepted fits are shown. Horizontal dotted lines represent the median ratio between fits performed using a given model and the median result for a given cluster.}
\label{fig}
\end{figure*}

\begin{figure*}[htb!]\centering
\subfloat[Age Residuals vs. Age]{\label{a}\includegraphics[width=.49\linewidth]{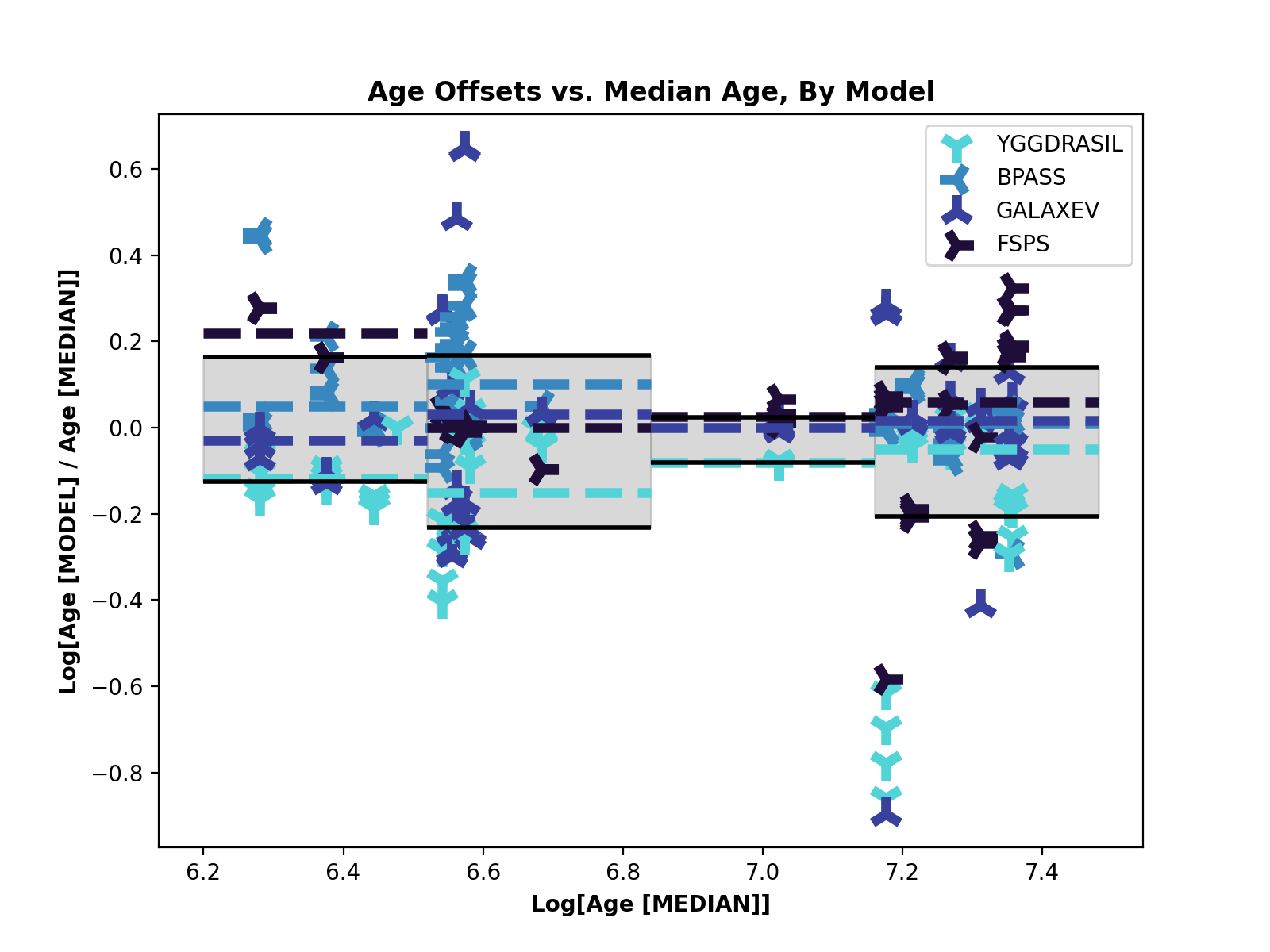}}\hfill
\subfloat[Mass Residuals vs. Age]{\label{b}\includegraphics[width=.49\linewidth]{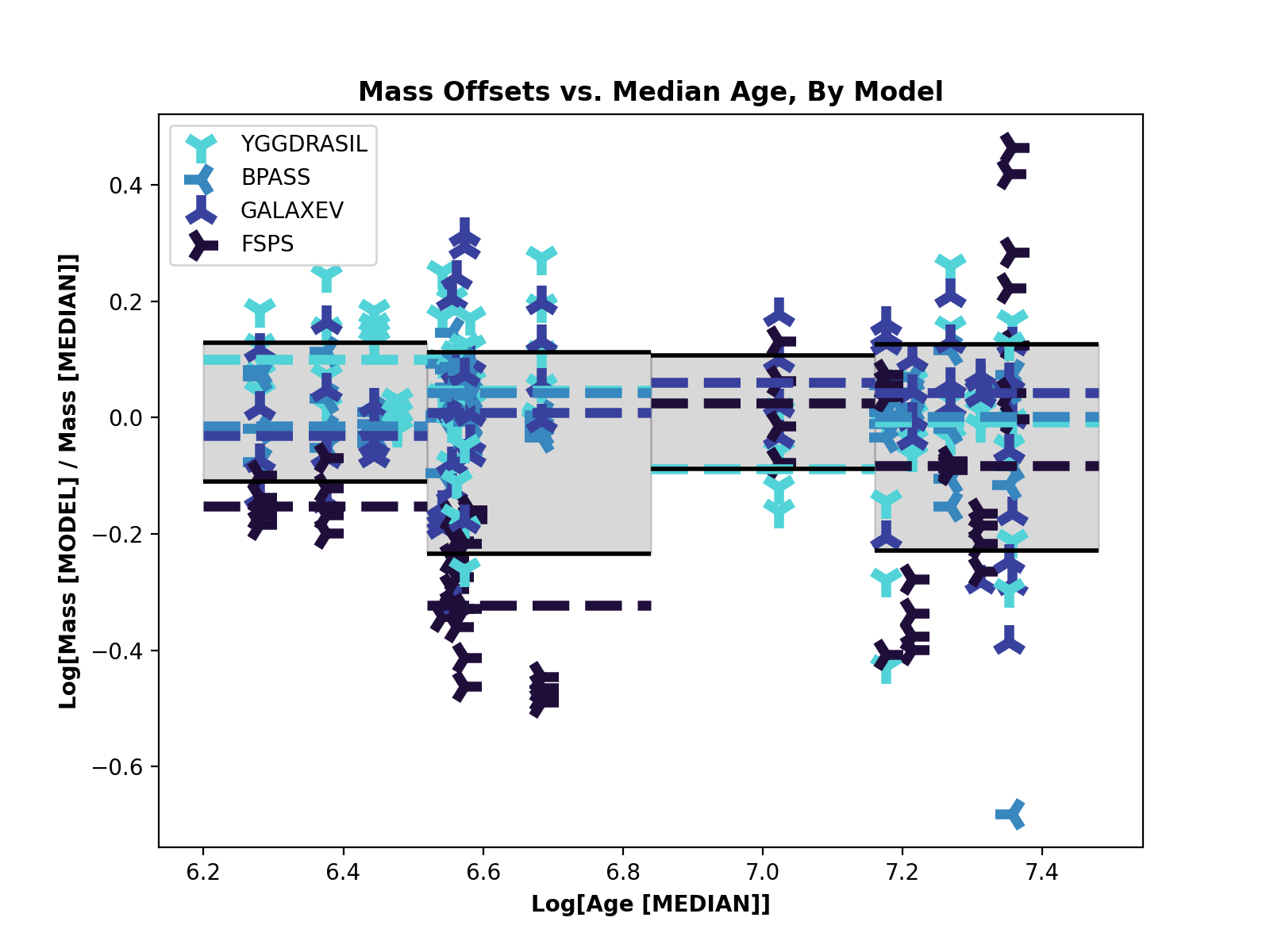}}\par 
\subfloat[Extinction Residuals vs. Age]{\label{c}\includegraphics[width=.49\linewidth]{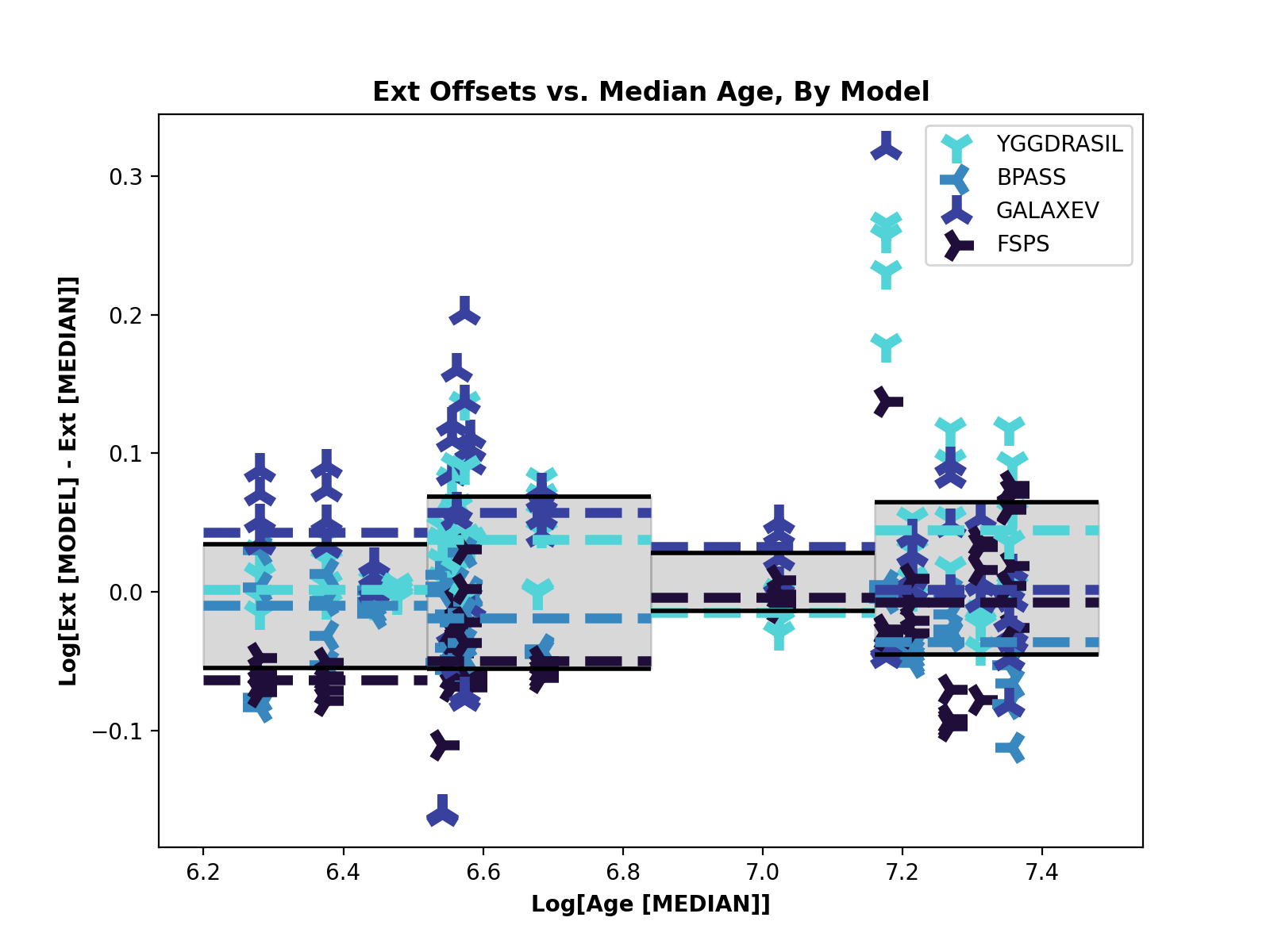}}
\caption{With the possible exception of FSPS, each model we tested appears to perform similarly as a function of cluster age. Here, we show residuals in age, mass, and extinction as a function of cluster median age for each model. The running medians are shown as dashed lines, while the region spanned by the 16th and 84th percentiles are shown as the shaded regions. FSPS seems to systematically find lower stellar masses than the other models, \textbf{particularly at younger ages and high metallicity.}}
\end{figure*}

\subsection{$R_V$-$R_V$ Comparison: Cluster-by-Cluster}
Analogous to the previous section, we plot the cluster properties determined using each $R_V$ relative to the per-cluster medians in Figure 14. The effect of $R_V$ on the results of SED fitting is of a similar magnitude to the effect of SPS model choice and is usually more predictable; however, it remains the case that extinction curve choice can introduce large offsets within individual clusters. 

$R_V$ does not strongly bias cluster age. As one pushes to higher $R_V$, however, one tends to recover higher masses and more total extinction. As discussed in Section 5.3, this is a consequence of how $R_V$ is defined. Keeping model choice constant, we find that extinction curve choice can introduce an age offset of up to 32.3 Myr in age, a factor of 10.4 in mass, and 0.41mag in extinction. This generally corresponds to cases where the underlying posterior distribution was multi-modal and the choice of extinction curve changed which mode was associated with the maximum-likelihood solution. Median offsets over all curves were a factor of 0.9 Myr in age, a factor of 1.47 in mass, and 0.05 mag in extinction. Although the curve-curve differences can be extreme in rare cases, the extinction curve choice introduces less scatter on average than the choice of SPS model.

\begin{figure*}[htb!]\centering
\subfloat[Cluster ages relative to ages derived with $R_V = 2.3$, for all clusters]{\label{a}\includegraphics[width=.49\linewidth]{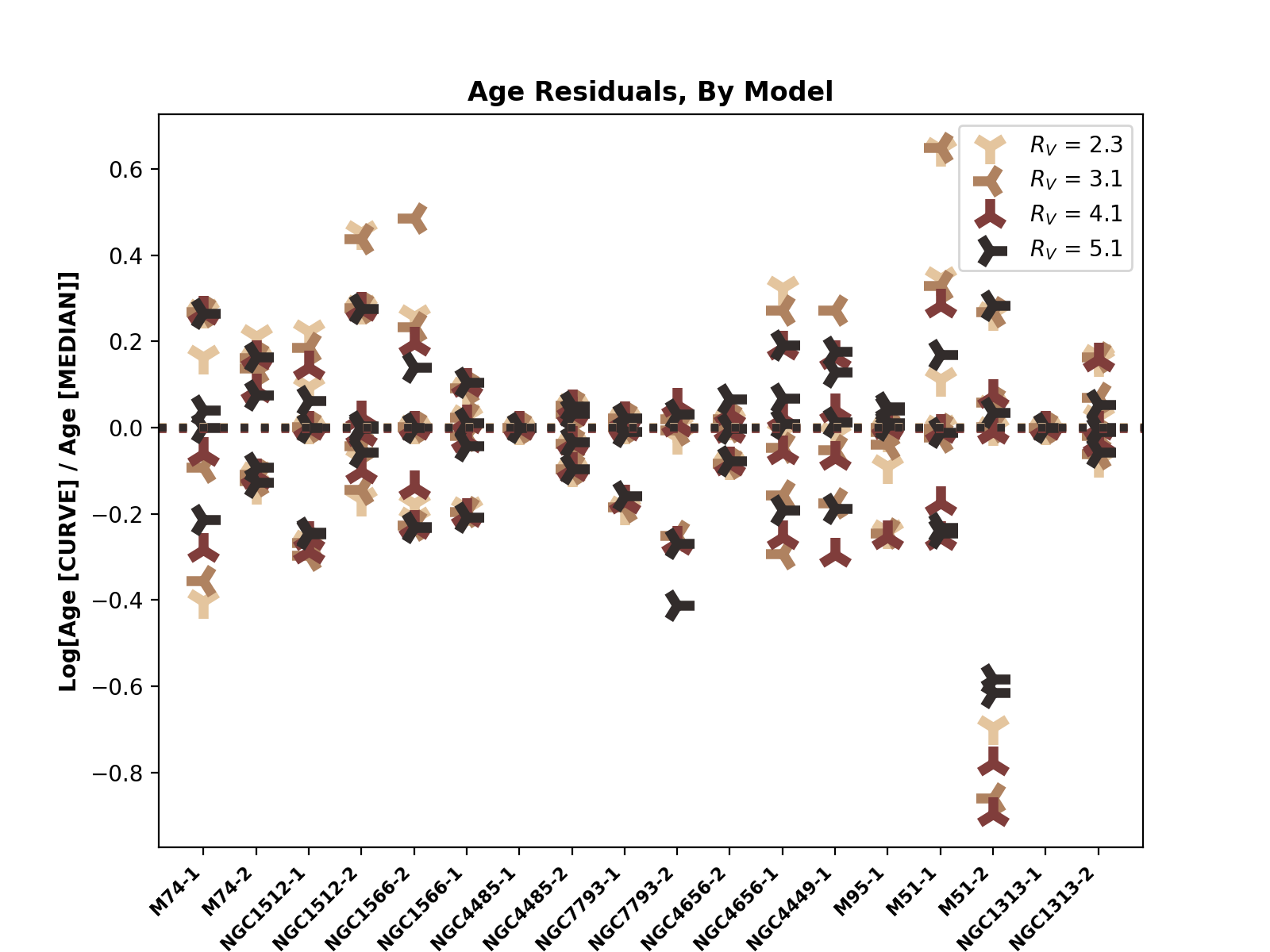}}\hfill
\subfloat[Cluster masses relative to masses derived with $R_V = 2.3$, for all clusters]{\label{b}\includegraphics[width=.49\linewidth]{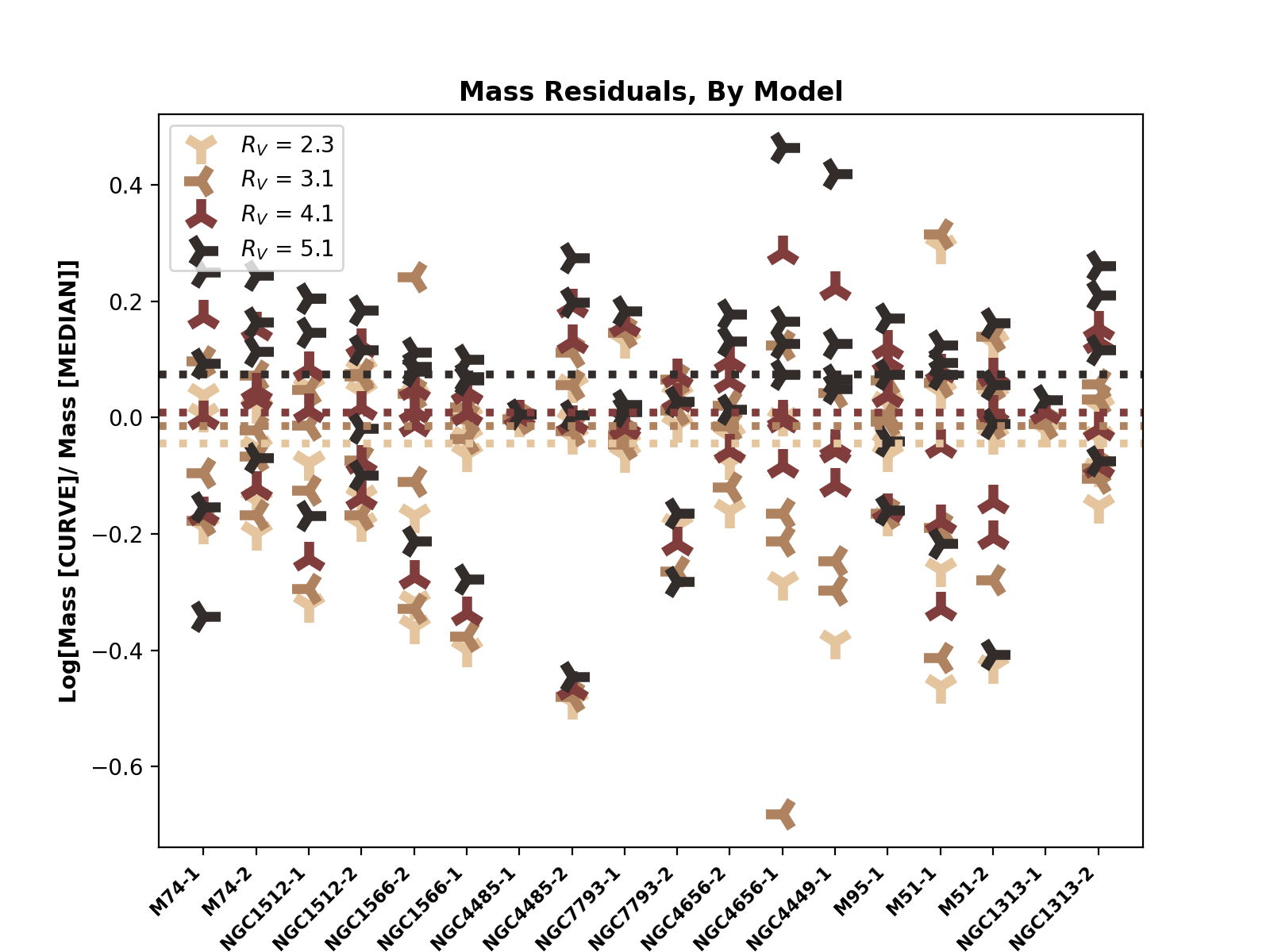}}\par 
\subfloat[Cluster E(V-B)s relative to extinctions derived with $R_V = 2.3$, for all clusters]{\label{c}\includegraphics[width=.49\linewidth]{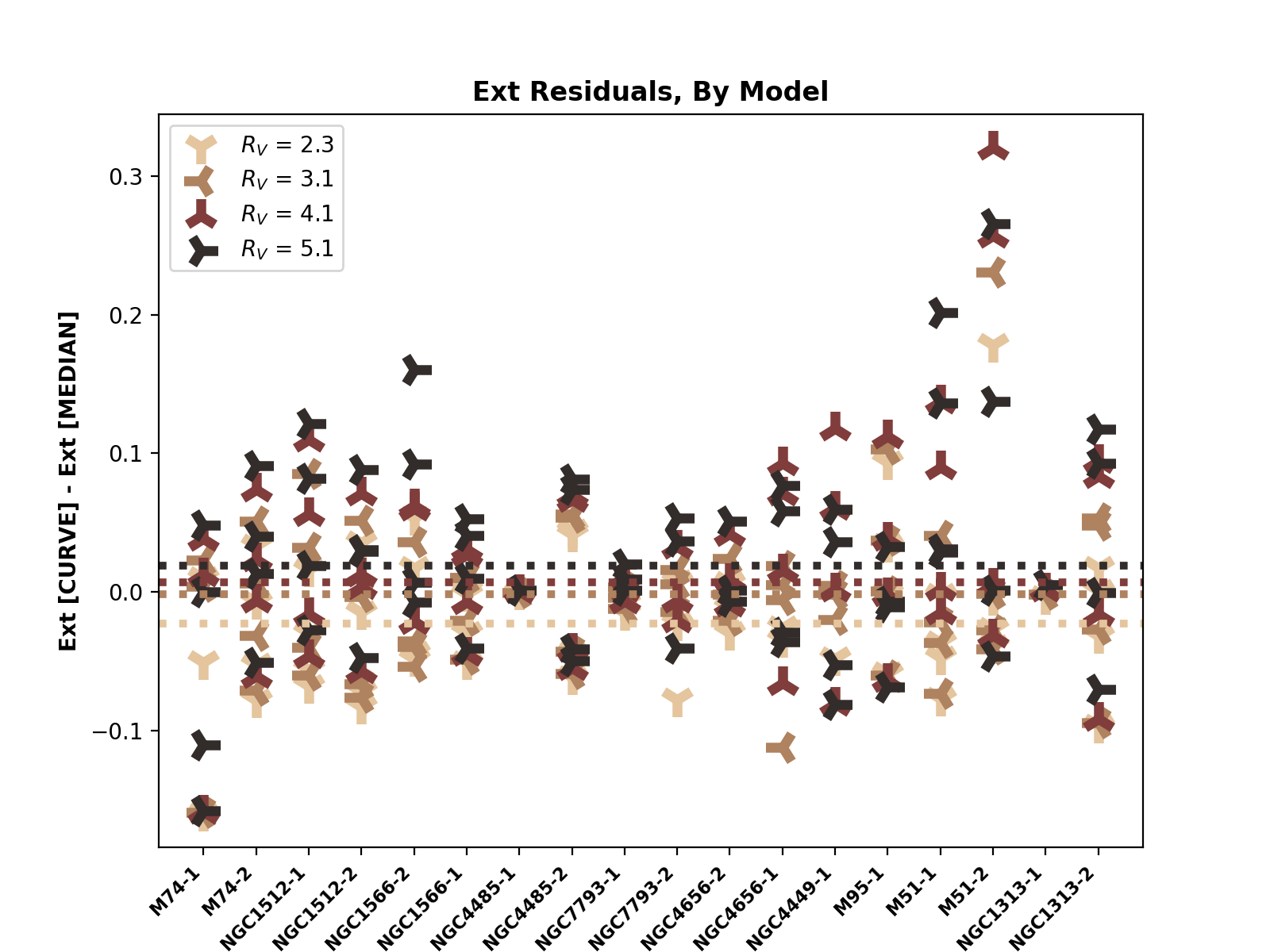}}
\caption{Panel (a) shows cluster ages (posterior median) relative to the cluster age determined using an extinction curve with $R_V = 2.3$, marginalized over choice of SPS model. Similarly, panel (b) shows cluster masses (posterior median) relative to the cluster mass determined with $R_V = 2.3$. Panel (c) is the same, this time in terms of extinction.}
\label{fig}
\end{figure*}

\subsection{How Important Is Nebular Emission?}
In this work, we have gauged the performance of a modified version of GALAXEV (which we call GALAXEVneb), that has been altered to include contributions to the SED from nebular emission. Given that we necessarily also have the unaltered (i.e. stellar-continuum-only) GALAXEV spectra, it is sensible to ask: how much does the inclusion (or lack thereof) of nebular emission alter the results of broadband SED fitting? Can reasonable fits be obtained for young systems without the inclusion of nebular emission?

We find that nebular emission must be included if one wishes to measure stellar population properties with accuracy. In particular, nebular emission predictions are required to accurately recover stellar population ages. The presence of nebular continuum and line emission causes optical colors (with the exception of U-B) to become redder at young ages \citep{Leitherer1995}. This means that SED fitting codes will underestimate the age of old systems and overestimate the dust content of young systems if using templates that do not include this emission. In these cases, the age-extinction degeneracy is important and the fit cannot distinguish between \textit{any} various solutions at young ages (Figure 15). This can significantly impact the average age of a given cluster population (Figure 16). If nebular emission is not accounted for, it becomes difficult to discern between the youngest systems. In turn, this tends to push average age within a given sample down relative to otherwise identical fits performed with nebular emission included.

\begin{figure}[htb!]\centering
\subfloat[1d-posterior comparison for M74-1]{\label{a}\includegraphics[width=.95\linewidth]{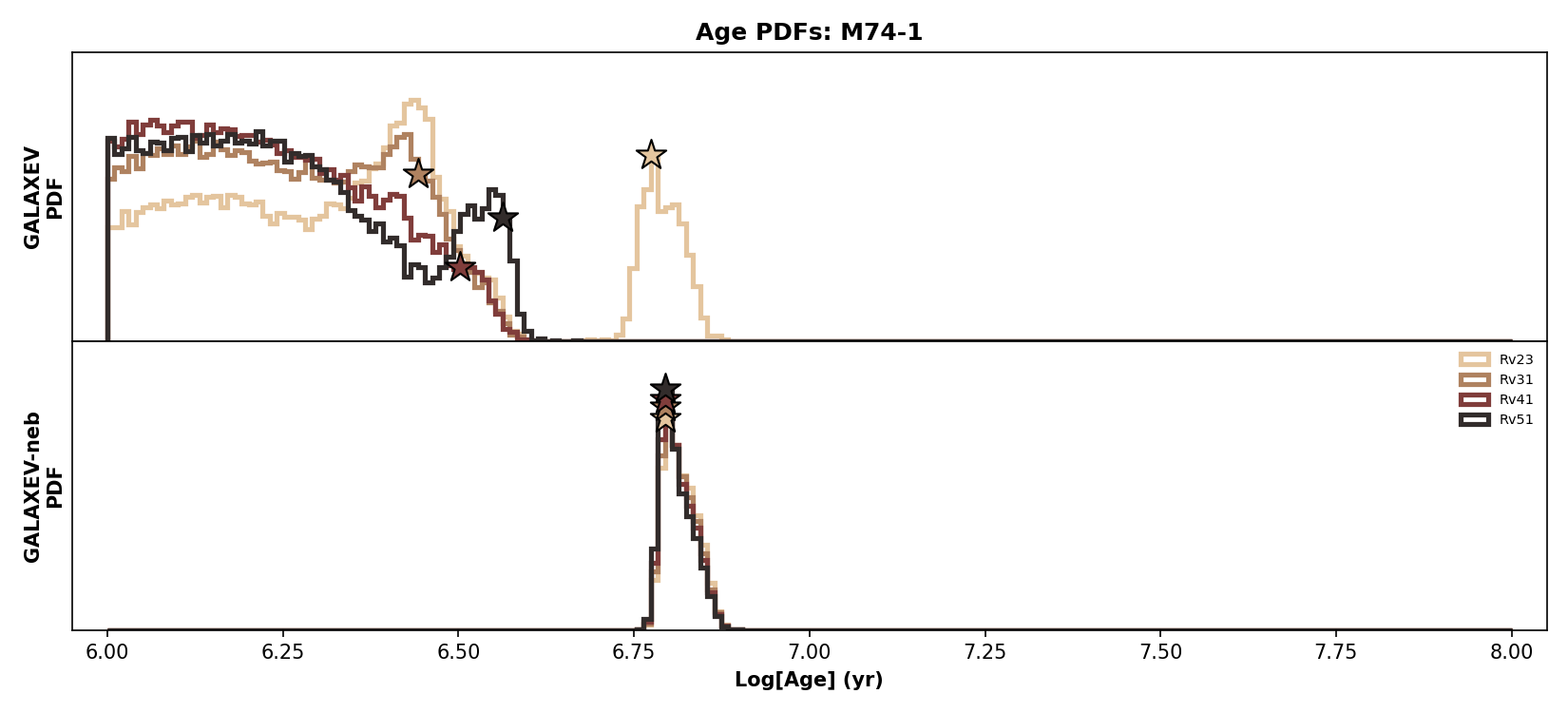}}\hfill

\subfloat[1d-posterior comparison for NGC7793-1]
{\label{b}\includegraphics[width=.95\linewidth]{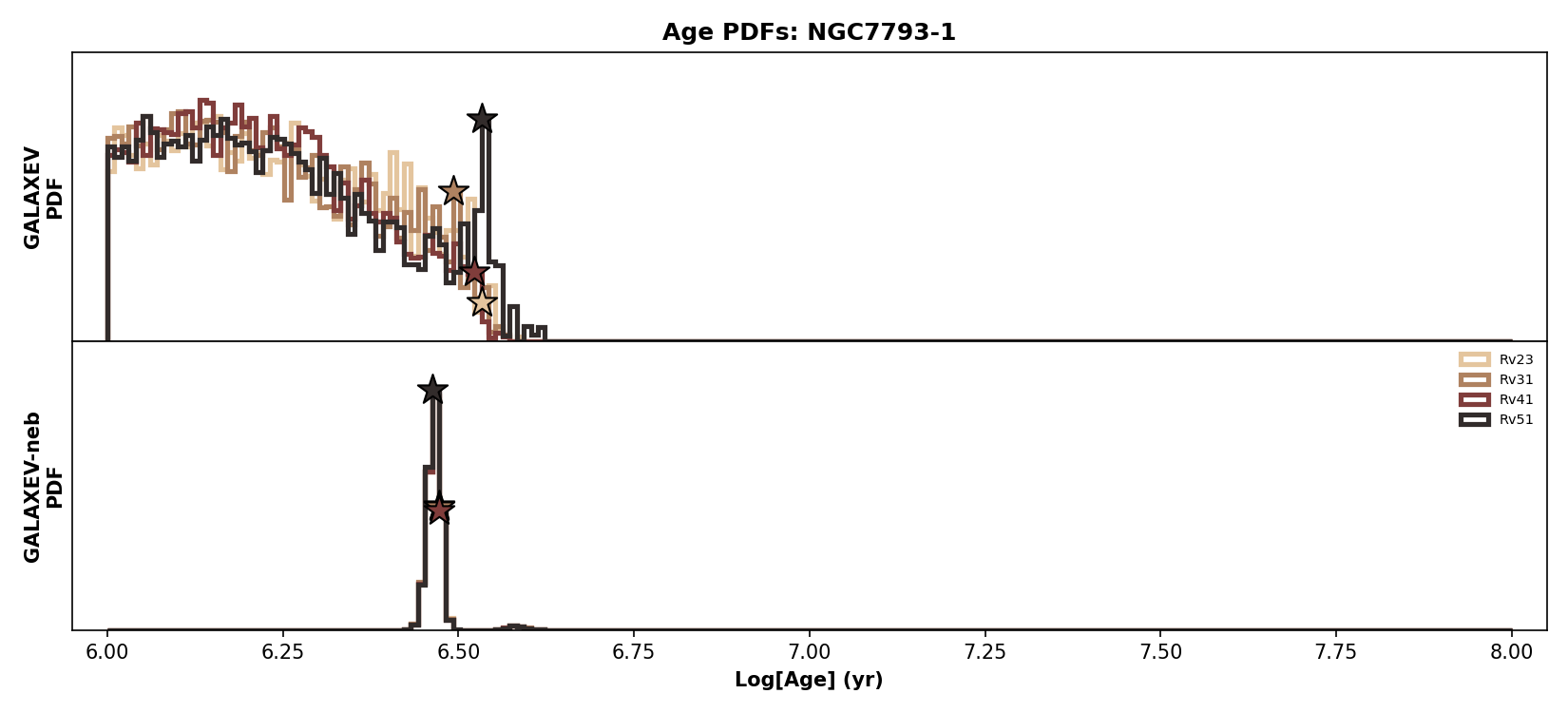}}\hfil\par 
\caption{1d-posterior comparisons for two example clusters: M74-1 and NGC7793-1. The top panel of each shows the age posterior obtained if nebular emission is not included; the bottom panel shows the result if nebular emission is included. In both cases, the lack of nebular emission in the runs using unmodified GALAXEV causes the code to (erroneously?) infer that these systems are younger on average than what we find when including nebular emission. Additionally, various young ages become difficult to distinguish without nebular emission in the models.}
\label{fig}
\end{figure}

This is all to say that if we are fitting a stellar population that might include a component young enough to produce nebular emission, nebular emission should be included in our spectral templates. For example, it has been found that omitting nebular emission can bias galaxy ages \textit{high} if the intrinsic optical colors are somewhat red \citep[][]{deBarros2014} -- the fits must push to old ages if they cannot account for this behavior via nebular lines. Although the precise impact nebular emission has on SED fitting varies as a function of the systems under examination, it absolutely must be considered. 

\begin{figure}[htb!]
\includegraphics[width=0.45\textwidth]{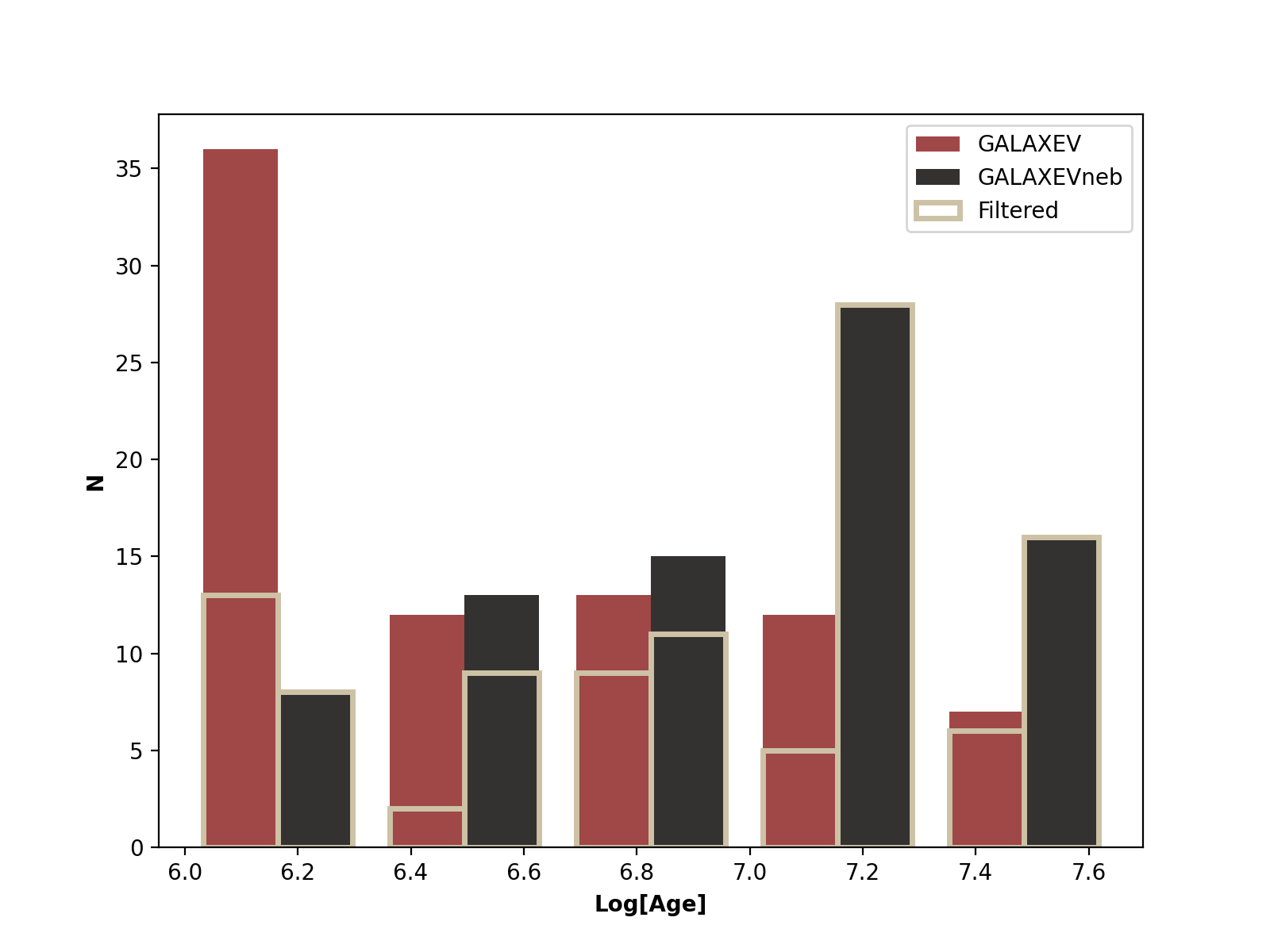}
\caption{This plot shows age histograms for our entire cluster sample, using our modified version of GALAXEV that includes nebular emission (black) as well as unmodified (i.e., stellar continuum only) GALAXEV (red). Fits that were otherwise rejected due to having reduced $\chi^2 > 10$ are included here as to better demonstrate this phenomenon. The grey open bars show the relative level in each bin if such fits are actually rejected. Note that although raw GALAXEV produced many fits with ages below $\sim 10^7$ yr, very few of these fits were good enough to be accepted.} 
\label{fig:nebHists}
\end{figure}


\subsection{Comparison With Previous Results}
We are by no means the first to benchmark the performance of SPS models using YSCs. The best and most recent outstanding example of such an analysis is \citep{Wofford}, who test 7 different SPS models against 8 clusters in the LEGUS sample with a focus on understanding the impact of massive-star prescriptions. Through a similar analysis to our own, they find that typical maximum differences in properties derived using different models for an individual cluster are 0.09 mag in E(V-B), a factor of 2.8 in mass, and a factor of 2.5 in age. These are generally smaller maximum offsets than we find, perhaps due to contamination in our photometry driven by our large apertures. The key advances provided by our work are the comparison with FUV spectroscopy and the analysis of how the choice of extinction curve affects SED fitting results as a function of $R_V$. We also use a larger cluster sample than existing studies. 

FUV spectra and extinction curves aside, our approach differs from \citep{Wofford} primarily in how we think about the SPS models themselves. \citet{Wofford} is interested in performing precise tests of how different ways of handling massive-star evolution effects the results of SED fitting - they, for example, test how using different input stellar tracks affects fits produced using SB99. We, on the other hand, are interested in testing SPS models that are publicly available and (with the exception of GALAXEVneb) are usable ``off-the-shelf". Our philosophy is that these codes are used commonly throughout the literature, and it is uncommon for teams to test how adopting a different SPS code would alter their results. 

\subsection{How Should These Systematics Be Managed When Performing And Interpreting SED Fits?}
We have demonstrated that the choice of both SPS model and extinction curve can introduce scatter into the results of SED fitting. This scatter is mostly random; model-model and curve-curve differences are more modest when averaged over our whole cluster sample and there is generally little evidence of ordered bias. However, we note that the choice of extinction curve introduces a subtle systematic such that increasing $R_V$ tends to increase the mass and total extinction one derives. 

Given that this scatter can be quite strong - over an order of magnitude in some cases - we suggest that the most conservative approach is to run multiple SED fits for each object of interest using various reasonable combinations of underlying SSP and extinction curve. Though more computationally expensive than a single run, this allows one to identify situations where the idiosyncrasies of model or data produce large scatter. This also allows one to propagate the scatter introduced by model and extinction curve choice throughout the rest of one`s analysis, if desired. We also suggest that great care be taken when comparing the properties of different objects if those properties were derived using different SPS models. After all, if the apparent difference between two objects could feasibly be attributed to scatter, it is difficult to convincingly interpret that difference as an actual result.

\subsection{How Broadly Applicable Are These Results?}
Our analysis here has some limitations that should be kept in mind. First, our sample is limited to young ($< \sim30$ Myr), relatively unobscured, stellar clusters. It is not clear how our results would change if extended to clusters with older ages and higher levels of dust attenuation, where we might expect the age-extinction degeneracy to play a more significant role. It is also unclear how our results would change were an analogous analysis to be performed for entire galaxies rather than individual star clusters. Such an experiment would likely be impacted by factors such as a more complicated star formation history. Still, in broad terms, we are able to show that careful SED fitting should account for the possibility of scatter introduced by SPS model and extinction curve choice. 

\section{SUMMARY AND CONCLUSIONS}
In this paper, we perform a comprehensive test of stellar population synthesis models and extinction curves via SED fitting of young stellar clusters in the CLUES sample. Our intention is to identify and quantify any systematic effects introduced into the results of SED fitting through SPS model or extinction curve choice. In particular, we test four commonly used SPS codes in ``off-the-shelf" forms that would likely be encountered by a typical end user: YGGDRASIL \citep{YGGDRASIL}, BPASS \citep{Eldridge2017}, FSPS \citep{Conroy2009,Conroy2010}, and a modified version of GALAXEV \citep[][2016 version]{GALAXEV}. Additionally, we test the extinction curve parameterized by \citet{Gordon2023} at four different values of $R_V$ spanning a reasonable range: 2.3, 3.1, 4.1, and 5.1.

We begin by performing aperture photometry of 18 young stellar clusters in the CLUES sample using HST ACS and WFC3 data from the LEGUS survey, being careful to use an aperture precisely matching that of HST's COS instrument. This is because the CLUES clusters are associated with COS FUV spectroscopy and spectral fits, which are useful as a baseline for comparison with our work in the broadband photometry.

We then perform SED fitting via a Monte Carlo Markov Chain approach, which gives us a reliable estimate of the posterior distributions on each physical parameter of interest (age, mass and extinction). 16 fits are performed for each cluster - one per unique combination of SPS model and extinction curve - for a total of 288. We reject any fits with a reduced  $\chi^2 > 10$, leaving us with a final sample of 243 fits suitable for further analysis. 

Averaged across all clusters, we find (via the KS test) that the samples of best-fit age and extinction are often drawn from different parent distributions on a model-model basis. This suggests that different models ``see'' different cluster populations given the same data. From this perspective, the effect of extinction curve is less important with only the derived extinctions often drawn from different parent distributions on a $R_V$-$R_V$ basis. 

We find that the agreement between broadband photometric SED fits and FUV-based spectral fits is generally poor. Depending on how exactly one defines ``agreement", broadband photometry agrees with the FUV in age terms between $32.1\% - 64.1\%$ of the time in age, $20.6\% - 64.1\%$ of the time in mass, and $18.5\% - 49.7\%$ in extinction. We find that different some SPS models perform better than others in terms of how well they agree with the FUV. The reason for these offsets is unclear, but are possibly due to the FUV and optical broadband data better tracing different underlying stellar populations. Differences between the model used to perform the FUV fits and the models tested here may also play a role.

Finally, we find that the scatter introduced by SPS model choice can be large, but model choice introduces minimal ordered bias. The model-derived scatter for an individual cluster can be quite large, up to a factor of 34.8 Myr in age, a factor of 8.3 in mass, and 0.21mag in extinction. The median scatter introduced by model choice is much smaller but still significant; it is 2.6 Myr in age, a factor of 1.9 in mass, and 0.08 mag in extinction. We find a similar picture for extinction curves, though we note that as one pushes to higher $R_V$ one tends to systematically recover higher masses and more total extinction. Large, scatter for single objects exist here, too; up to 16.3 Myr in age, a factor of 10.3 in mass, and 0.28mag in extinction. Median scatter
between extinction curves was a 0.48 Myr in age, a factor of 1.31 in mass, and 0.04 mag in extinction.

For individual sources, SPS model and extinction curve choice can introduce large scatter. We thus make two strong recommendations: First, we suggest that when performing SED fits, one should explore various reasonable combinations of SPS model and extinction curve. This should make any outliers obvious and provide a measure of the systematic uncertainty involved. Second, we caution that because of the potentially significant scatter introduced by model and extinction curve choice, one-to-one comparisons between the properties of individual objects derived using different SED fitting setups may not be meaningful.

We also suggest that more work be done to understand the interplay between nebular emission modeling and the results of SED fitting. Because we wanted to test various SPS+nebular emission models as they exist off-the shelf, a detailed examination of nebular emission modeling was out-of-scope for this paper. However, we have demonstrated that nebular emission is clearly an important component of any realistic SED model. A careful study of which nebular emission parameters are important in SED fitting, and over what range these parameters are physically reasonable, would be of clear value.

\section{Acknowledgments}
A.A and M.S. acknowledge support from the Swedish National Space Agency (SNSA) through the grant Dnr158/19. A. W. acknowledges UNAM's DGAPA for support during her sabbatical at UCSD, USA, through program PASPA; and additional support through project PAPIIT-DGAPA IN106922. KG is supported by the Australian Research Council through the Discovery Early Career Researcher Award (DECRA) Fellowship (project number DE220100766) funded by the Australian Government. KG is supported by the Australian Research Council Centre of Excellence for All Sky Astrophysics in 3 Dimensions (ASTRO~3D), through project number CE170100013. M.J.H. is fellow of the Knut \& Alice Wallenberg Foundation. Calculations were performed on the University of Massachusetts' Green High Performance Computing Cluster (MGHPCC) and Unity Cluster. The CLUES data presented in this article were obtained from the Mikulski Archive for Space Telescopes (MAST) at the Space Telescope Science Institute. The specific observations analyzed can be accessed via \dataset[doi: 0.17909/2bay-1d16]{https://doi.org/0.17909/2bay-1d16}. We thank our anonymous referee for comments and suggestions that greatly improved the quality of this manuscript.

\appendix
\section{Cluster Fitted Parameters}

\label{sec:appendixA}

\startlongtable
\begin{deluxetable*}{l|c||ccccccc}
\tablenum{3}
\tablecaption{Best-Fit Parameters: All Clusters, All Models, All Curves}
\tablehead{\colhead{Model} & \colhead{R$_V$} & \colhead{Age$_{ML}$} & \colhead{Age$_{MED}$} & \colhead{Mass$_{ML}$} & \colhead{Mass$_{MED}$} & \colhead{E(B-V)$_{ML}$} & \colhead{E(B-V)$_{MED}$} & \colhead{$\chi^2/\nu$}\\\colhead{} & \colhead{} & \colhead{Log$_{10}$[yr]} & \colhead{Log$_{10}$[yr]} & \colhead{Log$_{10}$[M$_{\odot}$]} & \colhead{Log$_{10}$[M$_{\odot}$]} & \colhead{Mag} & \colhead{Mag} & \colhead{}} 
\startdata
\sidehead{\textbf{M74-1}}
\hline\hline
YGGDRASIL & 2.3 & 6.053 & ${6.138}_{-0.092}^{+0.139}$ & 4.877 & ${4.865}_{-0.042}^{+0.029}$ & 0.192 & ${0.205}_{-0.021}^{+0.024}$ & 4.183\\
 & 3.1 & 6.093 & ${6.186}_{-0.12}^{+0.157}$ & 4.945 & ${4.925}_{-0.055}^{+0.038}$ & 0.202 & ${0.218}_{-0.024}^{+0.026}$ & 7.19\\
 & 4.1 & 6.348 & ${6.258}_{-0.161}^{+0.159}$ & 4.973 & ${5.001}_{-0.059}^{+0.047}$ & 0.249 & ${0.233}_{-0.027}^{+0.027}$ & 0.445\\
 & 5.1 & 6.5 & ${6.328}_{-0.194}^{+0.17}$ & 5.024 & ${5.076}_{-0.057}^{+0.055}$ & 0.258 & ${0.243}_{-0.029}^{+0.026}$ & 0.754\\
\hline\hline
BPASS & 2.3 & 6.735 & ${6.705}_{-0.448}^{+0.046}$ & 4.87 & ${4.832}_{-0.273}^{+0.047}$ & 0.114 & ${0.144}_{-0.048}^{+0.069}$ & 1.35\\
 & 3.1 & 6.724 & ${6.449}_{-0.153}^{+0.278}$ & 4.91 & ${4.731}_{-0.082}^{+0.172}$ & 0.128 & ${0.199}_{-0.077}^{+0.029}$ & 1.071\\
 & 4.1 & 6.465 & ${6.48}_{-0.115}^{+0.215}$ & 4.824 & ${4.827}_{-0.059}^{+0.105}$ & 0.223 & ${0.207}_{-0.062}^{+0.028}$ & 0.08\\
 & 5.1 & 6.522 & ${6.541}_{-0.094}^{+0.139}$ & 4.897 & ${4.921}_{-0.058}^{+0.073}$ & 0.206 & ${0.195}_{-0.044}^{+0.038}$ & 0.315\\
\hline\hline
GALAXEV & 2.3 & 6.797 & ${6.812}_{-0.02}^{+0.029}$ & 4.613 & ${4.64}_{-0.039}^{+0.038}$ & 0.039 & ${0.035}_{-0.018}^{+0.019}$ & 1.436\\
 & 3.1 & 6.797 & ${6.811}_{-0.019}^{+0.029}$ & 4.625 & ${4.65}_{-0.04}^{+0.039}$ & 0.04 & ${0.036}_{-0.019}^{+0.02}$ & 1.463\\
 & 4.1 & 6.796 & ${6.808}_{-0.018}^{+0.029}$ & 4.639 & ${4.662}_{-0.044}^{+0.042}$ & 0.041 & ${0.037}_{-0.02}^{+0.021}$ & 1.53\\
 & 5.1 & 6.795 & ${6.807}_{-0.017}^{+0.029}$ & 4.653 & ${4.673}_{-0.046}^{+0.046}$ & 0.041 & ${0.037}_{-0.02}^{+0.022}$ & 1.645\\
\hline\hline
FSPS & 2.3 & 6.582 & ${6.585}_{-0.01}^{+0.017}$ & 4.377 & ${4.366}_{-0.056}^{+0.049}$ & 0.067 & ${0.061}_{-0.034}^{+0.027}$ & 11.232 (*)\\
 & 3.1 & 6.581 & ${6.583}_{-0.01}^{+0.014}$ & 4.408 & ${4.401}_{-0.064}^{+0.053}$ & 0.073 & ${0.07}_{-0.031}^{+0.026}$ & 10.846 (*)\\
 & 4.1 & 6.581 & ${6.582}_{-0.009}^{+0.011}$ & 4.447 & ${4.444}_{-0.067}^{+0.057}$ & 0.08 & ${0.079}_{-0.028}^{+0.024}$ & 10.219 (*)\\
 & 5.1 & 6.582 & ${6.582}_{-0.008}^{+0.01}$ & 4.485 & ${4.485}_{-0.069}^{+0.064}$ & 0.085 & ${0.085}_{-0.025}^{+0.024}$ & 9.504 (*)\\
\hline\hline
\sidehead{\textbf{M74-2}}
\hline\hline
YGGDRASIL & 2.3 & 6.0 & ${6.237}_{-0.163}^{+0.235}$ & 4.343 & ${4.313}_{-0.093}^{+0.085}$ & 0.112 & ${0.133}_{-0.049}^{+0.047}$ & 0.195\\
 & 3.1 & 6.0 & ${6.252}_{-0.171}^{+0.229}$ & 4.388 & ${4.367}_{-0.102}^{+0.095}$ & 0.121 & ${0.147}_{-0.051}^{+0.049}$ & 0.11\\
 & 4.1 & 6.0 & ${6.265}_{-0.178}^{+0.218}$ & 4.451 & ${4.447}_{-0.12}^{+0.114}$ & 0.131 & ${0.165}_{-0.054}^{+0.053}$ & 0.032\\
 & 5.1 & 6.0 & ${6.283}_{-0.189}^{+0.218}$ & 4.517 & ${4.538}_{-0.138}^{+0.132}$ & 0.14 & ${0.181}_{-0.056}^{+0.055}$ & 0.016\\
\hline\hline
BPASS & 2.3 & 6.719 & ${6.588}_{-0.383}^{+0.267}$ & 4.35 & ${4.242}_{-0.186}^{+0.26}$ & 0.047 & ${0.088}_{-0.062}^{+0.07}$ & 0.409\\
 & 3.1 & 6.231 & ${6.513}_{-0.314}^{+0.314}$ & 4.118 & ${4.273}_{-0.164}^{+0.225}$ & 0.148 & ${0.109}_{-0.076}^{+0.07}$ & 0.247\\
 & 4.1 & 6.287 & ${6.462}_{-0.27}^{+0.322}$ & 4.226 & ${4.327}_{-0.156}^{+0.179}$ & 0.167 & ${0.134}_{-0.089}^{+0.068}$ & 0.138\\
 & 5.1 & 6.452 & ${6.452}_{-0.253}^{+0.303}$ & 4.399 & ${4.407}_{-0.16}^{+0.156}$ & 0.178 & ${0.154}_{-0.095}^{+0.069}$ & 0.102\\
\hline\hline
GALAXEV & 2.3 & 6.424 & ${6.269}_{-0.169}^{+0.133}$ & 4.154 & ${4.151}_{-0.079}^{+0.075}$ & 0.188 & ${0.174}_{-0.057}^{+0.05}$ & 0.628\\
 & 3.1 & 6.424 & ${6.266}_{-0.166}^{+0.13}$ & 4.231 & ${4.228}_{-0.098}^{+0.093}$ & 0.203 & ${0.192}_{-0.056}^{+0.052}$ & 0.385\\
 & 4.1 & 6.372 & ${6.261}_{-0.163}^{+0.126}$ & 4.348 & ${4.341}_{-0.12}^{+0.114}$ & 0.222 & ${0.214}_{-0.056}^{+0.054}$ & 0.21\\
 & 5.1 & 6.426 & ${6.248}_{-0.155}^{+0.131}$ & 4.442 & ${4.457}_{-0.145}^{+0.14}$ & 0.233 & ${0.232}_{-0.059}^{+0.057}$ & 0.077\\
\hline\hline
FSPS & 2.3 & 6.557 & ${6.537}_{-0.061}^{+0.033}$ & 4.114 & ${4.096}_{-0.075}^{+0.08}$ & 0.066 & ${0.062}_{-0.039}^{+0.047}$ & 0.851\\
 & 3.1 & 6.557 & ${6.538}_{-0.058}^{+0.032}$ & 4.145 & ${4.126}_{-0.09}^{+0.098}$ & 0.074 & ${0.07}_{-0.043}^{+0.05}$ & 0.79\\
 & 4.1 & 6.556 & ${6.539}_{-0.055}^{+0.031}$ & 4.187 & ${4.173}_{-0.114}^{+0.12}$ & 0.082 & ${0.08}_{-0.047}^{+0.053}$ & 0.716\\
 & 5.1 & 6.557 & ${6.539}_{-0.054}^{+0.031}$ & 4.233 & ${4.224}_{-0.142}^{+0.149}$ & 0.089 & ${0.09}_{-0.053}^{+0.058}$ & 0.65\\
\hline\hline
\sidehead{\textbf{NGC1512-1}}
\hline\hline
YGGDRASIL & 2.3 & 6.753 & ${6.289}_{-0.191}^{+0.322}$ & 4.207 & ${4.37}_{-0.099}^{+0.094}$ & 0.0 & ${0.144}_{-0.06}^{+0.054}$ & 0.169\\
 & 3.1 & 6.753 & ${6.287}_{-0.19}^{+0.3}$ & 4.207 & ${4.435}_{-0.115}^{+0.109}$ & 0.0 & ${0.16}_{-0.061}^{+0.058}$ & 0.169\\
 & 4.1 & 6.753 & ${6.299}_{-0.194}^{+0.275}$ & 4.207 & ${4.533}_{-0.137}^{+0.132}$ & 0.0 & ${0.184}_{-0.066}^{+0.063}$ & 0.169\\
 & 5.1 & 6.197 & ${6.311}_{-0.201}^{+0.261}$ & 4.649 & ${4.653}_{-0.167}^{+0.156}$ & 0.188 & ${0.21}_{-0.07}^{+0.066}$ & 0.751\\
\hline\hline
BPASS & 2.3 & 6.968 & ${6.777}_{-0.363}^{+0.213}$ & 4.616 & ${4.501}_{-0.274}^{+0.203}$ & 0.0 & ${0.072}_{-0.05}^{+0.082}$ & 0.048\\
 & 3.1 & 6.968 & ${6.74}_{-0.383}^{+0.23}$ & 4.616 & ${4.497}_{-0.232}^{+0.208}$ & 0.0 & ${0.088}_{-0.062}^{+0.091}$ & 0.048\\
 & 4.1 & 6.968 & ${6.694}_{-0.371}^{+0.256}$ & 4.616 & ${4.532}_{-0.195}^{+0.197}$ & 0.0 & ${0.112}_{-0.078}^{+0.101}$ & 0.048\\
 & 5.1 & 6.968 & ${6.617}_{-0.335}^{+0.303}$ & 4.616 & ${4.595}_{-0.183}^{+0.176}$ & 0.0 & ${0.147}_{-0.1}^{+0.101}$ & 0.049\\
\hline\hline
GALAXEV & 2.3 & 6.419 & ${6.649}_{-0.478}^{+0.601}$ & 4.245 & ${4.37}_{-0.182}^{+0.576}$ & 0.212 & ${0.099}_{-0.08}^{+0.121}$ & 0.572\\
 & 3.1 & 6.419 & ${6.259}_{-0.154}^{+0.132}$ & 4.333 & ${4.323}_{-0.114}^{+0.103}$ & 0.231 & ${0.213}_{-0.065}^{+0.059}$ & 0.375\\
 & 4.1 & 6.422 & ${6.268}_{-0.159}^{+0.13}$ & 4.444 & ${4.46}_{-0.145}^{+0.143}$ & 0.248 & ${0.238}_{-0.076}^{+0.064}$ & 0.177\\
 & 5.1 & 6.424 & ${6.308}_{-0.181}^{+0.771}$ & 4.565 & ${4.653}_{-0.192}^{+0.237}$ & 0.265 & ${0.249}_{-0.179}^{+0.078}$ & 0.08\\
\hline\hline
FSPS & 2.3 & 6.59 & ${6.557}_{-0.05}^{+0.035}$ & 4.061 & ${4.126}_{-0.074}^{+0.085}$ & 0.016 & ${0.06}_{-0.04}^{+0.052}$ & 0.636\\
 & 3.1 & 6.585 & ${6.556}_{-0.049}^{+0.035}$ & 4.095 & ${4.155}_{-0.09}^{+0.106}$ & 0.031 & ${0.068}_{-0.044}^{+0.057}$ & 0.659\\
 & 4.1 & 6.579 & ${6.555}_{-0.047}^{+0.034}$ & 4.147 & ${4.205}_{-0.118}^{+0.134}$ & 0.05 & ${0.082}_{-0.052}^{+0.061}$ & 0.66\\
 & 5.1 & 6.575 & ${6.554}_{-0.043}^{+0.033}$ & 4.211 & ${4.28}_{-0.157}^{+0.171}$ & 0.069 & ${0.101}_{-0.06}^{+0.069}$ & 0.626\\
\hline\hline
\sidehead{\textbf{NGC1512-2}}
\hline\hline
YGGDRASIL & 2.3 & 6.0 & ${6.115}_{-0.08}^{+0.141}$ & 4.678 & ${4.679}_{-0.044}^{+0.034}$ & 0.072 & ${0.084}_{-0.021}^{+0.023}$ & 3.68\\
 & 3.1 & 6.0 & ${6.137}_{-0.096}^{+0.158}$ & 4.712 & ${4.711}_{-0.052}^{+0.042}$ & 0.08 & ${0.096}_{-0.023}^{+0.026}$ & 3.127\\
 & 4.1 & 6.0 & ${6.18}_{-0.125}^{+0.162}$ & 4.76 & ${4.76}_{-0.058}^{+0.051}$ & 0.091 & ${0.112}_{-0.026}^{+0.028}$ & 2.373\\
 & 5.1 & 6.0 & ${6.223}_{-0.155}^{+0.155}$ & 4.816 & ${4.819}_{-0.063}^{+0.059}$ & 0.101 & ${0.128}_{-0.029}^{+0.03}$ & 1.609\\
\hline\hline
BPASS & 2.3 & 6.753 & ${6.733}_{-0.487}^{+0.047}$ & 4.716 & ${4.717}_{-0.299}^{+0.048}$ & 0.0 & ${0.016}_{-0.012}^{+0.055}$ & 1.9\\
 & 3.1 & 6.753 & ${6.719}_{-0.558}^{+0.052}$ & 4.716 & ${4.706}_{-0.291}^{+0.057}$ & 0.0 & ${0.023}_{-0.017}^{+0.082}$ & 1.901\\
 & 4.1 & 6.753 & ${6.305}_{-0.191}^{+0.441}$ & 4.716 & ${4.559}_{-0.114}^{+0.186}$ & 0.0 & ${0.102}_{-0.088}^{+0.032}$ & 1.899\\
 & 5.1 & 6.225 & ${6.283}_{-0.154}^{+0.228}$ & 4.57 & ${4.617}_{-0.102}^{+0.111}$ & 0.131 & ${0.129}_{-0.051}^{+0.027}$ & 0.91\\
\hline\hline
GALAXEV & 2.3 & 6.174 & ${6.209}_{-0.106}^{+0.143}$ & 4.488 & ${4.498}_{-0.038}^{+0.036}$ & 0.127 & ${0.134}_{-0.026}^{+0.026}$ & 4.879\\
 & 3.1 & 6.394 & ${6.238}_{-0.113}^{+0.135}$ & 4.59 & ${4.562}_{-0.046}^{+0.043}$ & 0.176 & ${0.15}_{-0.027}^{+0.026}$ & 4.15\\
 & 4.1 & 6.393 & ${6.266}_{-0.115}^{+0.116}$ & 4.683 & ${4.652}_{-0.056}^{+0.053}$ & 0.192 & ${0.17}_{-0.028}^{+0.027}$ & 2.338\\
 & 5.1 & 6.389 & ${6.278}_{-0.112}^{+0.104}$ & 4.781 & ${4.751}_{-0.069}^{+0.065}$ & 0.206 & ${0.187}_{-0.029}^{+0.028}$ & 0.873\\
\hline\hline
FSPS & 2.3 & 6.565 & ${6.559}_{-0.014}^{+0.012}$ & 4.436 & ${4.451}_{-0.029}^{+0.034}$ & 0.018 & ${0.027}_{-0.017}^{+0.021}$ & 9.947\\
 & 3.1 & 6.563 & ${6.558}_{-0.013}^{+0.012}$ & 4.457 & ${4.468}_{-0.037}^{+0.042}$ & 0.026 & ${0.033}_{-0.02}^{+0.023}$ & 9.914\\
 & 4.1 & 6.56 & ${6.557}_{-0.013}^{+0.011}$ & 4.49 & ${4.498}_{-0.051}^{+0.052}$ & 0.038 & ${0.042}_{-0.023}^{+0.024}$ & 9.697\\
 & 5.1 & 6.559 & ${6.556}_{-0.013}^{+0.011}$ & 4.531 & ${4.536}_{-0.064}^{+0.063}$ & 0.049 & ${0.051}_{-0.025}^{+0.026}$ & 9.288\\
\hline\hline
\sidehead{\textbf{NGC1566-2}}
\hline\hline
YGGDRASIL & 2.3 & 6.735 & ${6.35}_{-0.227}^{+0.331}$ & 5.202 & ${5.284}_{-0.087}^{+0.086}$ & 0.018 & ${0.117}_{-0.053}^{+0.042}$ & 0.199\\
 & 3.1 & 6.734 & ${6.335}_{-0.211}^{+0.325}$ & 5.212 & ${5.343}_{-0.104}^{+0.101}$ & 0.021 & ${0.136}_{-0.055}^{+0.044}$ & 0.197\\
 & 4.1 & 6.732 & ${6.332}_{-0.207}^{+0.299}$ & 5.232 & ${5.441}_{-0.125}^{+0.118}$ & 0.026 & ${0.163}_{-0.056}^{+0.048}$ & 0.196\\
 & 5.1 & 6.731 & ${6.332}_{-0.202}^{+0.274}$ & 5.252 & ${5.566}_{-0.153}^{+0.142}$ & 0.031 & ${0.192}_{-0.06}^{+0.054}$ & 0.195\\
\hline\hline
BPASS & 2.3 & 6.689 & ${6.819}_{-0.168}^{+0.165}$ & 5.333 & ${5.489}_{-0.211}^{+0.17}$ & 0.064 & ${0.052}_{-0.034}^{+0.045}$ & 0.171\\
 & 3.1 & 6.681 & ${6.796}_{-0.185}^{+0.176}$ & 5.347 & ${5.494}_{-0.215}^{+0.17}$ & 0.07 & ${0.061}_{-0.04}^{+0.054}$ & 0.17\\
 & 4.1 & 6.67 & ${6.757}_{-0.242}^{+0.2}$ & 5.371 & ${5.503}_{-0.19}^{+0.175}$ & 0.079 & ${0.078}_{-0.052}^{+0.076}$ & 0.155\\
 & 5.1 & 6.656 & ${6.702}_{-0.28}^{+0.226}$ & 5.4 & ${5.54}_{-0.166}^{+0.164}$ & 0.089 & ${0.107}_{-0.071}^{+0.099}$ & 0.126\\
\hline\hline
GALAXEV & 2.3 & 6.706 & ${6.386}_{-0.151}^{+0.3}$ & 5.057 & ${5.136}_{-0.071}^{+0.073}$ & 0.008 & ${0.153}_{-0.118}^{+0.053}$ & 0.743\\
 & 3.1 & 6.708 & ${7.048}_{-0.723}^{+0.182}$ & 5.056 & ${5.694}_{-0.51}^{+0.186}$ & 0.005 & ${0.046}_{-0.035}^{+0.157}$ & 0.738\\
 & 4.1 & 6.415 & ${6.424}_{-0.158}^{+0.774}$ & 5.395 & ${5.464}_{-0.188}^{+0.4}$ & 0.245 & ${0.16}_{-0.141}^{+0.096}$ & 0.612\\
 & 5.1 & 6.417 & ${6.33}_{-0.129}^{+0.09}$ & 5.555 & ${5.532}_{-0.177}^{+0.143}$ & 0.277 & ${0.26}_{-0.07}^{+0.055}$ & 0.473\\
\hline\hline
FSPS & 2.3 & 6.589 & ${6.565}_{-0.051}^{+0.026}$ & 5.029 & ${5.093}_{-0.065}^{+0.078}$ & 0.017 & ${0.056}_{-0.036}^{+0.042}$ & 0.557\\
 & 3.1 & 6.588 & ${6.563}_{-0.05}^{+0.027}$ & 5.044 & ${5.125}_{-0.084}^{+0.097}$ & 0.022 & ${0.065}_{-0.041}^{+0.047}$ & 0.562\\
 & 4.1 & 6.586 & ${6.562}_{-0.046}^{+0.025}$ & 5.072 & ${5.179}_{-0.113}^{+0.123}$ & 0.031 & ${0.079}_{-0.048}^{+0.052}$ & 0.567\\
 & 5.1 & 6.585 & ${6.562}_{-0.045}^{+0.025}$ & 5.108 & ${5.241}_{-0.146}^{+0.155}$ & 0.041 & ${0.093}_{-0.055}^{+0.057}$ & 0.564\\
\hline\hline
\sidehead{\textbf{NGC1566-1}}
\hline\hline
YGGDRASIL & 2.3 & 7.246 & ${7.201}_{-0.153}^{+0.103}$ & 5.791 & ${5.756}_{-0.229}^{+0.095}$ & 0.083 & ${0.081}_{-0.05}^{+0.058}$ & 0.243\\
 & 3.1 & 7.233 & ${7.195}_{-0.153}^{+0.104}$ & 5.838 & ${5.783}_{-0.231}^{+0.114}$ & 0.102 & ${0.092}_{-0.057}^{+0.065}$ & 0.261\\
 & 4.1 & 7.191 & ${7.183}_{-0.18}^{+0.105}$ & 5.91 & ${5.826}_{-0.219}^{+0.147}$ & 0.137 & ${0.113}_{-0.067}^{+0.078}$ & 0.118\\
 & 5.1 & 7.159 & ${7.17}_{-0.381}^{+0.113}$ & 5.939 & ${5.884}_{-0.202}^{+0.179}$ & 0.148 & ${0.135}_{-0.078}^{+0.1}$ & 0.099\\
\hline\hline
BPASS & 2.3 & 7.285 & ${7.303}_{-0.169}^{+0.109}$ & 5.743 & ${5.821}_{-0.151}^{+0.098}$ & 0.0 & ${0.031}_{-0.022}^{+0.034}$ & 0.634\\
 & 3.1 & 7.285 & ${7.306}_{-0.163}^{+0.107}$ & 5.743 & ${5.838}_{-0.147}^{+0.1}$ & 0.0 & ${0.034}_{-0.023}^{+0.037}$ & 0.634\\
 & 4.1 & 7.285 & ${7.312}_{-0.146}^{+0.102}$ & 5.743 & ${5.864}_{-0.139}^{+0.109}$ & 0.0 & ${0.038}_{-0.026}^{+0.04}$ & 0.633\\
 & 5.1 & 7.284 & ${7.317}_{-0.132}^{+0.099}$ & 5.743 & ${5.889}_{-0.141}^{+0.124}$ & 0.0 & ${0.042}_{-0.028}^{+0.043}$ & 0.633\\
\hline\hline
GALAXEV & 2.3 & 7.285 & ${7.239}_{-0.183}^{+0.157}$ & 5.797 & ${5.785}_{-0.236}^{+0.106}$ & 0.082 & ${0.084}_{-0.052}^{+0.056}$ & 0.059\\
 & 3.1 & 7.27 & ${7.236}_{-0.188}^{+0.158}$ & 5.846 & ${5.818}_{-0.252}^{+0.117}$ & 0.102 & ${0.093}_{-0.057}^{+0.061}$ & 0.065\\
 & 4.1 & 7.181 & ${7.228}_{-0.192}^{+0.159}$ & 5.883 & ${5.865}_{-0.253}^{+0.139}$ & 0.144 & ${0.109}_{-0.066}^{+0.068}$ & 0.117\\
 & 5.1 & 7.358 & ${7.225}_{-0.211}^{+0.16}$ & 5.875 & ${5.92}_{-0.246}^{+0.168}$ & 0.057 & ${0.123}_{-0.074}^{+0.076}$ & 0.146\\
\hline\hline
FSPS & 2.3 & 6.898 & ${7.024}_{-0.091}^{+0.126}$ & 5.119 & ${5.42}_{-0.139}^{+0.168}$ & 0.019 & ${0.053}_{-0.037}^{+0.049}$ & 0.337\\
 & 3.1 & 6.9 & ${7.018}_{-0.087}^{+0.101}$ & 5.144 & ${5.444}_{-0.144}^{+0.153}$ & 0.027 & ${0.062}_{-0.042}^{+0.054}$ & 0.328\\
 & 4.1 & 6.905 & ${7.01}_{-0.082}^{+0.091}$ & 5.199 & ${5.483}_{-0.158}^{+0.159}$ & 0.043 & ${0.075}_{-0.051}^{+0.06}$ & 0.302\\
 & 5.1 & 7.017 & ${7.005}_{-0.077}^{+0.087}$ & 5.499 & ${5.542}_{-0.174}^{+0.171}$ & 0.077 & ${0.092}_{-0.06}^{+0.065}$ & 0.22\\
\hline\hline
\sidehead{\textbf{NGC4485-1}}
\hline\hline
YGGDRASIL & 2.3 & 6.681 & ${6.677}_{-0.016}^{+0.013}$ & 4.652 & ${4.66}_{-0.014}^{+0.016}$ & 0.0 & ${0.005}_{-0.004}^{+0.008}$ & 8.708\\
 & 3.1 & 6.681 & ${6.677}_{-0.015}^{+0.013}$ & 4.652 & ${4.663}_{-0.015}^{+0.018}$ & 0.0 & ${0.006}_{-0.004}^{+0.008}$ & 8.701\\
 & 4.1 & 6.681 & ${6.677}_{-0.015}^{+0.013}$ & 4.652 & ${4.666}_{-0.016}^{+0.021}$ & 0.0 & ${0.006}_{-0.005}^{+0.009}$ & 8.7\\
 & 5.1 & 6.681 & ${6.677}_{-0.015}^{+0.013}$ & 4.652 & ${4.671}_{-0.019}^{+0.027}$ & 0.0 & ${0.007}_{-0.005}^{+0.011}$ & 8.7\\
\hline\hline
BPASS & 2.3 & 6.7 & ${6.7}_{-0.017}^{+0.016}$ & 4.635 & ${4.64}_{-0.028}^{+0.027}$ & 0.0 & ${0.002}_{-0.002}^{+0.004}$ & 61.28 (*)\\
 & 3.1 & 6.7 & ${6.7}_{-0.017}^{+0.016}$ & 4.635 & ${4.641}_{-0.028}^{+0.027}$ & 0.0 & ${0.002}_{-0.002}^{+0.004}$ & 61.28 (*)\\
 & 4.1 & 6.7 & ${6.699}_{-0.017}^{+0.016}$ & 4.635 & ${4.642}_{-0.028}^{+0.027}$ & 0.0 & ${0.003}_{-0.002}^{+0.004}$ & 61.28 (*)\\
 & 5.1 & 6.7 & ${6.7}_{-0.017}^{+0.016}$ & 4.635 & ${4.644}_{-0.028}^{+0.027}$ & 0.0 & ${0.003}_{-0.002}^{+0.004}$ & 61.28 (*)\\
\hline\hline
GALAXEV & 2.3 & 6.735 & ${6.734}_{-0.006}^{+0.006}$ & 4.618 & ${4.625}_{-0.014}^{+0.015}$ & 0.0 & ${0.004}_{-0.003}^{+0.006}$ & 12.225 (*)\\
 & 3.1 & 6.735 & ${6.734}_{-0.006}^{+0.006}$ & 4.618 & ${4.627}_{-0.015}^{+0.016}$ & 0.0 & ${0.005}_{-0.003}^{+0.007}$ & 12.214 (*)\\
 & 4.1 & 6.735 & ${6.734}_{-0.006}^{+0.006}$ & 4.618 & ${4.629}_{-0.016}^{+0.019}$ & 0.0 & ${0.005}_{-0.004}^{+0.008}$ & 12.214 (*)\\
 & 5.1 & 6.735 & ${6.734}_{-0.006}^{+0.006}$ & 4.618 & ${4.632}_{-0.017}^{+0.023}$ & 0.0 & ${0.006}_{-0.004}^{+0.009}$ & 12.236 (*)\\
\hline\hline
FSPS & 2.3 & 6.58 & ${6.58}_{-0.007}^{+0.007}$ & 4.308 & ${4.312}_{-0.02}^{+0.02}$ & 0.0 & ${0.002}_{-0.001}^{+0.003}$ & 81.386 (*)\\
 & 3.1 & 6.58 & ${6.58}_{-0.007}^{+0.007}$ & 4.308 & ${4.313}_{-0.02}^{+0.02}$ & 0.0 & ${0.002}_{-0.001}^{+0.003}$ & 81.386 (*)\\
 & 4.1 & 6.58 & ${6.58}_{-0.007}^{+0.007}$ & 4.308 & ${4.314}_{-0.02}^{+0.02}$ & 0.0 & ${0.002}_{-0.001}^{+0.003}$ & 81.386 (*)\\
 & 5.1 & 6.58 & ${6.58}_{-0.007}^{+0.007}$ & 4.308 & ${4.316}_{-0.02}^{+0.021}$ & 0.0 & ${0.002}_{-0.002}^{+0.003}$ & 81.386 (*)\\
\hline\hline
\sidehead{\textbf{NGC4485-2}}
\hline\hline
YGGDRASIL & 2.3 & 6.644 & ${6.643}_{-0.015}^{+0.013}$ & 4.856 & ${4.854}_{-0.041}^{+0.038}$ & 0.125 & ${0.124}_{-0.02}^{+0.019}$ & 6.775\\
 & 3.1 & 6.646 & ${6.646}_{-0.014}^{+0.012}$ & 4.913 & ${4.912}_{-0.048}^{+0.045}$ & 0.136 & ${0.136}_{-0.021}^{+0.021}$ & 5.442\\
 & 4.1 & 6.649 & ${6.649}_{-0.014}^{+0.012}$ & 4.992 & ${4.992}_{-0.056}^{+0.054}$ & 0.149 & ${0.15}_{-0.022}^{+0.022}$ & 3.863\\
 & 5.1 & 6.65 & ${6.65}_{-0.014}^{+0.012}$ & 5.071 & ${5.074}_{-0.066}^{+0.064}$ & 0.16 & ${0.161}_{-0.023}^{+0.023}$ & 2.687\\
\hline\hline
BPASS & 2.3 & 6.735 & ${6.735}_{-0.017}^{+0.017}$ & 4.765 & ${4.766}_{-0.024}^{+0.023}$ & 0.035 & ${0.036}_{-0.017}^{+0.017}$ & 1.744\\
 & 3.1 & 6.734 & ${6.734}_{-0.017}^{+0.017}$ & 4.775 & ${4.777}_{-0.027}^{+0.027}$ & 0.036 & ${0.037}_{-0.018}^{+0.018}$ & 1.916\\
 & 4.1 & 6.734 & ${6.733}_{-0.018}^{+0.017}$ & 4.789 & ${4.791}_{-0.032}^{+0.032}$ & 0.037 & ${0.038}_{-0.019}^{+0.02}$ & 2.171\\
 & 5.1 & 6.733 & ${6.732}_{-0.019}^{+0.018}$ & 4.8 & ${4.804}_{-0.037}^{+0.04}$ & 0.036 & ${0.039}_{-0.02}^{+0.021}$ & 2.497\\
\hline\hline
GALAXEV & 2.3 & 6.715 & ${6.714}_{-0.011}^{+0.009}$ & 4.797 & ${4.794}_{-0.076}^{+0.056}$ & 0.123 & ${0.121}_{-0.029}^{+0.023}$ & 4.121\\
 & 3.1 & 6.716 & ${6.716}_{-0.011}^{+0.009}$ & 4.856 & ${4.855}_{-0.079}^{+0.06}$ & 0.135 & ${0.134}_{-0.029}^{+0.024}$ & 2.587\\
 & 4.1 & 6.717 & ${6.717}_{-0.01}^{+0.009}$ & 4.928 & ${4.93}_{-0.087}^{+0.068}$ & 0.146 & ${0.146}_{-0.029}^{+0.024}$ & 1.203\\
 & 5.1 & 6.715 & ${6.716}_{-0.011}^{+0.009}$ & 4.989 & ${4.997}_{-0.102}^{+0.081}$ & 0.151 & ${0.154}_{-0.03}^{+0.026}$ & 0.692\\
\hline\hline
FSPS & 2.3 & 6.586 & ${6.587}_{-0.005}^{+0.005}$ & 4.305 & ${4.31}_{-0.016}^{+0.018}$ & 0.014 & ${0.018}_{-0.011}^{+0.014}$ & 3.376\\
 & 3.1 & 6.586 & ${6.587}_{-0.005}^{+0.005}$ & 4.314 & ${4.32}_{-0.02}^{+0.023}$ & 0.017 & ${0.021}_{-0.013}^{+0.015}$ & 3.324\\
 & 4.1 & 6.587 & ${6.587}_{-0.005}^{+0.005}$ & 4.328 & ${4.334}_{-0.027}^{+0.03}$ & 0.022 & ${0.025}_{-0.014}^{+0.016}$ & 3.212\\
 & 5.1 & 6.587 & ${6.588}_{-0.005}^{+0.005}$ & 4.347 & ${4.354}_{-0.035}^{+0.038}$ & 0.027 & ${0.03}_{-0.016}^{+0.018}$ & 3.03\\
\hline\hline
\sidehead{\textbf{NGC7793-1}}
\hline\hline
YGGDRASIL & 2.3 & 6.21 & ${6.258}_{-0.139}^{+0.198}$ & 4.025 & ${4.004}_{-0.101}^{+0.057}$ & 0.024 & ${0.029}_{-0.017}^{+0.019}$ & 6.893\\
 & 3.1 & 6.217 & ${6.259}_{-0.139}^{+0.188}$ & 4.034 & ${4.018}_{-0.098}^{+0.057}$ & 0.027 & ${0.032}_{-0.019}^{+0.021}$ & 6.271\\
 & 4.1 & 6.225 & ${6.273}_{-0.146}^{+0.188}$ & 4.049 & ${4.033}_{-0.092}^{+0.065}$ & 0.032 & ${0.038}_{-0.022}^{+0.025}$ & 5.481\\
 & 5.1 & 6.229 & ${6.285}_{-0.155}^{+0.185}$ & 4.07 & ${4.055}_{-0.089}^{+0.076}$ & 0.037 & ${0.046}_{-0.025}^{+0.028}$ & 5.112\\
\hline\hline
BPASS & 2.3 & 6.526 & ${6.446}_{-0.104}^{+0.083}$ & 3.838 & ${3.829}_{-0.051}^{+0.051}$ & 0.0 & ${0.021}_{-0.015}^{+0.019}$ & 0.627\\
 & 3.1 & 6.526 & ${6.446}_{-0.105}^{+0.086}$ & 3.838 & ${3.842}_{-0.054}^{+0.059}$ & 0.0 & ${0.024}_{-0.017}^{+0.022}$ & 0.627\\
 & 4.1 & 6.526 & ${6.442}_{-0.101}^{+0.083}$ & 3.838 & ${3.861}_{-0.057}^{+0.066}$ & 0.0 & ${0.03}_{-0.02}^{+0.025}$ & 0.627\\
 & 5.1 & 6.526 & ${6.434}_{-0.099}^{+0.075}$ & 3.838 & ${3.882}_{-0.064}^{+0.069}$ & 0.0 & ${0.038}_{-0.024}^{+0.028}$ & 0.627\\
\hline\hline
GALAXEV & 2.3 & 6.468 & ${6.466}_{-0.01}^{+0.009}$ & 3.805 & ${3.808}_{-0.036}^{+0.037}$ & 0.036 & ${0.036}_{-0.019}^{+0.019}$ & 8.024\\
 & 3.1 & 6.467 & ${6.466}_{-0.01}^{+0.009}$ & 3.824 & ${3.826}_{-0.044}^{+0.045}$ & 0.041 & ${0.04}_{-0.021}^{+0.021}$ & 8.02\\
 & 4.1 & 6.467 & ${6.466}_{-0.01}^{+0.009}$ & 3.853 & ${3.855}_{-0.057}^{+0.059}$ & 0.047 & ${0.048}_{-0.024}^{+0.024}$ & 7.951\\
 & 5.1 & 6.467 & ${6.466}_{-0.01}^{+0.009}$ & 3.89 & ${3.895}_{-0.075}^{+0.075}$ & 0.056 & ${0.057}_{-0.028}^{+0.028}$ & 7.882\\
\hline\hline
FSPS & 2.3 & 6.558 & ${6.557}_{-0.007}^{+0.006}$ & 3.779 & ${3.788}_{-0.013}^{+0.015}$ & 0.0 & ${0.004}_{-0.003}^{+0.006}$ & 15.905 (*)\\
 & 3.1 & 6.558 & ${6.557}_{-0.007}^{+0.006}$ & 3.779 & ${3.79}_{-0.014}^{+0.017}$ & 0.0 & ${0.004}_{-0.003}^{+0.006}$ & 15.9 (*)\\
 & 4.1 & 6.558 & ${6.557}_{-0.007}^{+0.006}$ & 3.779 & ${3.793}_{-0.016}^{+0.019}$ & 0.0 & ${0.005}_{-0.003}^{+0.007}$ & 15.89 (*)\\
 & 5.1 & 6.558 & ${6.557}_{-0.007}^{+0.006}$ & 3.779 & ${3.796}_{-0.017}^{+0.024}$ & 0.0 & ${0.005}_{-0.004}^{+0.008}$ & 15.897 (*)\\
\hline\hline
\sidehead{\textbf{NGC7793-2}}
\hline\hline
YGGDRASIL & 2.3 & 7.319 & ${7.316}_{-0.063}^{+0.063}$ & 4.734 & ${4.741}_{-0.027}^{+0.052}$ & 0.075 & ${0.078}_{-0.046}^{+0.06}$ & 3.799\\
 & 3.1 & 7.324 & ${7.305}_{-0.093}^{+0.067}$ & 4.753 & ${4.766}_{-0.043}^{+0.078}$ & 0.072 & ${0.086}_{-0.05}^{+0.073}$ & 3.993\\
 & 4.1 & 7.342 & ${7.317}_{-0.162}^{+0.063}$ & 4.762 & ${4.784}_{-0.051}^{+0.1}$ & 0.059 & ${0.08}_{-0.048}^{+0.102}$ & 5.041\\
 & 5.1 & 7.361 & ${7.342}_{-0.217}^{+0.052}$ & 4.762 & ${4.781}_{-0.046}^{+0.091}$ & 0.045 & ${0.06}_{-0.036}^{+0.113}$ & 6.059\\
\hline\hline
BPASS & 2.3 & 7.43 & ${7.446}_{-0.031}^{+0.164}$ & 4.836 & ${4.839}_{-0.016}^{+0.054}$ & 0.04 & ${0.03}_{-0.025}^{+0.02}$ & 15.227 (*)\\
 & 3.1 & 7.426 & ${7.437}_{-0.028}^{+0.084}$ & 4.852 & ${4.853}_{-0.02}^{+0.028}$ & 0.044 & ${0.038}_{-0.027}^{+0.019}$ & 14.487 (*)\\
 & 4.1 & 7.422 & ${7.429}_{-0.027}^{+0.042}$ & 4.874 & ${4.874}_{-0.025}^{+0.028}$ & 0.049 & ${0.045}_{-0.024}^{+0.019}$ & 13.502 (*)\\
 & 5.1 & 7.419 & ${7.424}_{-0.027}^{+0.036}$ & 4.898 & ${4.898}_{-0.032}^{+0.032}$ & 0.053 & ${0.051}_{-0.022}^{+0.019}$ & 12.575 (*)\\
\hline\hline
GALAXEV & 2.3 & 7.362 & ${7.328}_{-0.145}^{+0.077}$ & 4.802 & ${4.793}_{-0.034}^{+0.023}$ & 0.096 & ${0.108}_{-0.039}^{+0.056}$ & 4.109\\
 & 3.1 & 7.366 & ${7.33}_{-0.163}^{+0.083}$ & 4.831 & ${4.82}_{-0.046}^{+0.035}$ & 0.095 & ${0.107}_{-0.043}^{+0.063}$ & 4.696\\
 & 4.1 & 6.929 & ${7.364}_{-0.428}^{+0.093}$ & 4.409 & ${4.826}_{-0.403}^{+0.047}$ & 0.131 & ${0.095}_{-0.061}^{+0.05}$ & 5.001\\
 & 5.1 & 6.887 & ${6.898}_{-0.03}^{+0.038}$ & 4.488 & ${4.472}_{-0.03}^{+0.03}$ & 0.163 & ${0.154}_{-0.028}^{+0.021}$ & 5.314\\
\hline\hline
FSPS & 2.3 & 7.3 & ${7.289}_{-0.233}^{+0.011}$ & 4.588 & ${4.569}_{-0.145}^{+0.029}$ & 0.006 & ${0.023}_{-0.016}^{+0.095}$ & 4.272\\
 & 3.1 & 7.3 & ${7.061}_{-0.034}^{+0.231}$ & 4.588 & ${4.49}_{-0.032}^{+0.088}$ & 0.005 & ${0.117}_{-0.096}^{+0.025}$ & 4.249\\
 & 4.1 & 7.043 & ${7.044}_{-0.035}^{+0.04}$ & 4.549 & ${4.538}_{-0.031}^{+0.028}$ & 0.135 & ${0.133}_{-0.032}^{+0.023}$ & 2.766\\
 & 5.1 & 7.049 & ${7.042}_{-0.048}^{+0.041}$ & 4.599 & ${4.589}_{-0.036}^{+0.029}$ & 0.134 & ${0.137}_{-0.032}^{+0.028}$ & 2.304\\
\hline\hline
\sidehead{\textbf{NGC4656-2}}
\hline\hline
YGGDRASIL & 2.3 & 6.922 & ${6.943}_{-0.075}^{+0.069}$ & 5.709 & ${5.732}_{-0.083}^{+0.057}$ & 0.105 & ${0.103}_{-0.012}^{+0.011}$ & 6.979\\
 & 3.1 & 6.919 & ${6.94}_{-0.071}^{+0.065}$ & 5.752 & ${5.773}_{-0.079}^{+0.056}$ & 0.114 & ${0.112}_{-0.012}^{+0.011}$ & 6.408\\
 & 4.1 & 6.92 & ${6.94}_{-0.067}^{+0.061}$ & 5.816 & ${5.835}_{-0.074}^{+0.056}$ & 0.126 & ${0.124}_{-0.013}^{+0.012}$ & 6.294\\
 & 5.1 & 6.929 & ${6.945}_{-0.065}^{+0.058}$ & 5.892 & ${5.906}_{-0.071}^{+0.055}$ & 0.135 & ${0.135}_{-0.014}^{+0.013}$ & 7.419\\
\hline\hline
BPASS & 2.3 & 7.299 & ${7.298}_{-0.018}^{+0.017}$ & 5.884 & ${5.885}_{-0.024}^{+0.022}$ & 0.0 & ${0.001}_{-0.001}^{+0.001}$ & 70.089 (*)\\
 & 3.1 & 7.299 & ${7.298}_{-0.018}^{+0.017}$ & 5.884 & ${5.885}_{-0.024}^{+0.022}$ & 0.0 & ${0.001}_{-0.001}^{+0.002}$ & 70.083 (*)\\
 & 4.1 & 7.299 & ${7.298}_{-0.018}^{+0.017}$ & 5.884 & ${5.886}_{-0.024}^{+0.022}$ & 0.0 & ${0.001}_{-0.001}^{+0.002}$ & 70.071 (*)\\
 & 5.1 & 7.299 & ${7.298}_{-0.018}^{+0.016}$ & 5.884 & ${5.887}_{-0.024}^{+0.022}$ & 0.0 & ${0.001}_{-0.001}^{+0.002}$ & 70.071 (*)\\
\hline\hline
GALAXEV & 2.3 & 7.051 & ${7.041}_{-0.041}^{+0.028}$ & 5.875 & ${5.861}_{-0.035}^{+0.024}$ & 0.133 & ${0.139}_{-0.026}^{+0.017}$ & 1.432\\
 & 3.1 & 7.036 & ${7.025}_{-0.04}^{+0.031}$ & 5.936 & ${5.914}_{-0.043}^{+0.03}$ & 0.155 & ${0.157}_{-0.023}^{+0.015}$ & 1.76\\
 & 4.1 & 7.02 & ${7.015}_{-0.039}^{+0.033}$ & 6.015 & ${5.992}_{-0.047}^{+0.037}$ & 0.178 & ${0.175}_{-0.021}^{+0.015}$ & 2.915\\
 & 5.1 & 7.013 & ${7.021}_{-0.041}^{+0.045}$ & 6.1 & ${6.07}_{-0.061}^{+0.045}$ & 0.193 & ${0.184}_{-0.039}^{+0.018}$ & 7.32\\
\hline\hline
FSPS & 2.3 & 6.92 & ${7.034}_{-0.071}^{+0.07}$ & 5.472 & ${5.815}_{-0.139}^{+0.058}$ & 0.085 & ${0.122}_{-0.02}^{+0.013}$ & 1.88\\
 & 3.1 & 6.984 & ${7.043}_{-0.064}^{+0.063}$ & 5.773 & ${5.878}_{-0.115}^{+0.045}$ & 0.138 & ${0.132}_{-0.02}^{+0.013}$ & 1.769\\
 & 4.1 & 7.049 & ${7.056}_{-0.061}^{+0.068}$ & 5.965 & ${5.956}_{-0.088}^{+0.039}$ & 0.149 & ${0.142}_{-0.024}^{+0.015}$ & 2.872\\
 & 5.1 & 6.836 & ${7.089}_{-0.182}^{+0.099}$ & 5.122 & ${6.024}_{-0.588}^{+0.044}$ & 0.017 & ${0.126}_{-0.068}^{+0.033}$ & 4.607\\
\hline\hline
\sidehead{\textbf{NGC4656-1}}
\hline\hline
YGGDRASIL & 2.3 & 7.201 & ${7.265}_{-0.097}^{+0.122}$ & 5.906 & ${5.961}_{-0.054}^{+0.061}$ & 0.378 & ${0.368}_{-0.034}^{+0.037}$ & 11.433 (*)\\
 & 3.1 & 7.158 & ${7.199}_{-0.11}^{+0.136}$ & 6.048 & ${6.092}_{-0.054}^{+0.058}$ & 0.42 & ${0.406}_{-0.04}^{+0.083}$ & 6.804\\
 & 4.1 & 7.118 & ${7.102}_{-0.151}^{+0.168}$ & 6.284 & ${6.299}_{-0.079}^{+0.086}$ & 0.478 & ${0.493}_{-0.09}^{+0.078}$ & 2.444\\
 & 5.1 & 7.157 & ${7.164}_{-0.184}^{+0.185}$ & 6.413 & ${6.47}_{-0.08}^{+0.123}$ & 0.456 & ${0.459}_{-0.062}^{+0.123}$ & 0.449\\
\hline\hline
BPASS & 2.3 & 6.852 & ${6.931}_{-0.08}^{+0.079}$ & 5.134 & ${5.321}_{-0.191}^{+0.109}$ & 0.22 & ${0.25}_{-0.032}^{+0.027}$ & 13.728 (*)\\
 & 3.1 & 6.999 & ${7.063}_{-0.128}^{+0.36}$ & 5.499 & ${5.623}_{-0.207}^{+0.488}$ & 0.272 & ${0.288}_{-0.034}^{+0.037}$ & 9.085\\
 & 4.1 & 7.402 & ${7.378}_{-0.242}^{+0.049}$ & 6.246 & ${6.221}_{-0.299}^{+0.058}$ & 0.332 & ${0.335}_{-0.037}^{+0.034}$ & 5.102\\
 & 5.1 & 7.395 & ${7.365}_{-0.117}^{+0.059}$ & 6.397 & ${6.378}_{-0.097}^{+0.064}$ & 0.35 & ${0.364}_{-0.037}^{+0.031}$ & 2.239\\
\hline\hline
GALAXEV & 2.3 & 7.378 & ${7.348}_{-0.099}^{+0.094}$ & 6.038 & ${6.019}_{-0.036}^{+0.033}$ & 0.353 & ${0.366}_{-0.051}^{+0.039}$ & 5.522\\
 & 3.1 & 7.287 & ${7.31}_{-0.106}^{+0.105}$ & 6.137 & ${6.14}_{-0.04}^{+0.039}$ & 0.4 & ${0.395}_{-0.05}^{+0.043}$ & 2.617\\
 & 4.1 & 7.352 & ${7.298}_{-0.14}^{+0.124}$ & 6.328 & ${6.306}_{-0.05}^{+0.053}$ & 0.398 & ${0.416}_{-0.059}^{+0.06}$ & 0.557\\
 & 5.1 & 7.458 & ${7.424}_{-0.159}^{+0.135}$ & 6.395 & ${6.431}_{-0.071}^{+0.075}$ & 0.349 & ${0.371}_{-0.064}^{+0.071}$ & 0.105\\
\hline\hline
FSPS & 2.3 & 7.658 & ${7.679}_{-0.072}^{+0.126}$ & 6.277 & ${6.302}_{-0.033}^{+0.032}$ & 0.38 & ${0.374}_{-0.048}^{+0.043}$ & 8.396\\
 & 3.1 & 7.639 & ${7.629}_{-0.176}^{+0.096}$ & 6.418 & ${6.429}_{-0.062}^{+0.045}$ & 0.409 & ${0.419}_{-0.051}^{+0.054}$ & 4.441\\
 & 4.1 & 7.615 & ${7.543}_{-0.338}^{+0.146}$ & 6.63 & ${6.588}_{-0.099}^{+0.073}$ & 0.447 & ${0.471}_{-0.067}^{+0.067}$ & 1.076\\
 & 5.1 & 7.691 & ${7.548}_{-0.218}^{+0.2}$ & 6.756 & ${6.768}_{-0.079}^{+0.085}$ & 0.417 & ${0.477}_{-0.082}^{+0.046}$ & 0.034\\
\hline\hline
\sidehead{\textbf{NGC4449-1}}
\hline\hline
YGGDRASIL & 2.3 & 7.197 & ${7.247}_{-0.079}^{+0.108}$ & 6.149 & ${6.191}_{-0.043}^{+0.066}$ & 0.462 & ${0.453}_{-0.027}^{+0.031}$ & 17.531 (*)\\
 & 3.1 & 7.149 & ${7.176}_{-0.086}^{+0.124}$ & 6.319 & ${6.346}_{-0.046}^{+0.057}$ & 0.509 & ${0.494}_{-0.038}^{+0.075}$ & 9.257\\
 & 4.1 & 7.101 & ${7.058}_{-0.156}^{+0.144}$ & 6.595 & ${6.595}_{-0.087}^{+0.083}$ & 0.574 & ${0.614}_{-0.114}^{+0.043}$ & 2.279\\
 & 5.1 & 7.17 & ${7.164}_{-0.208}^{+0.191}$ & 6.702 & ${6.77}_{-0.074}^{+0.118}$ & 0.523 & ${0.532}_{-0.057}^{+0.128}$ & 0.244\\
\hline\hline
BPASS & 2.3 & 6.852 & ${6.887}_{-0.042}^{+0.106}$ & 5.367 & ${5.464}_{-0.117}^{+0.178}$ & 0.297 & ${0.32}_{-0.029}^{+0.027}$ & 22.272 (*)\\
 & 3.1 & 6.997 & ${6.975}_{-0.093}^{+0.047}$ & 5.756 & ${5.729}_{-0.174}^{+0.072}$ & 0.352 & ${0.35}_{-0.026}^{+0.021}$ & 13.563 (*)\\
 & 4.1 & 7.402 & ${7.393}_{-0.065}^{+0.036}$ & 6.535 & ${6.527}_{-0.061}^{+0.041}$ & 0.412 & ${0.415}_{-0.031}^{+0.029}$ & 5.967\\
 & 5.1 & 7.392 & ${7.366}_{-0.106}^{+0.055}$ & 6.709 & ${6.693}_{-0.082}^{+0.053}$ & 0.43 & ${0.444}_{-0.035}^{+0.028}$ & 2.105\\
\hline\hline
GALAXEV & 2.3 & 7.372 & ${7.338}_{-0.084}^{+0.082}$ & 6.278 & ${6.258}_{-0.032}^{+0.029}$ & 0.436 & ${0.448}_{-0.042}^{+0.032}$ & 8.55\\
 & 3.1 & 7.28 & ${7.3}_{-0.092}^{+0.09}$ & 6.389 & ${6.396}_{-0.034}^{+0.034}$ & 0.48 & ${0.476}_{-0.04}^{+0.035}$ & 3.261\\
 & 4.1 & 7.306 & ${7.282}_{-0.138}^{+0.119}$ & 6.594 & ${6.586}_{-0.042}^{+0.046}$ & 0.489 & ${0.498}_{-0.052}^{+0.06}$ & 0.22\\
 & 5.1 & 7.483 & ${7.48}_{-0.133}^{+0.129}$ & 6.681 & ${6.709}_{-0.054}^{+0.067}$ & 0.41 & ${0.415}_{-0.046}^{+0.072}$ & 0.269\\
\hline\hline
FSPS & 2.3 & 7.658 & ${7.675}_{-0.052}^{+0.095}$ & 6.518 & ${6.541}_{-0.026}^{+0.027}$ & 0.461 & ${0.456}_{-0.039}^{+0.032}$ & 13.201 (*)\\
 & 3.1 & 6.908 & ${7.625}_{-0.339}^{+0.079}$ & 5.96 & ${6.686}_{-0.117}^{+0.043}$ & 0.55 & ${0.501}_{-0.043}^{+0.057}$ & 6.238\\
 & 4.1 & 7.606 & ${7.515}_{-0.354}^{+0.153}$ & 6.925 & ${6.865}_{-0.106}^{+0.073}$ & 0.532 & ${0.557}_{-0.064}^{+0.073}$ & 0.91\\
 & 5.1 & 7.345 & ${7.528}_{-0.202}^{+0.233}$ & 6.977 & ${7.062}_{-0.069}^{+0.082}$ & 0.567 & ${0.556}_{-0.089}^{+0.04}$ & 0.104\\
\hline\hline
\sidehead{\textbf{M95-1}}
\hline\hline
YGGDRASIL & 2.3 & 6.582 & ${6.491}_{-0.287}^{+0.134}$ & 5.512 & ${5.553}_{-0.059}^{+0.074}$ & 0.129 & ${0.137}_{-0.029}^{+0.025}$ & 0.371\\
 & 3.1 & 6.594 & ${6.543}_{-0.304}^{+0.093}$ & 5.568 & ${5.593}_{-0.054}^{+0.073}$ & 0.132 & ${0.14}_{-0.031}^{+0.028}$ & 0.317\\
 & 4.1 & 6.615 & ${6.587}_{-0.255}^{+0.065}$ & 5.642 & ${5.65}_{-0.056}^{+0.066}$ & 0.132 & ${0.141}_{-0.034}^{+0.031}$ & 0.946\\
 & 5.1 & 6.634 & ${6.622}_{-0.063}^{+0.046}$ & 5.71 & ${5.7}_{-0.065}^{+0.063}$ & 0.13 & ${0.136}_{-0.038}^{+0.033}$ & 2.625\\
\hline\hline
BPASS & 2.3 & 6.586 & ${6.565}_{-0.117}^{+0.054}$ & 5.482 & ${5.484}_{-0.043}^{+0.041}$ & 0.086 & ${0.103}_{-0.034}^{+0.044}$ & 0.646\\
 & 3.1 & 6.588 & ${6.572}_{-0.095}^{+0.048}$ & 5.512 & ${5.522}_{-0.043}^{+0.043}$ & 0.088 & ${0.103}_{-0.034}^{+0.045}$ & 0.952\\
 & 4.1 & 6.592 & ${6.582}_{-0.06}^{+0.041}$ & 5.548 & ${5.568}_{-0.05}^{+0.051}$ & 0.089 & ${0.101}_{-0.033}^{+0.042}$ & 1.461\\
 & 5.1 & 6.598 & ${6.592}_{-0.04}^{+0.035}$ & 5.58 & ${5.603}_{-0.061}^{+0.062}$ & 0.087 & ${0.096}_{-0.031}^{+0.037}$ & 2.119\\
\hline\hline
GALAXEV & 2.3 & 6.424 & ${6.34}_{-0.082}^{+0.066}$ & 5.452 & ${5.465}_{-0.033}^{+0.032}$ & 0.198 & ${0.196}_{-0.022}^{+0.022}$ & 1.328\\
 & 3.1 & 6.426 & ${6.336}_{-0.08}^{+0.069}$ & 5.519 & ${5.535}_{-0.041}^{+0.04}$ & 0.206 & ${0.206}_{-0.023}^{+0.023}$ & 2.007\\
 & 4.1 & 6.429 & ${6.328}_{-0.079}^{+0.077}$ & 5.598 & ${5.625}_{-0.052}^{+0.05}$ & 0.211 & ${0.215}_{-0.025}^{+0.024}$ & 3.491\\
 & 5.1 & 6.645 & ${6.63}_{-0.331}^{+0.029}$ & 5.429 & ${5.489}_{-0.103}^{+0.238}$ & 0.065 & ${0.092}_{-0.048}^{+0.134}$ & 5.126\\
\hline\hline
FSPS & 2.3 & 6.586 & ${6.581}_{-0.014}^{+0.012}$ & 5.331 & ${5.355}_{-0.048}^{+0.054}$ & 0.03 & ${0.044}_{-0.027}^{+0.032}$ & 2.301\\
 & 3.1 & 6.589 & ${6.582}_{-0.014}^{+0.012}$ & 5.328 & ${5.364}_{-0.054}^{+0.065}$ & 0.025 & ${0.043}_{-0.027}^{+0.033}$ & 2.341\\
 & 4.1 & 6.592 & ${6.584}_{-0.013}^{+0.011}$ & 5.315 & ${5.37}_{-0.059}^{+0.077}$ & 0.015 & ${0.039}_{-0.026}^{+0.034}$ & 2.351\\
 & 5.1 & 6.594 & ${6.587}_{-0.012}^{+0.011}$ & 5.307 & ${5.37}_{-0.061}^{+0.086}$ & 0.01 & ${0.035}_{-0.023}^{+0.033}$ & 2.37\\
\hline\hline
\sidehead{\textbf{M51-1}}
\hline\hline
YGGDRASIL & 2.3 & 6.737 & ${6.685}_{-0.52}^{+0.059}$ & 5.183 & ${5.224}_{-0.074}^{+0.07}$ & 0.063 & ${0.11}_{-0.055}^{+0.052}$ & 0.589\\
 & 3.1 & 6.732 & ${6.552}_{-0.41}^{+0.173}$ & 5.224 & ${5.297}_{-0.091}^{+0.086}$ & 0.075 & ${0.153}_{-0.068}^{+0.042}$ & 0.562\\
 & 4.1 & 6.723 & ${6.318}_{-0.188}^{+0.345}$ & 5.302 & ${5.437}_{-0.111}^{+0.098}$ & 0.098 & ${0.202}_{-0.054}^{+0.037}$ & 0.542\\
 & 5.1 & 6.561 & ${6.329}_{-0.195}^{+0.274}$ & 5.505 & ${5.61}_{-0.129}^{+0.111}$ & 0.236 & ${0.248}_{-0.049}^{+0.041}$ & 0.784\\
\hline\hline
BPASS & 2.3 & 6.823 & ${6.919}_{-0.15}^{+0.133}$ & 5.427 & ${5.53}_{-0.148}^{+0.122}$ & 0.083 & ${0.065}_{-0.037}^{+0.036}$ & 0.13\\
 & 3.1 & 6.805 & ${6.903}_{-0.16}^{+0.146}$ & 5.452 & ${5.547}_{-0.145}^{+0.125}$ & 0.096 & ${0.076}_{-0.044}^{+0.041}$ & 0.175\\
 & 4.1 & 6.683 & ${6.857}_{-0.174}^{+0.179}$ & 5.409 & ${5.567}_{-0.162}^{+0.128}$ & 0.13 & ${0.097}_{-0.057}^{+0.05}$ & 0.305\\
 & 5.1 & 6.654 & ${6.742}_{-0.214}^{+0.263}$ & 5.452 & ${5.581}_{-0.147}^{+0.141}$ & 0.149 & ${0.141}_{-0.083}^{+0.083}$ & 0.275\\
\hline\hline
GALAXEV & 2.3 & 6.714 & ${7.22}_{-0.159}^{+0.134}$ & 5.025 & ${5.78}_{-0.138}^{+0.11}$ & 0.043 & ${0.035}_{-0.024}^{+0.03}$ & 1.063\\
 & 3.1 & 6.711 & ${7.223}_{-0.16}^{+0.133}$ & 5.043 & ${5.801}_{-0.137}^{+0.103}$ & 0.048 & ${0.039}_{-0.027}^{+0.036}$ & 1.122\\
 & 4.1 & 6.708 & ${6.398}_{-0.096}^{+0.302}$ & 5.072 & ${5.308}_{-0.203}^{+0.113}$ & 0.057 & ${0.249}_{-0.177}^{+0.048}$ & 1.227\\
 & 5.1 & 6.425 & ${6.34}_{-0.119}^{+0.074}$ & 5.56 & ${5.561}_{-0.134}^{+0.113}$ & 0.319 & ${0.313}_{-0.051}^{+0.042}$ & 0.406\\
\hline\hline
FSPS & 2.3 & 6.583 & ${6.575}_{-0.036}^{+0.022}$ & 5.014 & ${5.025}_{-0.066}^{+0.065}$ & 0.067 & ${0.075}_{-0.038}^{+0.035}$ & 0.86\\
 & 3.1 & 6.58 & ${6.572}_{-0.036}^{+0.023}$ & 5.059 & ${5.074}_{-0.084}^{+0.08}$ & 0.081 & ${0.091}_{-0.042}^{+0.038}$ & 0.798\\
 & 4.1 & 6.577 & ${6.567}_{-0.034}^{+0.023}$ & 5.136 & ${5.157}_{-0.112}^{+0.1}$ & 0.103 & ${0.114}_{-0.047}^{+0.041}$ & 0.707\\
 & 5.1 & 6.574 & ${6.562}_{-0.034}^{+0.023}$ & 5.234 & ${5.269}_{-0.144}^{+0.128}$ & 0.128 & ${0.143}_{-0.053}^{+0.047}$ & 0.643\\
\hline\hline
\sidehead{\textbf{M51-2}}
\hline\hline
YGGDRASIL & 2.3 & 6.766 & ${6.48}_{-0.355}^{+0.29}$ & 4.734 & ${4.892}_{-0.165}^{+0.111}$ & 0.121 & ${0.253}_{-0.139}^{+0.065}$ & 0.439\\
 & 3.1 & 6.761 & ${6.316}_{-0.196}^{+0.421}$ & 4.793 & ${5.042}_{-0.158}^{+0.079}$ & 0.135 & ${0.305}_{-0.123}^{+0.041}$ & 0.311\\
 & 4.1 & 6.747 & ${6.397}_{-0.239}^{+0.286}$ & 4.909 & ${5.177}_{-0.105}^{+0.083}$ & 0.165 & ${0.331}_{-0.064}^{+0.04}$ & 0.209\\
 & 5.1 & 6.71 & ${6.561}_{-0.317}^{+0.125}$ & 5.175 & ${5.31}_{-0.11}^{+0.085}$ & 0.243 & ${0.34}_{-0.063}^{+0.042}$ & 0.282\\
\hline\hline
BPASS & 2.3 & 7.252 & ${7.174}_{-0.29}^{+0.171}$ & 5.351 & ${5.286}_{-0.192}^{+0.123}$ & 0.047 & ${0.07}_{-0.046}^{+0.086}$ & 0.138\\
 & 3.1 & 7.26 & ${7.178}_{-0.354}^{+0.169}$ & 5.376 & ${5.31}_{-0.199}^{+0.117}$ & 0.049 & ${0.074}_{-0.048}^{+0.106}$ & 0.153\\
 & 4.1 & 7.272 & ${7.166}_{-0.517}^{+0.18}$ & 5.409 & ${5.331}_{-0.209}^{+0.119}$ & 0.052 & ${0.08}_{-0.053}^{+0.178}$ & 0.158\\
 & 5.1 & 7.286 & ${7.211}_{-0.598}^{+0.149}$ & 5.444 & ${5.377}_{-0.155}^{+0.106}$ & 0.054 & ${0.075}_{-0.05}^{+0.211}$ & 0.137\\
\hline\hline
GALAXEV & 2.3 & 7.493 & ${7.441}_{-0.078}^{+0.051}$ & 5.438 & ${5.449}_{-0.052}^{+0.037}$ & 0.0 & ${0.035}_{-0.025}^{+0.038}$ & 1.051\\
 & 3.1 & 7.493 & ${7.445}_{-0.066}^{+0.05}$ & 5.438 & ${5.46}_{-0.047}^{+0.043}$ & 0.0 & ${0.034}_{-0.023}^{+0.037}$ & 1.052\\
 & 4.1 & 6.42 & ${6.28}_{-0.155}^{+0.101}$ & 5.127 & ${5.116}_{-0.074}^{+0.07}$ & 0.407 & ${0.395}_{-0.036}^{+0.035}$ & 0.913\\
 & 5.1 & 7.493 & ${7.459}_{-0.055}^{+0.043}$ & 5.438 & ${5.483}_{-0.047}^{+0.059}$ & 0.0 & ${0.028}_{-0.02}^{+0.033}$ & 1.051\\
\hline\hline
FSPS & 2.3 & 7.304 & ${7.223}_{-0.115}^{+0.117}$ & 5.368 & ${5.357}_{-0.059}^{+0.069}$ & 0.016 & ${0.047}_{-0.033}^{+0.045}$ & 1.377\\
 & 3.1 & 7.306 & ${7.235}_{-0.121}^{+0.108}$ & 5.371 & ${5.377}_{-0.061}^{+0.066}$ & 0.013 & ${0.047}_{-0.033}^{+0.048}$ & 1.413\\
 & 4.1 & 7.31 & ${7.25}_{-0.139}^{+0.101}$ & 5.371 & ${5.395}_{-0.072}^{+0.072}$ & 0.01 & ${0.043}_{-0.031}^{+0.054}$ & 1.448\\
 & 5.1 & 6.593 & ${6.593}_{-0.022}^{+0.032}$ & 4.906 & ${4.913}_{-0.191}^{+0.139}$ & 0.21 & ${0.212}_{-0.075}^{+0.054}$ & 1.414\\
\hline\hline
\sidehead{\textbf{NGC1313-1}}
\hline\hline
YGGDRASIL & 2.3 & 6.466 & ${6.476}_{-0.033}^{+0.194}$ & 4.479 & ${4.507}_{-0.032}^{+0.072}$ & 0.013 & ${0.019}_{-0.013}^{+0.018}$ & 6.95\\
 & 3.1 & 6.466 & ${6.477}_{-0.033}^{+0.194}$ & 4.486 & ${4.519}_{-0.039}^{+0.069}$ & 0.015 & ${0.021}_{-0.014}^{+0.02}$ & 6.951\\
 & 4.1 & 6.466 & ${6.477}_{-0.033}^{+0.195}$ & 4.497 & ${4.539}_{-0.051}^{+0.067}$ & 0.018 & ${0.025}_{-0.016}^{+0.022}$ & 6.94\\
 & 5.1 & 6.466 & ${6.478}_{-0.033}^{+0.194}$ & 4.511 & ${4.559}_{-0.062}^{+0.071}$ & 0.021 & ${0.028}_{-0.018}^{+0.024}$ & 6.922\\
\hline\hline
BPASS & 2.3 & 6.728 & ${6.728}_{-0.018}^{+0.019}$ & 4.571 & ${4.579}_{-0.031}^{+0.032}$ & 0.0 & ${0.003}_{-0.002}^{+0.005}$ & 57.249 (*)\\
 & 3.1 & 6.728 & ${6.728}_{-0.018}^{+0.019}$ & 4.571 & ${4.58}_{-0.031}^{+0.032}$ & 0.0 & ${0.003}_{-0.002}^{+0.005}$ & 57.367 (*)\\
 & 4.1 & 6.728 & ${6.728}_{-0.018}^{+0.019}$ & 4.571 & ${4.582}_{-0.031}^{+0.032}$ & 0.0 & ${0.003}_{-0.003}^{+0.005}$ & 57.367 (*)\\
 & 5.1 & 6.728 & ${6.728}_{-0.018}^{+0.019}$ & 4.571 & ${4.584}_{-0.032}^{+0.033}$ & 0.0 & ${0.004}_{-0.003}^{+0.006}$ & 57.357 (*)\\
\hline\hline
GALAXEV & 2.3 & 6.488 & ${6.488}_{-0.006}^{+0.005}$ & 4.368 & ${4.368}_{-0.037}^{+0.038}$ & 0.044 & ${0.045}_{-0.021}^{+0.022}$ & 14.648 (*)\\
 & 3.1 & 6.488 & ${6.489}_{-0.006}^{+0.243}$ & 4.393 & ${4.406}_{-0.052}^{+0.119}$ & 0.051 & ${0.045}_{-0.031}^{+0.026}$ & 14.631 (*)\\
 & 4.1 & 6.488 & ${6.489}_{-0.006}^{+0.007}$ & 4.432 & ${4.444}_{-0.061}^{+0.079}$ & 0.061 & ${0.059}_{-0.033}^{+0.027}$ & 14.526 (*)\\
 & 5.1 & 6.489 & ${6.489}_{-0.006}^{+0.006}$ & 4.482 & ${4.495}_{-0.074}^{+0.072}$ & 0.073 & ${0.073}_{-0.032}^{+0.029}$ & 14.238 (*)\\
\hline\hline
FSPS & 2.3 & 6.567 & ${6.566}_{-0.008}^{+0.008}$ & 4.247 & ${4.256}_{-0.023}^{+0.021}$ & 0.0 & ${0.003}_{-0.002}^{+0.005}$ & 26.574 (*)\\
 & 3.1 & 6.567 & ${6.566}_{-0.008}^{+0.008}$ & 4.247 & ${4.257}_{-0.023}^{+0.022}$ & 0.0 & ${0.003}_{-0.003}^{+0.006}$ & 26.567 (*)\\
 & 4.1 & 6.567 & ${6.566}_{-0.008}^{+0.008}$ & 4.247 & ${4.26}_{-0.024}^{+0.024}$ & 0.0 & ${0.004}_{-0.003}^{+0.006}$ & 26.579 (*)\\
 & 5.1 & 6.567 & ${6.566}_{-0.008}^{+0.008}$ & 4.247 & ${4.263}_{-0.024}^{+0.026}$ & 0.0 & ${0.005}_{-0.003}^{+0.007}$ & 26.565 (*)\\
\hline\hline
\sidehead{\textbf{NGC1313-2}}
\hline\hline
YGGDRASIL & 2.3 & 7.004 & ${7.294}_{-0.274}^{+0.18}$ & 4.704 & ${5.027}_{-0.286}^{+0.058}$ & 0.147 & ${0.138}_{-0.086}^{+0.061}$ & 2.27\\
 & 3.1 & 7.005 & ${7.251}_{-0.222}^{+0.125}$ & 4.756 & ${5.094}_{-0.281}^{+0.083}$ & 0.153 & ${0.175}_{-0.073}^{+0.064}$ & 1.62\\
 & 4.1 & 7.233 & ${7.229}_{-0.092}^{+0.081}$ & 5.265 & ${5.217}_{-0.178}^{+0.081}$ & 0.237 & ${0.215}_{-0.07}^{+0.052}$ & 2.258\\
 & 5.1 & 7.221 & ${7.211}_{-0.088}^{+0.078}$ & 5.376 & ${5.323}_{-0.194}^{+0.088}$ & 0.254 & ${0.238}_{-0.073}^{+0.046}$ & 1.341\\
\hline\hline
BPASS & 2.3 & 7.218 & ${7.194}_{-0.082}^{+0.047}$ & 4.943 & ${4.909}_{-0.104}^{+0.08}$ & 0.088 & ${0.089}_{-0.025}^{+0.026}$ & 3.923\\
 & 3.1 & 7.225 & ${7.208}_{-0.085}^{+0.05}$ & 4.984 & ${4.957}_{-0.113}^{+0.1}$ & 0.094 & ${0.094}_{-0.027}^{+0.028}$ & 3.802\\
 & 4.1 & 7.235 & ${7.233}_{-0.085}^{+0.121}$ & 5.041 & ${5.043}_{-0.136}^{+0.143}$ & 0.104 & ${0.105}_{-0.031}^{+0.034}$ & 3.62\\
 & 5.1 & 7.246 & ${7.269}_{-0.067}^{+0.148}$ & 5.11 & ${5.178}_{-0.17}^{+0.117}$ & 0.115 & ${0.121}_{-0.037}^{+0.037}$ & 3.407\\
\hline\hline
GALAXEV & 2.3 & 7.036 & ${7.427}_{-0.314}^{+0.127}$ & 4.789 & ${5.08}_{-0.159}^{+0.046}$ & 0.182 & ${0.121}_{-0.063}^{+0.078}$ & 1.447\\
 & 3.1 & 7.04 & ${7.338}_{-0.247}^{+0.179}$ & 4.863 & ${5.119}_{-0.166}^{+0.056}$ & 0.192 & ${0.169}_{-0.091}^{+0.062}$ & 0.735\\
 & 4.1 & 7.045 & ${7.268}_{-0.166}^{+0.175}$ & 4.957 & ${5.194}_{-0.137}^{+0.083}$ & 0.203 & ${0.205}_{-0.085}^{+0.055}$ & 0.336\\
 & 5.1 & 7.27 & ${7.259}_{-0.153}^{+0.165}$ & 5.352 & ${5.273}_{-0.145}^{+0.108}$ & 0.237 & ${0.214}_{-0.078}^{+0.055}$ & 0.19\\
\hline\hline
FSPS & 2.3 & 7.324 & ${7.433}_{-0.113}^{+0.045}$ & 4.894 & ${4.972}_{-0.081}^{+0.04}$ & 0.037 & ${0.024}_{-0.017}^{+0.027}$ & 0.492\\
 & 3.1 & 7.323 & ${7.432}_{-0.114}^{+0.048}$ & 4.905 & ${4.975}_{-0.076}^{+0.047}$ & 0.04 & ${0.027}_{-0.018}^{+0.031}$ & 0.509\\
 & 4.1 & 7.322 & ${7.426}_{-0.113}^{+0.053}$ & 4.921 & ${4.98}_{-0.073}^{+0.053}$ & 0.042 & ${0.03}_{-0.02}^{+0.039}$ & 0.533\\
 & 5.1 & 7.321 & ${7.321}_{-0.321}^{+0.147}$ & 4.939 & ${4.987}_{-0.07}^{+0.058}$ & 0.045 & ${0.051}_{-0.037}^{+0.182}$ & 0.561\\
\hline\hline
\enddata

\tablecomments{This table reports the fitted parameters produced during this work. Each row represents a fit produced using a unique combination of cluster, extinction curve, and model SSP. Column 1 denotes which cluster and model were used in a given row. Column 2 shows the the $R_V$ used in a given row. Column 3 is the maximum-likelihood ("most probable") age for a given fit. Column 4 is the median ("most typical") age for a given fit, along with approximate uncertainties (calculated as the 16th and 84th percentiles within the posterior). Column 5 is the maximum likelihood stellar mass. Column 6 is the median stellar mass and associated uncertainties. Column 7 is the maximum likelihood extinction in magnitudes. Column 8 is the median extinction and associated uncertainties. Column 9 reports the reduced $\chi^2$ associated with the maximum-likelihood fit in each row. Rejected fits are included in this table and are denoted with an asterisk.}
\end{deluxetable*}

\section{Chi-Square Per Filter}
It is possible for SPS models to perform better in some regions of the spectrum than others due to effects such as the presence (or lack) of strong spectral lines that may not be properly accounted for in the nebular emission models. This behavior may elide important information about the performance of the models if only the reduced Chi-Squared of the entire fit is considered. As such, we calculate the Chi-Square for each band in each (accepted) fit.
\begin{figure}[htb!]
\includegraphics[width=1.0\textwidth]{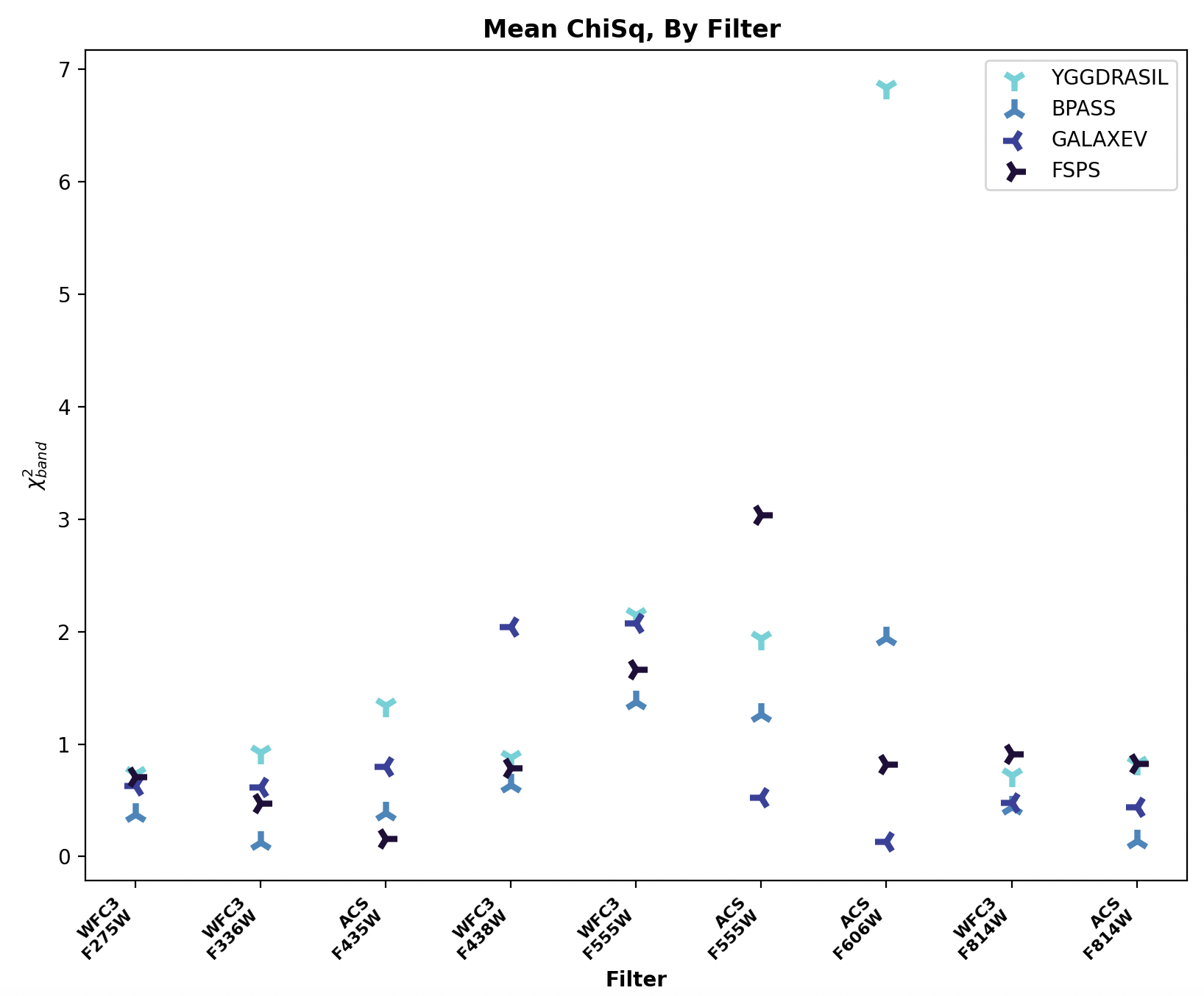}
\caption{The mean Chi-Squared in each filter for each model, marginalized over all clusters and extinction curves. Only accepted fits are included here. The models appear to have comparable performance, with the performance of YGGDRASIL in F606W the only potential outlier.}
We find that by this metric, the models perform similarly well to one another. There are no strong biases as a function of filter, though all models are marginally worse at predicting the emission around $\sim 5000 - 7000$ \AA. The only significant outlier is YGGDRASIL, which seems to perform poorly in F606W. However, given that we only have 2 clusters (out of 18) with an F606W measurement, this may be the result of low number statistics and so is not a cause for concern.
\end{figure}

\bibliography{sample631}{}

\begin{thebibliography}{}
\expandafter\ifx\csname natexlab\endcsname\relax\def\natexlab#1{#1}\fi
\providecommand{\url}[1]{\href{#1}{#1}}
\providecommand{\dodoi}[1]{doi:~\href{http://doi.org/#1}{\nolinkurl{#1}}}
\providecommand{\doeprint}[1]{\href{http://ascl.net/#1}{\nolinkurl{http://ascl.net/#1}}}
\providecommand{\doarXiv}[1]{\href{https://arxiv.org/abs/#1}{\nolinkurl{https://arxiv.org/abs/#1}}}

\bibitem[{{Adams} {et~al.}(2023){Adams}, {Conselice}, {Austin}, {Harvey}, {Ferreira}, {Trussler}, {Juodzbalis}, {Li}, {Windhorst}, {Cohen}, {Jansen}, {Summers}, {Tompkins}, {Driver}, {Robotham}, {D'Silva}, {Yan}, {Coe}, {Frye}, {Grogin}, {Koekemoer}, {Marshall}, {Pirzkal}, {Ryan}, {Maksym}, {Rutkowski}, {Willmer}, {Hammel}, {Nonino}, {Bhatawdekar}, {Wilkins}, {Willner}, {Bradley}, {Broadhurst}, {Cheng}, {Dole}, {Hathi}, \& {Zitrin}}]{Adams2023}
{Adams}, N.~J., {Conselice}, C.~J., {Austin}, D., {et~al.} 2023, arXiv e-prints, arXiv:2304.13721, \dodoi{10.48550/arXiv.2304.13721}

\bibitem[{{Alongi} {et~al.}(1993){Alongi}, {Bertelli}, {Bressan}, {Chiosi}, {Fagotto}, {Greggio}, \& {Nasi}}]{Alongi1993}
{Alongi}, M., {Bertelli}, G., {Bressan}, A., {et~al.} 1993, \aaps, 97, 851

\bibitem[{{Anders} {et~al.}(2004){Anders}, {Bissantz}, {Fritze-v. Alvensleben}, \& {de Grijs}}]{Anders2004}
{Anders}, P., {Bissantz}, N., {Fritze-v. Alvensleben}, U., \& {de Grijs}, R. 2004, \mnras, 347, 196, \dodoi{10.1111/j.1365-2966.2004.07197.x}

\bibitem[{{Astropy Collaboration} {et~al.}(2013){Astropy Collaboration}, {Robitaille}, {Tollerud}, {Greenfield}, {Droettboom}, {Bray}, {Aldcroft}, {Davis}, {Ginsburg}, {Price-Whelan}, {Kerzendorf}, {Conley}, {Crighton}, {Barbary}, {Muna}, {Ferguson}, {Grollier}, {Parikh}, {Nair}, {Unther}, {Deil}, {Woillez}, {Conseil}, {Kramer}, {Turner}, {Singer}, {Fox}, {Weaver}, {Zabalza}, {Edwards}, {Azalee Bostroem}, {Burke}, {Casey}, {Crawford}, {Dencheva}, {Ely}, {Jenness}, {Labrie}, {Lim}, {Pierfederici}, {Pontzen}, {Ptak}, {Refsdal}, {Servillat}, \& {Streicher}}]{astropy}
{Astropy Collaboration}, {Robitaille}, T.~P., {Tollerud}, E.~J., {et~al.} 2013, \aap, 558, A33, \dodoi{10.1051/0004-6361/201322068}

\bibitem[{{Berg} {et~al.}(2024){Berg}, {Skillman}, {Chisholm}, {Pogge}, {Gazagnes}, {Rogers}, {Erb}, {Arellano-C{\'o}rdova}, {Leitherer}, {Appel}, \& {Moustakas}}]{Berg2024}
{Berg}, D.~A., {Skillman}, E.~D., {Chisholm}, J., {et~al.} 2024, \apj, 971, 87, \dodoi{10.3847/1538-4357/ad5292}

\bibitem[{{Boquien} {et~al.}(2019){Boquien}, {Burgarella}, {Roehlly}, {Buat}, {Ciesla}, {Corre}, {Inoue}, \& {Salas}}]{Boquien2019}
{Boquien}, M., {Burgarella}, D., {Roehlly}, Y., {et~al.} 2019, \aap, 622, A103, \dodoi{10.1051/0004-6361/201834156}

\bibitem[{{Bradley} {et~al.}(2016){Bradley}, {Sipocz}, {Robitaille}, {Tollerud}, {Deil}, {Vin{\'\i}cius}, {Barbary}, {G{\"u}nther}, {Bostroem}, {Droettboom}, {Bray}, {Bratholm}, {Pickering}, {Craig}, {Pascual}, {Greco}, {Donath}, {Kerzendorf}, {Littlefair}, {Barentsen}, {D'Eugenio}, \& {Weaver}}]{photutils}
{Bradley}, L., {Sipocz}, B., {Robitaille}, T., {et~al.} 2016, {Photutils: Photometry tools}.
\newblock \doeprint{1609.011}

\bibitem[{{Brammer} {et~al.}(2008){Brammer}, {van Dokkum}, \& {Coppi}}]{Brammer2008}
{Brammer}, G.~B., {van Dokkum}, P.~G., \& {Coppi}, P. 2008, \apj, 686, 1503, \dodoi{10.1086/591786}

\bibitem[{{Bressan} {et~al.}(1993){Bressan}, {Fagotto}, {Bertelli}, \& {Chiosi}}]{Bressan1993}
{Bressan}, A., {Fagotto}, F., {Bertelli}, G., \& {Chiosi}, C. 1993, \aaps, 100, 647

\bibitem[{{Brid{\v{z}}ius} {et~al.}(2008){Brid{\v{z}}ius}, {Narbutis}, {Stonkut{\.{e}}}, {Deveikis}, \& {Vansevi{\v{c}}ius}}]{Bridzius2008}
{Brid{\v{z}}ius}, A., {Narbutis}, D., {Stonkut{\.{e}}}, R., {Deveikis}, V., \& {Vansevi{\v{c}}ius}, V. 2008, Baltic Astronomy, 17, 337, \dodoi{10.48550/arXiv.0902.3167}

\bibitem[{{Bruzual} \& {Charlot}(2003)}]{GALAXEV}
{Bruzual}, G., \& {Charlot}, S. 2003, \mnras, 344, 1000, \dodoi{10.1046/j.1365-8711.2003.06897.x}

\bibitem[{{Bruzual A.} \& {Charlot}(1993)}]{BC_old}
{Bruzual A.}, G., \& {Charlot}, S. 1993, \apj, 405, 538, \dodoi{10.1086/172385}

\bibitem[{{Byler} {et~al.}(2017){Byler}, {Dalcanton}, {Conroy}, \& {Johnson}}]{Byler2017}
{Byler}, N., {Dalcanton}, J.~J., {Conroy}, C., \& {Johnson}, B.~D. 2017, \apj, 840, 44, \dodoi{10.3847/1538-4357/aa6c66}

\bibitem[{{Calzetti} {et~al.}(2000){Calzetti}, {Armus}, {Bohlin}, {Kinney}, {Koornneef}, \& {Storchi-Bergmann}}]{CalzettiDust}
{Calzetti}, D., {Armus}, L., {Bohlin}, R.~C., {et~al.} 2000, \apj, 533, 682, \dodoi{10.1086/308692}

\bibitem[{{Calzetti} {et~al.}(2015{\natexlab{a}}){Calzetti}, {Lee}, {Sabbi}, {Adamo}, {Smith}, {Andrews}, {Ubeda}, {Bright}, {Thilker}, {Aloisi}, {Brown}, {Chandar}, {Christian}, {Cignoni}, {Clayton}, {da Silva}, {de Mink}, {Dobbs}, {Elmegreen}, {Elmegreen}, {Evans}, {Fumagalli}, {Gallagher}, {Gouliermis}, {Grebel}, {Herrero}, {Hunter}, {Johnson}, {Kennicutt}, {Kim}, {Krumholz}, {Lennon}, {Levay}, {Martin}, {Nair}, {Nota}, {{\"O}stlin}, {Pellerin}, {Prieto}, {Regan}, {Ryon}, {Schaerer}, {Schiminovich}, {Tosi}, {Van Dyk}, {Walterbos}, {Whitmore}, \& {Wofford}}]{Calzetti2015}
{Calzetti}, D., {Lee}, J.~C., {Sabbi}, E., {et~al.} 2015{\natexlab{a}}, \aj, 149, 51, \dodoi{10.1088/0004-6256/149/2/51}

\bibitem[{{Calzetti} {et~al.}(2015{\natexlab{b}}){Calzetti}, {Johnson}, {Adamo}, {Gallagher}, {Andrews}, {Smith}, {Clayton}, {Lee}, {Sabbi}, {Ubeda}, {Kim}, {Ryon}, {Thilker}, {Bright}, {Zackrisson}, {Kennicutt}, {de Mink}, {Whitmore}, {Aloisi}, {Chandar}, {Cignoni}, {Cook}, {Dale}, {Elmegreen}, {Elmegreen}, {Evans}, {Fumagalli}, {Gouliermis}, {Grasha}, {Grebel}, {Krumholz}, {Walterbos}, {Wofford}, {Brown}, {Christian}, {Dobbs}, {Herrero}, {Kahre}, {Messa}, {Nair}, {Nota}, {{\"O}stlin}, {Pellerin}, {Sacchi}, {Schaerer}, \& {Tosi}}]{Calzetti2015b}
{Calzetti}, D., {Johnson}, K.~E., {Adamo}, A., {et~al.} 2015{\natexlab{b}}, \apj, 811, 75, \dodoi{10.1088/0004-637X/811/2/75}

\bibitem[{{Carnall} {et~al.}(2018){Carnall}, {McLure}, {Dunlop}, \& {Dav{\'e}}}]{Carnall2018}
{Carnall}, A.~C., {McLure}, R.~J., {Dunlop}, J.~S., \& {Dav{\'e}}, R. 2018, \mnras, 480, 4379, \dodoi{10.1093/mnras/sty2169}

\bibitem[{{Chen} {et~al.}(2010){Chen}, {Liang}, {Hammer}, {Prugniel}, {Zhong}, {Rodrigues}, {Zhao}, \& {Flores}}]{Chen2010}
{Chen}, X.~Y., {Liang}, Y.~C., {Hammer}, F., {et~al.} 2010, \aap, 515, A101, \dodoi{10.1051/0004-6361/200913894}

\bibitem[{{Chisholm} {et~al.}(2019){Chisholm}, {Rigby}, {Bayliss}, {Berg}, {Dahle}, {Gladders}, \& {Sharon}}]{Chisholm2019}
{Chisholm}, J., {Rigby}, J.~R., {Bayliss}, M., {et~al.} 2019, \apj, 882, 182, \dodoi{10.3847/1538-4357/ab3104}

\bibitem[{{Choi} {et~al.}(2016){Choi}, {Dotter}, {Conroy}, {Cantiello}, {Paxton}, \& {Johnson}}]{Choi2016}
{Choi}, J., {Dotter}, A., {Conroy}, C., {et~al.} 2016, \apj, 823, 102, \dodoi{10.3847/0004-637X/823/2/102}

\bibitem[{{Conroy}(2013{\natexlab{a}})}]{Conroy2013}
{Conroy}, C. 2013{\natexlab{a}}, \araa, 51, 393, \dodoi{10.1146/annurev-astro-082812-141017}

\bibitem[{{Conroy}(2013{\natexlab{b}})}]{Conroy2013new}
---. 2013{\natexlab{b}}, \araa, 51, 393, \dodoi{10.1146/annurev-astro-082812-141017}

\bibitem[{{Conroy} \& {Gunn}(2010)}]{Conroy2010}
{Conroy}, C., \& {Gunn}, J.~E. 2010, \apj, 712, 833, \dodoi{10.1088/0004-637X/712/2/833}

\bibitem[{{Conroy} {et~al.}(2009){Conroy}, {Gunn}, \& {White}}]{Conroy2009}
{Conroy}, C., {Gunn}, J.~E., \& {White}, M. 2009, \apj, 699, 486, \dodoi{10.1088/0004-637X/699/1/486}

\bibitem[{{Czekala} {et~al.}(2021){Czekala}, {Loomis}, {Teague}, {Booth}, {Huang}, {Cataldi}, {Ilee}, {Law}, {Walsh}, {Bosman}, {Guzm{\'a}n}, {Le Gal}, {{\"O}berg}, {Yamato}, {Aikawa}, {Andrews}, {Bae}, {Bergin}, {Bergner}, {Cleeves}, {Kurtovic}, {M{\'e}nard}, {Nomura}, {P{\'e}rez}, {Qi}, {Schwarz}, {Tsukagoshi}, {Waggoner}, {Wilner}, \& {Zhang}}]{Czekala2021}
{Czekala}, I., {Loomis}, R.~A., {Teague}, R., {et~al.} 2021, \apjs, 257, 2, \dodoi{10.3847/1538-4365/ac1430}

\bibitem[{{da Cunha} {et~al.}(2008){da Cunha}, {Charlot}, \& {Elbaz}}]{daCunha2008}
{da Cunha}, E., {Charlot}, S., \& {Elbaz}, D. 2008, \mnras, 388, 1595, \dodoi{10.1111/j.1365-2966.2008.13535.x}

\bibitem[{{da Silva} {et~al.}(2012){da Silva}, {Fumagalli}, \& {Krumholz}}]{daSilva2012}
{da Silva}, R.~L., {Fumagalli}, M., \& {Krumholz}, M. 2012, \apj, 745, 145, \dodoi{10.1088/0004-637X/745/2/145}

\bibitem[{{de Barros} {et~al.}(2014){de Barros}, {Schaerer}, \& {Stark}}]{deBarros2014}
{de Barros}, S., {Schaerer}, D., \& {Stark}, D.~P. 2014, \aap, 563, A81, \dodoi{10.1051/0004-6361/201220026}

\bibitem[{{de Jager} {et~al.}(1988){de Jager}, {Nieuwenhuijzen}, \& {van der Hucht}}]{deJagerMassLoss}
{de Jager}, C., {Nieuwenhuijzen}, H., \& {van der Hucht}, K.~A. 1988, \aaps, 72, 259

\bibitem[{{Eggleton}(1971)}]{EggletonCode}
{Eggleton}, P.~P. 1971, \mnras, 151, 351, \dodoi{10.1093/mnras/151.3.351}

\bibitem[{{Eldridge} {et~al.}(2017){Eldridge}, {Stanway}, {Xiao}, {McClelland}, {Taylor}, {Ng}, {Greis}, \& {Bray}}]{Eldridge2017}
{Eldridge}, J.~J., {Stanway}, E.~R., {Xiao}, L., {et~al.} 2017, \pasa, 34, e058, \dodoi{10.1017/pasa.2017.51}

\bibitem[{{Endsley} {et~al.}(2023){Endsley}, {Stark}, {Whitler}, {Topping}, {Chen}, {Plat}, {Chisholm}, \& {Charlot}}]{Endsley2023}
{Endsley}, R., {Stark}, D.~P., {Whitler}, L., {et~al.} 2023, \mnras, 524, 2312, \dodoi{10.1093/mnras/stad1919}

\bibitem[{{Fagotto} {et~al.}(1994{\natexlab{a}}){Fagotto}, {Bressan}, {Bertelli}, \& {Chiosi}}]{Fagotto1994a}
{Fagotto}, F., {Bressan}, A., {Bertelli}, G., \& {Chiosi}, C. 1994{\natexlab{a}}, \aaps, 104, 365

\bibitem[{{Fagotto} {et~al.}(1994{\natexlab{b}}){Fagotto}, {Bressan}, {Bertelli}, \& {Chiosi}}]{Fagotto1994b}
---. 1994{\natexlab{b}}, \aaps, 105, 29

\bibitem[{{Ferland} {et~al.}(2017){Ferland}, {Chatzikos}, {Guzm{\'a}n}, {Lykins}, {van Hoof}, {Williams}, {Abel}, {Badnell}, {Keenan}, {Porter}, \& {Stancil}}]{CLOUDY}
{Ferland}, G.~J., {Chatzikos}, M., {Guzm{\'a}n}, F., {et~al.} 2017, \rmxaa, 53, 385.
\newblock \doarXiv{1705.10877}

\bibitem[{{Fitzpatrick}(1999)}]{Fitz99}
{Fitzpatrick}, E.~L. 1999, \pasp, 111, 63, \dodoi{10.1086/316293}

\bibitem[{{Fitzpatrick} \& {Massa}(2007)}]{Fitzpatrick2007}
{Fitzpatrick}, E.~L., \& {Massa}, D. 2007, \apj, 663, 320, \dodoi{10.1086/518158}

\bibitem[{{Foreman-Mackey} {et~al.}(2013){Foreman-Mackey}, {Hogg}, {Lang}, \& {Goodman}}]{emcee}
{Foreman-Mackey}, D., {Hogg}, D.~W., {Lang}, D., \& {Goodman}, J. 2013, \pasp, 125, 306, \dodoi{10.1086/670067}

\bibitem[{{Foreman-Mackey} {et~al.}(2014){Foreman-Mackey}, {Sick}, \& {Johnson}}]{FSPS_python}
{Foreman-Mackey}, D., {Sick}, J., \& {Johnson}, B. 2014, {python-fsps: Python bindings to FSPS (v0.1.1)}, v0.1.1,  Zenodo, \dodoi{10.5281/zenodo.12157}

\bibitem[{{Fouesneau} {et~al.}(2012){Fouesneau}, {Lan{\c{c}}on}, {Chandar}, \& {Whitmore}}]{Fouesneau2012}
{Fouesneau}, M., {Lan{\c{c}}on}, A., {Chandar}, R., \& {Whitmore}, B.~C. 2012, \apj, 750, 60, \dodoi{10.1088/0004-637X/750/1/60}

\bibitem[{{Girardi} {et~al.}(2000){Girardi}, {Bressan}, {Bertelli}, \& {Chiosi}}]{Girardi2000}
{Girardi}, L., {Bressan}, A., {Bertelli}, G., \& {Chiosi}, C. 2000, \aaps, 141, 371, \dodoi{10.1051/aas:2000126}

\bibitem[{{Gordon} {et~al.}(2023){Gordon}, {Clayton}, {Decleir}, {Fitzpatrick}, {Massa}, {Misselt}, \& {Tollerud}}]{Gordon2023}
{Gordon}, K.~D., {Clayton}, G.~C., {Decleir}, M., {et~al.} 2023, \apj, 950, 86, \dodoi{10.3847/1538-4357/accb59}

\bibitem[{{Goudfrooij} {et~al.}(2010){Goudfrooij}, {Burgh}, {Aloisi}, {Hartig}, \& {Penton}}]{Goudfrooij2010}
{Goudfrooij}, P., {Burgh}, E., {Aloisi}, A., {Hartig}, G., \& {Penton}, S. 2010, {SMOV: COS NUV Imaging Performance}, Instrument Science Report COS 2010-10(v1), 17 pages

\bibitem[{{Groves} {et~al.}(2004){Groves}, {Dopita}, \& {Sutherland}}]{Groves2004}
{Groves}, B.~A., {Dopita}, M.~A., \& {Sutherland}, R.~S. 2004, \apjs, 153, 9, \dodoi{10.1086/421113}

\bibitem[{{Gutkin} {et~al.}(2016){Gutkin}, {Charlot}, \& {Bruzual}}]{Gutkin2016}
{Gutkin}, J., {Charlot}, S., \& {Bruzual}, G. 2016, \mnras, 462, 1757, \dodoi{10.1093/mnras/stw1716}

\bibitem[{{Henyey} {et~al.}(1964){Henyey}, {Forbes}, \& {Gould}}]{HenyeyMethod}
{Henyey}, L.~G., {Forbes}, J.~E., \& {Gould}, N.~L. 1964, \apj, 139, 306, \dodoi{10.1086/147754}

\bibitem[{{Hurley} {et~al.}(2002){Hurley}, {Tout}, \& {Pols}}]{HurleyBinaries}
{Hurley}, J.~R., {Tout}, C.~A., \& {Pols}, O.~R. 2002, \mnras, 329, 897, \dodoi{10.1046/j.1365-8711.2002.05038.x}

\bibitem[{{James} {et~al.}(2022){James}, {Berg}, {King}, {Sahnow}, {Mingozzi}, {Chisholm}, {Heckman}, {Martin}, {Stark}, {Aloisi}, {Amor{\'\i}n}, {Arellano-C{\'o}rdova}, {Bayliss}, {Bordoloi}, {Brinchmann}, {Charlot}, {Chen}, {Chevallard}, {Clark}, {Erb}, {Feltre}, {Hayes}, {Henry}, {Hernandez}, {Jaskot}, {Kewley}, {Kumari}, {Leitherer}, {Llerena}, {Maseda}, {Nanayakkara}, {Ouchi}, {Plat}, {Pogge}, {Ravindranath}, {Rigby}, {Scarlata}, {Senchyna}, {Skillman}, {Steidel}, {Strom}, {Sugahara}, {Wilkins}, {Wofford}, {Xu}, \& {Classy Team}}]{Bethan2022}
{James}, B.~L., {Berg}, D.~A., {King}, T., {et~al.} 2022, \apjs, 262, 37, \dodoi{10.3847/1538-4365/ac8008}

\bibitem[{{Johnson} {et~al.}(2021){Johnson}, {Leja}, {Conroy}, \& {Speagle}}]{Johnson2021}
{Johnson}, B.~D., {Leja}, J., {Conroy}, C., \& {Speagle}, J.~S. 2021, \apjs, 254, 22, \dodoi{10.3847/1538-4365/abef67}

\bibitem[{{Kroupa}(2001)}]{Kroupa2001}
{Kroupa}, P. 2001, \mnras, 322, 231, \dodoi{10.1046/j.1365-8711.2001.04022.x}

\bibitem[{{Kroupa} {et~al.}(1993){Kroupa}, {Tout}, \& {Gilmore}}]{Kroupa1993}
{Kroupa}, P., {Tout}, C.~A., \& {Gilmore}, G. 1993, \mnras, 262, 545, \dodoi{10.1093/mnras/262.3.545}

\bibitem[{{Krumholz} {et~al.}(2015){Krumholz}, {Adamo}, {Fumagalli}, {Wofford}, {Calzetti}, {Lee}, {Whitmore}, {Bright}, {Grasha}, {Gouliermis}, {Kim}, {Nair}, {Ryon}, {Smith}, {Thilker}, {Ubeda}, \& {Zackrisson}}]{Krumholz2015}
{Krumholz}, M.~R., {Adamo}, A., {Fumagalli}, M., {et~al.} 2015, \apj, 812, 147, \dodoi{10.1088/0004-637X/812/2/147}

\bibitem[{{Le Borgne} {et~al.}(2003){Le Borgne}, {Bruzual}, {Pell{\'o}}, {Lan{\c{c}}on}, {Rocca-Volmerange}, {Sanahuja}, {Schaerer}, {Soubiran}, \& {V{\'\i}lchez-G{\'o}mez}}]{LeBorgne}
{Le Borgne}, J.~F., {Bruzual}, G., {Pell{\'o}}, R., {et~al.} 2003, \aap, 402, 433, \dodoi{10.1051/0004-6361:20030243}

\bibitem[{{Leitherer} {et~al.}(2014){Leitherer}, {Ekstr{\"o}m}, {Meynet}, {Schaerer}, {Agienko}, \& {Levesque}}]{Leitherer2014}
{Leitherer}, C., {Ekstr{\"o}m}, S., {Meynet}, G., {et~al.} 2014, \apjs, 212, 14, \dodoi{10.1088/0067-0049/212/1/14}

\bibitem[{{Leitherer} \& {Heckman}(1995)}]{Leitherer1995}
{Leitherer}, C., \& {Heckman}, T.~M. 1995, \apjs, 96, 9, \dodoi{10.1086/192112}

\bibitem[{{Leitherer} {et~al.}(1999){Leitherer}, {Schaerer}, {Goldader}, {Delgado}, {Robert}, {Kune}, {de Mello}, {Devost}, \& {Heckman}}]{Sb99}
{Leitherer}, C., {Schaerer}, D., {Goldader}, J.~D., {et~al.} 1999, \apjs, 123, 3, \dodoi{10.1086/313233}

\bibitem[{{Lejeune} {et~al.}(1997{\natexlab{a}}){Lejeune}, {Cuisinier}, \& {Buser}}]{Lejeune1997}
{Lejeune}, T., {Cuisinier}, F., \& {Buser}, R. 1997{\natexlab{a}}, \aaps, 125, 229, \dodoi{10.1051/aas:1997373}

\bibitem[{{Lejeune} {et~al.}(1997{\natexlab{b}}){Lejeune}, {Cuisinier}, \& {Buser}}]{LejeuneBaSeL}
---. 1997{\natexlab{b}}, \aaps, 125, 229, \dodoi{10.1051/aas:1997373}

\bibitem[{{Lejeune} {et~al.}(1998){Lejeune}, {Cuisinier}, \& {Buser}}]{Lejeune1998}
---. 1998, \aaps, 130, 65, \dodoi{10.1051/aas:1998405}

\bibitem[{{Linden} {et~al.}(2023){Linden}, {Evans}, {Armus}, {Rich}, {Larson}, {Lai}, {Privon}, {U}, {Inami}, {Bohn}, {Song}, {Barcos-Mu{\~n}oz}, {Charmandaris}, {Medling}, {Stierwalt}, {Diaz-Santos}, {B{\"o}ker}, {van der Werf}, {Aalto}, {Appleton}, {Brown}, {Hayward}, {Howell}, {Iwasawa}, {Kemper}, {Frayer}, {Law}, {Malkan}, {Marshall}, {Mazzarella}, {Murphy}, {Sanders}, \& {Surace}}]{Linden2023}
{Linden}, S.~T., {Evans}, A.~S., {Armus}, L., {et~al.} 2023, \apjl, 944, L55, \dodoi{10.3847/2041-8213/acb335}

\bibitem[{{Maraston}(2005)}]{Maraston2005}
{Maraston}, C. 2005, \mnras, 362, 799, \dodoi{10.1111/j.1365-2966.2005.09270.x}

\bibitem[{{Marigo} {et~al.}(2008){Marigo}, {Girardi}, {Bressan}, {Groenewegen}, {Silva}, \& {Granato}}]{Marigo2008}
{Marigo}, P., {Girardi}, L., {Bressan}, A., {et~al.} 2008, \aap, 482, 883, \dodoi{10.1051/0004-6361:20078467}

\bibitem[{{Massa} {et~al.}(2020){Massa}, {Fitzpatrick}, \& {Gordon}}]{Massa2020}
{Massa}, D., {Fitzpatrick}, E.~L., \& {Gordon}, K.~D. 2020, \apj, 891, 67, \dodoi{10.3847/1538-4357/ab6f01}

\bibitem[{{Meynet} {et~al.}(1994){Meynet}, {Maeder}, {Schaller}, {Schaerer}, \& {Charbonnel}}]{MeynetMassLoss}
{Meynet}, G., {Maeder}, A., {Schaller}, G., {Schaerer}, D., \& {Charbonnel}, C. 1994, \aaps, 103, 97

\bibitem[{{Muzzin} {et~al.}(2009){Muzzin}, {Marchesini}, {van Dokkum}, {Labb{\'e}}, {Kriek}, \& {Franx}}]{Muzzin2009}
{Muzzin}, A., {Marchesini}, D., {van Dokkum}, P.~G., {et~al.} 2009, \apj, 701, 1839, \dodoi{10.1088/0004-637X/701/2/1839}

\bibitem[{{Nelson} {et~al.}(2014){Nelson}, {Ford}, \& {Payne}}]{DEMove}
{Nelson}, B., {Ford}, E.~B., \& {Payne}, M.~J. 2014, \apjs, 210, 11, \dodoi{10.1088/0067-0049/210/1/11}

\bibitem[{{Oey} {et~al.}(2010){Oey}, {Hanish}, {Rigby}, \& {de Mello}}]{Oey}
{Oey}, M.~S., {Hanish}, D., {Rigby}, J., \& {de Mello}, D. 2010, in American Astronomical Society Meeting Abstracts, Vol. 215, American Astronomical Society Meeting Abstracts \#215, 309.01

\bibitem[{{Orozco-Duarte} {et~al.}(2022){Orozco-Duarte}, {Wofford}, {Vidal-Garc{\'\i}a}, {Bruzual}, {Charlot}, {Krumholz}, {Hannon}, {Lee}, {Wofford}, {Fumagalli}, {Dale}, {Messa}, {Grebel}, {Smith}, {Grasha}, \& {Cook}}]{Orozco-Duarte2022b}
{Orozco-Duarte}, R., {Wofford}, A., {Vidal-Garc{\'\i}a}, A., {et~al.} 2022, \mnras, 509, 522, \dodoi{10.1093/mnras/stab2988}

\bibitem[{{Osterbrock}(1974)}]{Osterbrock}
{Osterbrock}, D.~E. 1974, {Astrophysics of gaseous nebulae}

\bibitem[{{Pacifici} {et~al.}(2023){Pacifici}, {Iyer}, {Mobasher}, {da Cunha}, {Acquaviva}, {Burgarella}, {Calistro Rivera}, {Carnall}, {Chang}, {Chartab}, {Cooke}, {Fairhurst}, {Kartaltepe}, {Leja}, {Ma{\l}ek}, {Salmon}, {Torelli}, {Vidal-Garc{\'\i}a}, {Boquien}, {Brammer}, {Brown}, {Capak}, {Chevallard}, {Circosta}, {Croton}, {Davidzon}, {Dickinson}, {Duncan}, {Faber}, {Ferguson}, {Fontana}, {Guo}, {Haeussler}, {Hemmati}, {Jafariyazani}, {Kassin}, {Larson}, {Lee}, {Mantha}, {Marchi}, {Nayyeri}, {Newman}, {Pandya}, {Pforr}, {Reddy}, {Sanders}, {Shah}, {Shahidi}, {Stevans}, {Triani}, {Tyler}, {Vanderhoof}, {de la Vega}, {Wang}, \& {Weston}}]{Pacifici2023}
{Pacifici}, C., {Iyer}, K.~G., {Mobasher}, B., {et~al.} 2023, \apj, 944, 141, \dodoi{10.3847/1538-4357/acacff}

\bibitem[{{P{\'e}rez-Gonz{\'a}lez} {et~al.}(2023){P{\'e}rez-Gonz{\'a}lez}, {Barro}, {Annunziatella}, {Costantin}, {Garc{\'\i}a-Argum{\'a}nez}, {McGrath}, {M{\'e}rida}, {Zavala}, {Arrabal Haro}, {Bagley}, {Backhaus}, {Behroozi}, {Bell}, {Bisigello}, {Buat}, {Calabr{\`o}}, {Casey}, {Cleri}, {Coogan}, {Cooper}, {Cooray}, {Dekel}, {Dickinson}, {Elbaz}, {Ferguson}, {Finkelstein}, {Fontana}, {Franco}, {Gardner}, {Giavalisco}, {G{\'o}mez-Guijarro}, {Grazian}, {Grogin}, {Guo}, {Huertas-Company}, {Jogee}, {Kartaltepe}, {Kewley}, {Kirkpatrick}, {Kocevski}, {Koekemoer}, {Long}, {Lotz}, {Lucas}, {Papovich}, {Pirzkal}, {Ravindranath}, {Somerville}, {Tacchella}, {Trump}, {Wang}, {Wilkins}, {Wuyts}, {Yang}, \& {Yung}}]{Gonzalez2023}
{P{\'e}rez-Gonz{\'a}lez}, P.~G., {Barro}, G., {Annunziatella}, M., {et~al.} 2023, \apjl, 946, L16, \dodoi{10.3847/2041-8213/acb3a5}

\bibitem[{{Portegies Zwart} {et~al.}(2010){Portegies Zwart}, {McMillan}, \& {Gieles}}]{Zwart2010}
{Portegies Zwart}, S.~F., {McMillan}, S. L.~W., \& {Gieles}, M. 2010, \araa, 48, 431, \dodoi{10.1146/annurev-astro-081309-130834}

\bibitem[{{Renzini} \& {Fusi Pecci}(1988)}]{Renzini1988}
{Renzini}, A., \& {Fusi Pecci}, F. 1988, \araa, 26, 199, \dodoi{10.1146/annurev.aa.26.090188.001215}

\bibitem[{{Sana} {et~al.}(2012){Sana}, {de Mink}, {de Koter}, {Langer}, {Evans}, {Gieles}, {Gosset}, {Izzard}, {Le Bouquin}, \& {Schneider}}]{SanaReview}
{Sana}, H., {de Mink}, S.~E., {de Koter}, A., {et~al.} 2012, Science, 337, 444, \dodoi{10.1126/science.1223344}

\bibitem[{{Schlafly} \& {Finkbeiner}(2011)}]{Schlafly2011}
{Schlafly}, E.~F., \& {Finkbeiner}, D.~P. 2011, \apj, 737, 103, \dodoi{10.1088/0004-637X/737/2/103}

\bibitem[{{Sirressi} {et~al.}(2022){Sirressi}, {Adamo}, {Hayes}, {Osborne}, {Hernandez}, {Chisholm}, {Messa}, {Smith}, {Aloisi}, {Wofford}, {Fox}, {Mizener}, {Usher}, {Bik}, {Calzetti}, {Sabbi}, {Schinnerer}, {{\"O}stlin}, {Grasha}, {Cignoni}, \& {Fumagalli}}]{Sirressi2022}
{Sirressi}, M., {Adamo}, A., {Hayes}, M., {et~al.} 2022, \aj, 164, 208, \dodoi{10.3847/1538-3881/ac9311}

\bibitem[{{Sirressi} {et~al.}(2024){Sirressi}, {Adamo}, {Hayes}, {Rivera-Thorsen}, {Aloisi}, {Bik}, {Calzetti}, {Chisholm}, {Fox}, {Fumagalli}, {Grasha}, {Hernandez}, {Messa}, {Osborne}, {{\"O}stlin}, {Sabbi}, {Schinnerer}, {Smith}, {Usher}, \& {Wofford}}]{Sirressi2024}
---. 2024, \aj, 167, 166, \dodoi{10.3847/1538-3881/ad29f9}

\bibitem[{{Smith} {et~al.}(2002){Smith}, {Norris}, \& {Crowther}}]{Smith2002}
{Smith}, L.~J., {Norris}, R. P.~F., \& {Crowther}, P.~A. 2002, \mnras, 337, 1309, \dodoi{10.1046/j.1365-8711.2002.06042.x}

\bibitem[{{Tacchella} {et~al.}(2022){Tacchella}, {Finkelstein}, {Bagley}, {Dickinson}, {Ferguson}, {Giavalisco}, {Graziani}, {Grogin}, {Hathi}, {Hutchison}, {Jung}, {Koekemoer}, {Larson}, {Papovich}, {Pirzkal}, {Rojas-Ruiz}, {Song}, {Schneider}, {Somerville}, {Wilkins}, \& {Yung}}]{Tachella2022}
{Tacchella}, S., {Finkelstein}, S.~L., {Bagley}, M., {et~al.} 2022, \apj, 927, 170, \dodoi{10.3847/1538-4357/ac4cad}

\bibitem[{{Tang} {et~al.}(2024){Tang}, {Grasha}, \& {Krumholz}}]{Tang2024}
{Tang}, J., {Grasha}, K., \& {Krumholz}, M.~R. 2024, \mnras, 532, 4583, \dodoi{10.1093/mnras/stae1799}

\bibitem[{ter Braak \& Vrugt(2008)}]{snookerMove}
ter Braak, C. J.~F., \& Vrugt, J.~A. 2008, Statistics and Computing, 18, 435

\bibitem[{{Tinsley}(1968)}]{Tinsley1}
{Tinsley}, B.~M. 1968, \apj, 151, 547, \dodoi{10.1086/149455}

\bibitem[{{Tinsley}(1973)}]{Tinsley2}
---. 1973, \aap, 24, 89

\bibitem[{{Tinsley} \& {Gunn}(1976)}]{Tinsley3}
{Tinsley}, B.~M., \& {Gunn}, J.~E. 1976, \apj, 203, 52, \dodoi{10.1086/154046}

\bibitem[{{Vazdekis} {et~al.}(2010){Vazdekis}, {S{\'a}nchez-Bl{\'a}zquez}, {Falc{\'o}n-Barroso}, {Cenarro}, {Beasley}, {Cardiel}, {Gorgas}, \& {Peletier}}]{Vazdekis2010}
{Vazdekis}, A., {S{\'a}nchez-Bl{\'a}zquez}, P., {Falc{\'o}n-Barroso}, J., {et~al.} 2010, \mnras, 404, 1639, \dodoi{10.1111/j.1365-2966.2010.16407.x}

\bibitem[{{V{\'a}zquez} \& {Leitherer}(2005{\natexlab{a}})}]{Vazquez2005}
{V{\'a}zquez}, G.~A., \& {Leitherer}, C. 2005{\natexlab{a}}, \apj, 621, 695, \dodoi{10.1086/427866}

\bibitem[{{V{\'a}zquez} \& {Leitherer}(2005{\natexlab{b}})}]{Sb99Params}
---. 2005{\natexlab{b}}, \apj, 621, 695, \dodoi{10.1086/427866}

\bibitem[{{Vink} {et~al.}(2001){Vink}, {de Koter}, \& {Lamers}}]{VinkMassLoss}
{Vink}, J.~S., {de Koter}, A., \& {Lamers}, H.~J.~G.~L.~M. 2001, \aap, 369, 574, \dodoi{10.1051/0004-6361:20010127}

\bibitem[{{Walcher} {et~al.}(2011){Walcher}, {Groves}, {Budav{\'a}ri}, \& {Dale}}]{Walcher2011}
{Walcher}, J., {Groves}, B., {Budav{\'a}ri}, T., \& {Dale}, D. 2011, \apss, 331, 1, \dodoi{10.1007/s10509-010-0458-z}

\bibitem[{{Wang} {et~al.}(2024){Wang}, {Leja}, {Atek}, {Labb{\'e}}, {Li}, {Bezanson}, {Brammer}, {Cutler}, {Dayal}, {Furtak}, {Greene}, {Kokorev}, {Pan}, {Price}, {Suess}, {Weaver}, {Whitaker}, \& {Williams}}]{Wang2024}
{Wang}, B., {Leja}, J., {Atek}, H., {et~al.} 2024, \apj, 963, 74, \dodoi{10.3847/1538-4357/ad187c}

\bibitem[{{Westera} {et~al.}(2002){Westera}, {Lejeune}, {Buser}, {Cuisinier}, \& {Bruzual}}]{WesteraLib}
{Westera}, P., {Lejeune}, T., {Buser}, R., {Cuisinier}, F., \& {Bruzual}, G. 2002, \aap, 381, 524, \dodoi{10.1051/0004-6361:20011493}

\bibitem[{{Whitler} {et~al.}(2023){Whitler}, {Stark}, {Endsley}, {Leja}, {Charlot}, \& {Chevallard}}]{Whitler2023}
{Whitler}, L., {Stark}, D.~P., {Endsley}, R., {et~al.} 2023, \mnras, 519, 5859, \dodoi{10.1093/mnras/stad004}

\bibitem[{{Wofford} {et~al.}(2011){Wofford}, {Leitherer}, \& {Chandar}}]{Wofford2011}
{Wofford}, A., {Leitherer}, C., \& {Chandar}, R. 2011, \apj, 727, 100, \dodoi{10.1088/0004-637X/727/2/100}

\bibitem[{{Wofford} {et~al.}(2013){Wofford}, {Leitherer}, \& {Salzer}}]{Wofford2013}
{Wofford}, A., {Leitherer}, C., \& {Salzer}, J. 2013, \apj, 765, 118, \dodoi{10.1088/0004-637X/765/2/118}

\bibitem[{{Wofford} {et~al.}(2016){Wofford}, {Charlot}, {Bruzual}, {Eldridge}, {Calzetti}, {Adamo}, {Cignoni}, {de Mink}, {Gouliermis}, {Grasha}, {Grebel}, {Lee}, {{\"O}stlin}, {Smith}, {Ubeda}, \& {Zackrisson}}]{Wofford}
{Wofford}, A., {Charlot}, S., {Bruzual}, G., {et~al.} 2016, \mnras, 457, 4296, \dodoi{10.1093/mnras/stw150}

\bibitem[{{Xiao} {et~al.}(2018){Xiao}, {Stanway}, \& {Eldridge}}]{Xiao2018}
{Xiao}, L., {Stanway}, E.~R., \& {Eldridge}, J.~J. 2018, \mnras, 477, 904, \dodoi{10.1093/mnras/sty646}

\bibitem[{{Zackrisson} {et~al.}(2001){Zackrisson}, {Bergvall}, {Olofsson}, \& {Siebert}}]{Zackrisson2001}
{Zackrisson}, E., {Bergvall}, N., {Olofsson}, K., \& {Siebert}, A. 2001, \aap, 375, 814, \dodoi{10.1051/0004-6361:20010912}

\bibitem[{{Zackrisson} {et~al.}(2011){Zackrisson}, {Rydberg}, {Schaerer}, {{\"O}stlin}, \& {Tuli}}]{YGGDRASIL}
{Zackrisson}, E., {Rydberg}, C.-E., {Schaerer}, D., {{\"O}stlin}, G., \& {Tuli}, M. 2011, \apj, 740, 13, \dodoi{10.1088/0004-637X/740/1/13}

\end{thebibliography}
\bibliographystyle{aasjournal}

\end{document}